%% file: main_arkiv.tex
\documentclass[twocolumn]{aastex631}
\usepackage{amsmath}
\begin{document}
\title{The Galactic Center in Color: Measuring Extinction with High-Proper-Motion Stars}
\author[0009-0004-0026-7757]{Zo\"e Haggard}
\affiliation{UCLA Galactic Center Group, Physics and Astronomy Department, University of California, Los Angeles, CA 90095, USA}
\author[0000-0003-3230-5055]{Andrea M. Ghez}
\affiliation{UCLA Galactic Center Group, Physics and Astronomy Department, University of California, Los Angeles, CA 90095, USA}
\author[0000-0001-5972-663X]{Shoko Sakai}
\affiliation{UCLA Galactic Center Group, Physics and Astronomy Department, University of California, Los Angeles, CA 90095, USA}
\author[0000-0002-2836-117X]{Abhimat K. Gautam}
\affiliation{UCLA Galactic Center Group, Physics and Astronomy Department, University of California, Los Angeles, CA 90095, USA}
\author[0000-0001-9554-6062]{Tuan Do}
\affiliation{UCLA Galactic Center Group, Physics and Astronomy Department, University of California, Los Angeles, CA 90095, USA}
\author[0000-0001-9611-0009]{Jessica R. Lu}
\affiliation{University of California, Berkeley, Department of Astronomy, Berkeley, CA 94720, USA}
\author[0000-0003-2874-1196]{Matthew Hosek}
\affiliation{UCLA Galactic Center Group, Physics and Astronomy Department, University of California, Los Angeles, CA 90095, USA}
\author[0000-0002-6753-2066]{Mark R. Morris}
\affiliation{UCLA Galactic Center Group, Physics and Astronomy Department, University of California, Los Angeles, CA 90095, USA}
\author[0009-0005-0419-4038]{Sean Granados}
\affiliation{UCLA Galactic Center Group, Physics and Astronomy Department, University of California, Los Angeles, CA 90095, USA}

\begin{abstract}

The Milky Way's central parsec is a highly extinguished region with a population of high-proper-motion stars. We have tracked 145 stars for $\sim$10 years at wavelengths between 1 and 4 microns to analyze extinction effects in color-magnitude space. Approximately $30\%$ of this sample dims and reddens over the course of years, likely from the motion of sources relative to an inhomogeneous screen of dust. We correct previous measurements of the intrinsic variability fraction for differential extinction effects, resulting in a reduced stellar variability fraction of $34\%$. The extinction variability sub-sample shows that the extinguishing material has sub-arcsecond scales, much smaller variations than previously reported. The observed extinction events imply a typical cross-section of 500 AU and a density of around $3 \times 10^{4} \ \mathrm{atoms/cm^{3}}$ for the extinguishing material, which are consistent with measurements of filamentary dust and gas at the Galactic Center. Furthermore, given that the stars showing extinction variability tend to be more highly reddened than the rest of the sample, the extinction changes are likely due to material localized to the Galactic Center region. We estimate the relative extinction between 1 and 4 microns as, $\mathrm{A}_{\mathrm{H}}:\mathrm{A}_{\mathrm{K'}}:\mathrm{A}_{\mathrm{L'}} = 1.67 \pm 0.05:1:0.69 \pm 0.03$. Our measurement of extinction at longer wavelengths -- L' (3.8 $\mu$m) -- is inconsistent with recent estimations of the integrated extinction towards the central parsec. One interpretation of this difference is that the dust variations this experiment is sensitive to -- which are local to the Galactic Center -- are dominated by grains of larger radius than the foreground.
\end{abstract}

\section{Introduction} \label{sec:intro}

The Galactic Center (GC) is a greatly extinguished region ($A_{V} > 30$, e.g. \citealt{1989_Rieke}, \citealt{2003ApJ...594..294S}, \citealt{ext_gc_visual}). Only in the near-infrared, where extinction is an order of magnitude smaller, have observations revealed the presence of high-proper-motion stars in central parsec. Tracking the movement of these stars over decades has resulted in various discoveries, such as direct evidence for a $4 \times 10^6M_{\odot}$ supermassive black hole (e.g. \citealt{Ghez_2008}, \citealt{2009ApJ...692.1075G}), the dynamical structure of the nuclear star cluster (e.g. \citealt{2009_Lu}, \citealt{2006_Paumard}), and the detection of GR effects (e.g. \citealt{GRAVITY_2018}, \citealt{Do_2019}). Recent work has even suggested that the GC environment -- the dust and extinguishing material -- could be studied using these high-proper-motion stars.

With astrometric imaging data sets from the Galactic Center Orbits Initiative (GCOI), \citealt{Gautam_2019} found that stars at the GC tend to be more photometrically variable than those in the solar neighborhood. They proposed that the high variability fraction could be due to inhomogeneities in the extinction screen as fast-moving stars travel behind it. While the bulk of the extinction toward the GC is due to non-local foreground material (e.g. \citealt{NL_2021}), studies of the central region have found that the extinguishing material has variable structure on arcsecond scales (e.g. \citealt{Schodel2010}, \citealt{Nogueras-Lara_2019}). In the central half parsec, large-scale dusty and gaseous features such as the \textit{Minispiral} and the IRS 13 complex are apparent into the mid-infrared (e.g. \citealt{2005_Moultaka}). Smaller scale features are also seen. For instance, \citealt{Mu_i__2007} observed fast-moving filaments with sizes of thousands of AU; \citealt{Ciurlo_2023} tracked the motion of X7, a gaseous and dusty bow shock source. If a fast-moving star at the GC were to pass behind one of these many features, its light would be dimmed and reddened over the span of years. Nevertheless, with only single-bandpass measurements in K', it was not possible for \citealt{Gautam_2019} to test this extinction variability hypothesis.

In this paper, we report long time-baseline color measurements (H-K' and K'-L') of high-proper-motion stars. This dataset allows us to examine not only the impact of extinction on the long-term photometric variability, but also the densities of the extinguishing features and the extinction law localized to the GC. The organization of this paper is as follows: Section \ref{sec:data} outlines the data, Section \ref{sec:sample} details the stellar sample selection, Section \ref{sec:results} describes the methodology for measuring color variability, Sections \ref{sec:discussion} examines the implications of our color variability measurements, and finally, Section \ref{sec:summary} summarizes our work. 

\section{Data} \label{sec:data}

\begin{figure*}[ht]
\includegraphics[width=\textwidth]{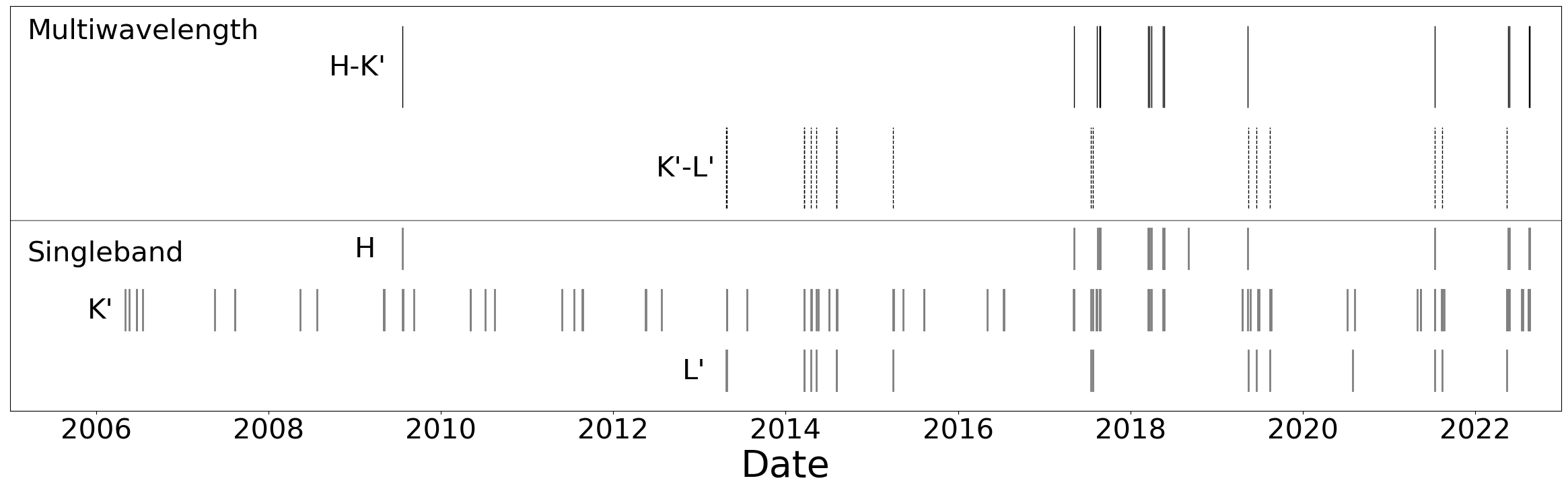}
\caption{The timing of the photometric measurements. The top panel shows the dates of the color measurements and the bottom marks the dates of both the single-band measurements that go into the color measurements as well as the supplemental K' photometry.}
\label{fig:observing_cadence}
\end{figure*}

The fundamental data used in this study are photometric measurements derived from images taken as part of W. M. Keck Observatory's Galactic Center Orbits Initiative (GCOI; PI Ghez). All of the GCOI images are obtained with the laser guide star adaptive optics system (\citealt{Wizinowich_2006}, \citealt{vanDam_2006}) and the facility near-infrared camera (NIRC2) in its narrow field mode (10'' x 10'') using one of the following three band-pass filters: H ($\mathrm{\lambda}_{0}=1.633$ $\mathrm{\mu}$m, $\mathrm{\Delta\lambda}_{0}=0.296$ $\mathrm{\mu}$m), K' ($\mathrm{\lambda}_{0}=2.124$ $\mathrm{\mu}$m, $\mathrm{\Delta\lambda}_{0}=0.351$ $\mathrm{\mu}$m), and L' ($\mathrm{\lambda}_{0}=3.776$ $\mathrm{\mu}$m, $\mathrm{\Delta\lambda}_{0}=0.776$ $\mathrm{\mu}$m). 
In this paper we report both new- and a re-analyses of L' observations (Section \ref{sec:lp_data}) and we combine the resulting L' photometry with photometric measurements through the H and K’ filters from \citealt{gautam2024} (see Section \ref{sec:data_multi-wavelength}). Figure \ref{fig:observing_cadence} provides an overview of the timing of these photometric observations.

\subsection{L' Image Analysis \& Photometry} \label{sec:lp_data}
The L' observations span 17 nights (see Table \ref{table:KL_measurements}), of which 10 are newly reported. The remaining seven were first reported by \citealt{Witzel_2017} and \citealt{Ciurlo_2023}, and are reprocessed here such that the analysis is consistent across our entire data set.

Our data processing and source extraction follows \citealt{Stolte_2010} with a few minor modifications. In the image calibration stage, we add non-linearity corrections, as described in \citealt{Metchev_2009}; the impact on photometry is smaller than the photometric errors. Stars are identified and characterized using the iterative PSF fitting code StarFinder (see \citealt{Diolaiti_2000}) as implemented in \citealt{gautam2024}. For the L’ data, we reduce the PSF support size from 3" to 0.9" to reduce the effect of the structured background on the PSF construction. We also lower the correlation threshold by 0.1 to 0.6 and 0.7 for the submaps and combination image, respectively. Together, these changes improve the quality of source detections, as measured by image residuals and photometric stability.  

In this analysis, a final check is implemented on the robustness of the individual L' detections by checking the L' positions against the K’ proper motions catalog associated with \citealt{gautam2024}. To do this, the L’ source measurements across individual epochs are placed in a common reference frame using the techniques outlined in \citealt{Jia_2019} and the reference frame constructed by \citealt{Sakai_2019}. To avoid measurements that are matched incorrectly, any L’ detection with astrometry values differing by more than a 0.5 pixel (5 mas) offset and 5$\mathrm{\sigma}$ from the K’ model is dropped. This removes only a small percentage ($\sim5$\%) of the L’ detections. 

\begin{figure}[ht]
\centering
\includegraphics[width=0.45\textwidth]{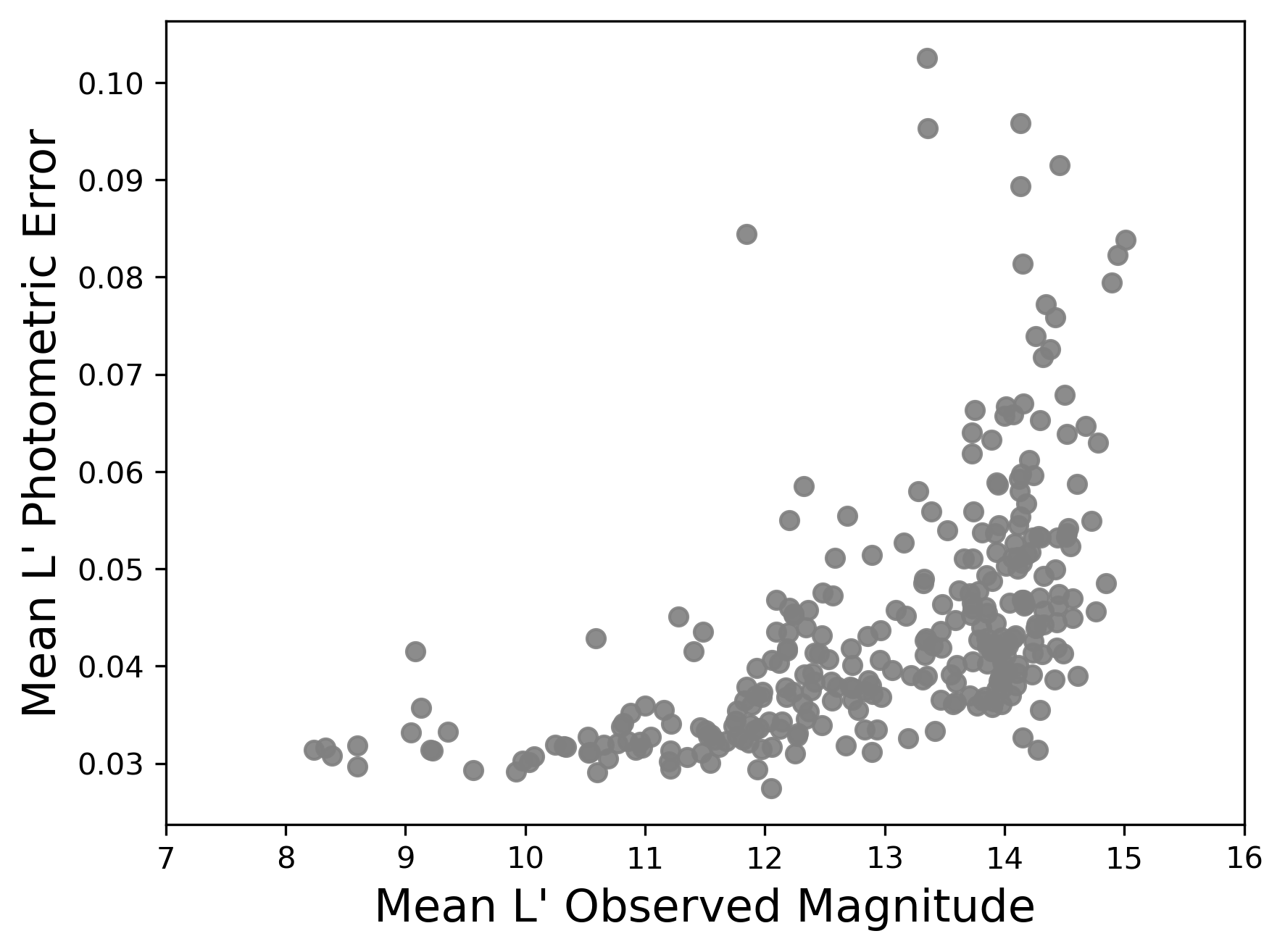}
\caption{Average L' relative photometric uncertainty as a function of stellar brightness. The points plotted are the average uncertainties for each star's relative brightness across all epochs and include both the statistical and zeropoint uncertainties. Only the non-variable stars highlighted in Table \ref{table:non_var_lp} are shown.}
\label{fig:Lp_photometric_quality}
\end{figure}

% The photometric uncertainties for these stars from \citealt{Schodel2010} is $\sim$0.04 mag (statistical) and $\sim$0.15 mag (zero-point). 
Our relative photometric calibration is based on the magnitude of six stars measured with NACO by \citealt{Schodel2010}. We select these stars because they are non-variable across our L' baseline based on the same reduced chi-squared metric used in \citealt{Gautam_2019}, are detected in all of our L' images, and are spread across the field-of-view (see Table \ref{table:lp_photometry} for relative calibrator information). Because our measurements are made through the NIRC2 L’-bandpass filter, which has slightly different characteristics than the NACO L’-bandpass filter, a bandpass correction (L’ NACO - L’ NIRC2) of $\sim$0.04 mag is applied, which just results in a bulk offset to the relative photometry. The photometric error floor for our calibration is $\sim$0.03 magnitudes and is dominated by the zero-point error for the brightest stars. While we do apply a local photometric correction as described in \citealt{Gautam_2019}, on average, it is smaller than our photometric uncertainties ($\sim$0.01 mag). The photometric uncertainty for all non-variable stars (see Table \ref{table:non_var_lp}) as a function of L' magnitude is shown in Figure \ref{fig:Lp_photometric_quality}.

\input{lp_phot.tex}

\subsection{Color measurements} \label{sec:data_multi-wavelength}

Our color measurements are composed of multi-wavelength photometric measurements taken through three different band-pass filters, either H\&K' or K'\&L'. We allow a maximum time difference ($\mathrm{\Delta} \mathrm{t_{max}}$) of six days between observations to derive each color measurement. For this dataset, there are 16 nights of H/K' same-night interleaved observations and 2 nights of nearest matches ($\mathrm{\Delta} \mathrm{t_{max}}$ = 2 day) for a total of 18 H-K' epochs (see Table \ref{table:HK_measurements}). For K'/L', there are 11 nights of same-night interleaved measurements and 6 nearest-night matches (all but one have $\mathrm{\Delta} \mathrm{t_{max}}$ = 2 day) for a total of 17 K'-L' epochs (see Table \ref{table:KL_measurements}). In addition to the 34 K' photometric measurements used in our color data sets, we also make use of the 66 additional K' measurements from \citealt{gautam2024}. For the brightest stars, the color uncertainties are 0.06 mags for H-K' and 0.08 mags for K'-L', and are dominated by the H and L' photometric errors respectively. 

\section{Sample Selection} \label{sec:sample}

Our primary sample is composed of stars that are photometrically variable at K' and have enough multi-wavelength data to study their color variations. These stars are a subset of the 1129 stars detected at K' in \citealt{gautam2024} and are selected based on the following two criteria:

\begin{itemize}
    \item At least 4 H-K' measurements. A total of 981 stars satisfy this first criteria, of which 517 are brighter than K' = 16 mag.  
    \item Variable flux at K' based on a reduced chi-squared ($\mathrm{\chi}_{\mathrm{\nu}}^2$) metric using the time baseline sampling of the H-K' data:
        \begin{equation}
        \centering
        \mathrm{\chi}_{\mathrm{\nu}}^2 = \frac{1}{\mathrm{\nu}}\Sigma{\frac{(\mathrm{F}_{i}-\mathrm{\overline{F}})^2}{\mathrm{\sigma}_{i}^2}}
    \end{equation}
    where $\mathrm{\nu}$ the the number of degrees of freedom for each source, $\mathrm{\overline{F}}$ is the weighted average flux, and $\mathrm{F}_{i}$ and $\mathrm{\sigma}_{i}$ are the flux and error on each individual measurement. We define variable sources as those that have a probability of being non-variable with a significance of at least $5\mathrm{\sigma}$. This produces the final sample of variable stars, of which 90 are brighter than K' = 16 mag\footnote{The variability fraction implied from our bright star ($\mathrm{K'} \leq 16$) sample is smaller than presented in \citealt{Gautam_2019} (20\% vs. 50\%) owing to the less frequent sampling as well as the effective shorter time baseline in this study.}.
  
\end{itemize}

This results in a final sample of 145 stars. These variable stars experience magnitude changes over their H-K' color sampling that are as large as $\mathrm{\Delta}\mathrm{{K'}} = 1.6$ magnitudes, where $\mathrm{\Delta}{\mathrm{K'}} = \mathrm{K'}_{\mathrm{min}} - \mathrm{K'}_{\mathrm{max}}$. The stars in the sample and their characteristics are detailed in Table \ref{table:extinction} and Table \ref{table:non-extinction} in Appendix \ref{appendix:non-extinction_results}. The criterion that sorts the sample into two different tables is explained in Section \ref{sec:GMM} 

\section{Results} \label{sec:results}
\subsection{Characterizing color variability for individual stars} \label{sec:line_fitting}
\begin{figure*}[ht]
\centering
\includegraphics[width=0.85\textwidth]{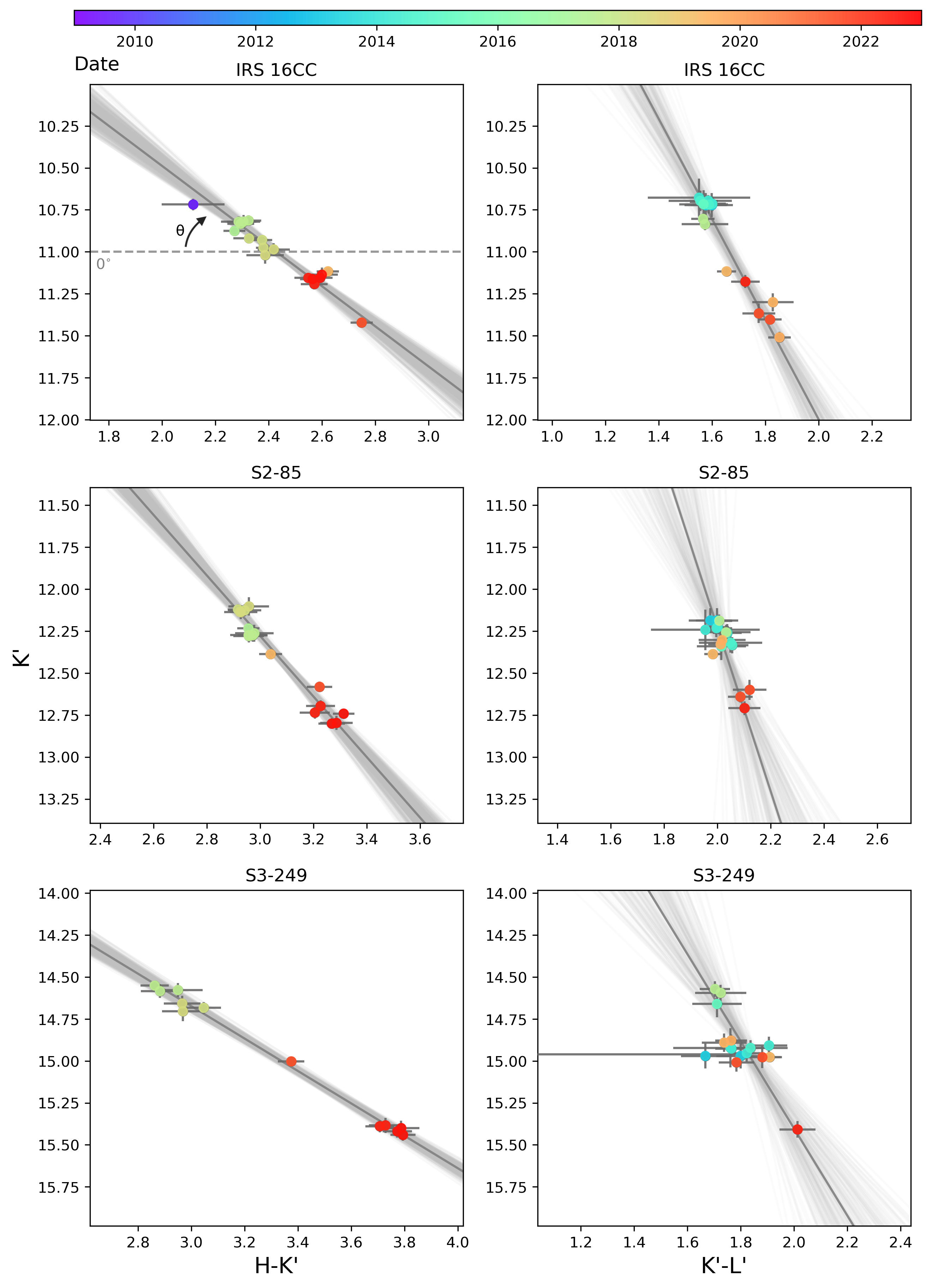}
\caption{Color-magnitude relationships for several stars that show significant color variations. On the left are the K'/H-K' color-magnitude, and on the right are the K'/K'-L' diagrams. 
Plotted are individual measurements for each star as well as the best-fit linear models; the linear models are parameterized by an angle $\mathrm{\theta}$ from the horizontal. The over-plotted grey lines are 100 randomly sampled points from the MCMC chains.}
\label{fig:fitting_example}
\end{figure*}

Changes in the colors of individual stars in our sample are generally well described by a line in the color vs. magnitude diagram (CMD), as shown for three exemplar stars in Figure \ref{fig:fitting_example}. We fit the measurements in the CMDs with a linear model following the approach described in \citealt{Hogg2010DataAR}.

This method, described in detail in Appendix \ref{sec:linear_fitting_results}, has two important attributes. First, it accounts for uncertainties in both the color- and magnitude-measurements along with correlations between the two. Second, this method parameterizes the linear model ($\mathrm{y=sx+b}$) with two variables, $\mathrm{\theta}$ and $\mathrm{b}$. The slope, $\mathrm{s}$, and $\mathrm{\theta}$ are related by the following relationship:
\begin{equation}
    \centering
    \mathbf{\hat{v}} = \frac{1}{\sqrt{1+\mathrm{s}^2}} \begin{bmatrix} -\mathrm{s}\\ 1\end{bmatrix} = \begin{bmatrix} -\mathrm{sin{\theta}}\\ \mathrm{cos{\theta}} \end{bmatrix}
    \label{equation:angle_to_slope}
\end{equation}
The resulting linear models are over-plotted for example stars in Figure \ref{fig:fitting_example} as well as in Appendix \ref{sec:linear_fitting_results}. At best, the slope uncertainties are $\sim$ 1 degree for K'/H-K' and $\sim$ 3 degrees for K'/K'-L'. We also use this line-fit to provide the expected color value at the median K'. Table \ref{table:extinction} and Table \ref{table:non-extinction} provide the resulting line-fit values.

\subsection{Identifying color variability caused primarily by extinction} \label{sec:GMM}
\begin{figure*}[ht]
\centering
\includegraphics[width=0.99\textwidth]{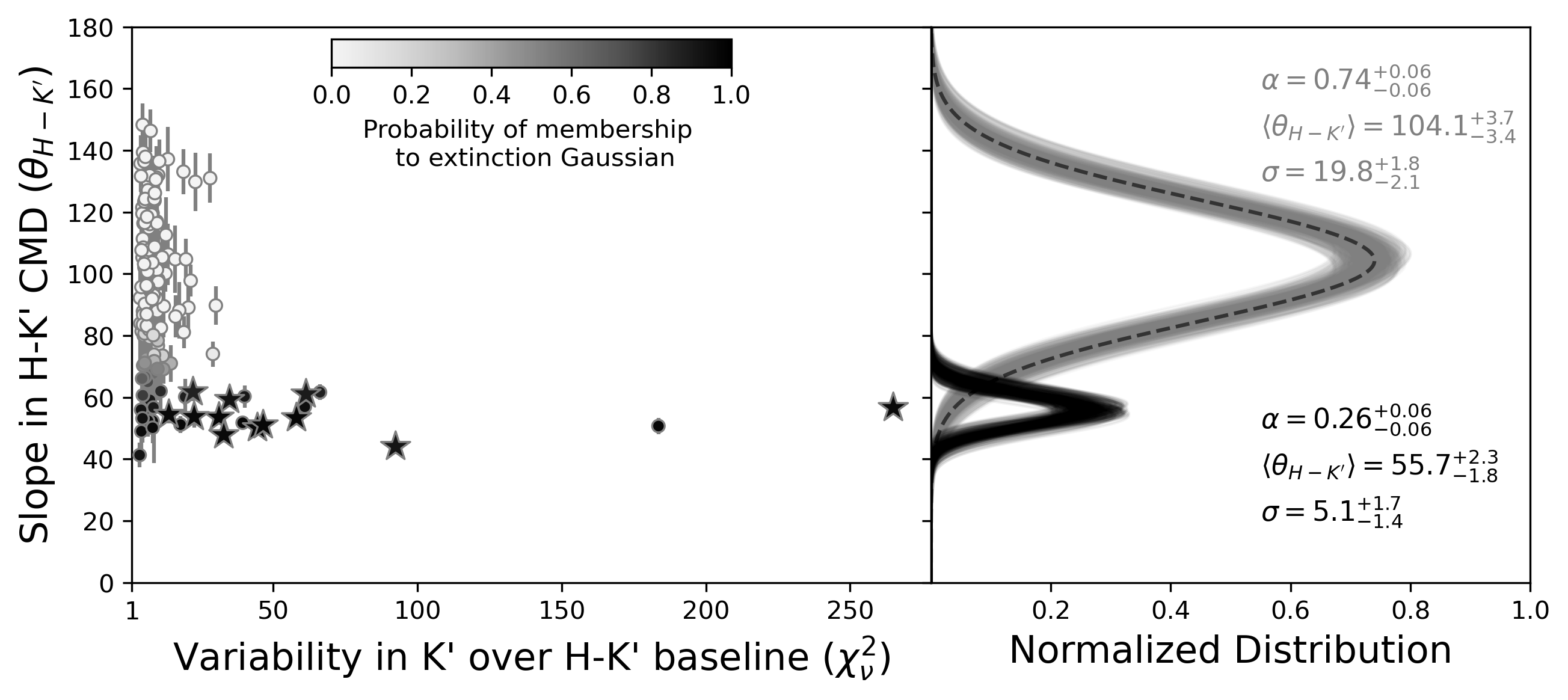}
\caption{(Left) Slopes ($\mathrm{\theta}_{\mathrm{H-K'}}$) in CMD space against K' variability metric ($\mathrm{\chi}^2_{\mathrm{\nu}}$) over the H-K' color baseline. Sources more likely to belong to the extinction Gaussian are colored in black, while those that are less likely are colored in white. As a reference, points marked with a star are considered in the K'-L' modeling. (Right) Results of error-corrected Gaussian mixture model (ecGMM) for the entire H-K' population of slopes in the CMD. The model prefers two groupings. In H-K', the extinction population -- in black -- has slopes around $55 \pm 5$ degrees and this group comprises around 30\% of the sample. The rest of the population -- represented by the grey Gaussian -- has $\langle{\mathrm{\theta}_{\mathrm{H-K'}}}\rangle$ inconsistent with extinction variability effects.}
\label{fig:slope_results_hkp}
\end{figure*}

Since extinction effects are most apparent at shorter wavelengths, we use the H-K' results in order to identify sources with variability that arises from extinction. Extinction effects -- dimming and reddening -- will manifest as slopes less than 90 degrees in the CMD. For instance, towards the Galactic Center, the results of \citealt{Fritz_2011} imply an extinction CMD slope of $53 \pm 3$ degrees in K'/H-K'. The left panel of Figure \ref{fig:slope_results_hkp} plots the slopes for the sample estimated from the H-K'/K' CMD ($\mathrm{\theta}_{\mathrm{H-K'}}$) compared to the significance of each source's K' variability  ($\mathrm{\chi}_{\mathrm{\nu}}^{2}$, see Section \ref{sec:sample}). A common slope value of $\sim 50^{\circ}$ is clearly apparent for the most significantly varying stars. This suggests that the color variability for the most photometrically variable sources is due to changing levels of extinction.

To independently derive the characteristic slope of the sub-population consistent with extinction effects, we model the sample's slope measurements with an error-corrected Gaussian mixture model (ecGMM; \citealt{Hao_2009}, see Appendix \ref{sec:ecGMM}). The best-fit model is a mixture of two Gaussians, which are described by the following 5 parameters: the mean ($\langle{\mathrm{\theta}}\rangle$) and standard deviation ($\mathrm{\sigma}$) for each Gaussian and the weight ($\mathrm{\alpha}$) associated with the extinction-like population. The right panel of Figure \ref{fig:slope_results_hkp} shows final best fit ecGMM and its uncertainties which are derived through a bootstrap process. The Gaussian component associated with extinction effects has a mean slope value of $\langle{\mathrm{\theta}_{\mathrm{H-K'}}\rangle} = 56 \pm 2$ degrees and its weight implies that $26\% \pm 6\%$ of the total variability sample is dominated by extinction effects.

To study extinction as a function of wavelength, we select sources whose variability is likely dominated by extinction based on their shorter-wavelength color variations. To pass the criterion for selection, the star's probability of belonging to the H-K' extinction Gaussian must be greater than 50\%. In total, we find 36 sources consistent with extinction variability in H-K' and they are reported in Table \ref{table:extinction}.

\begin{figure*}[ht]
\centering
\includegraphics[width=0.99\textwidth]{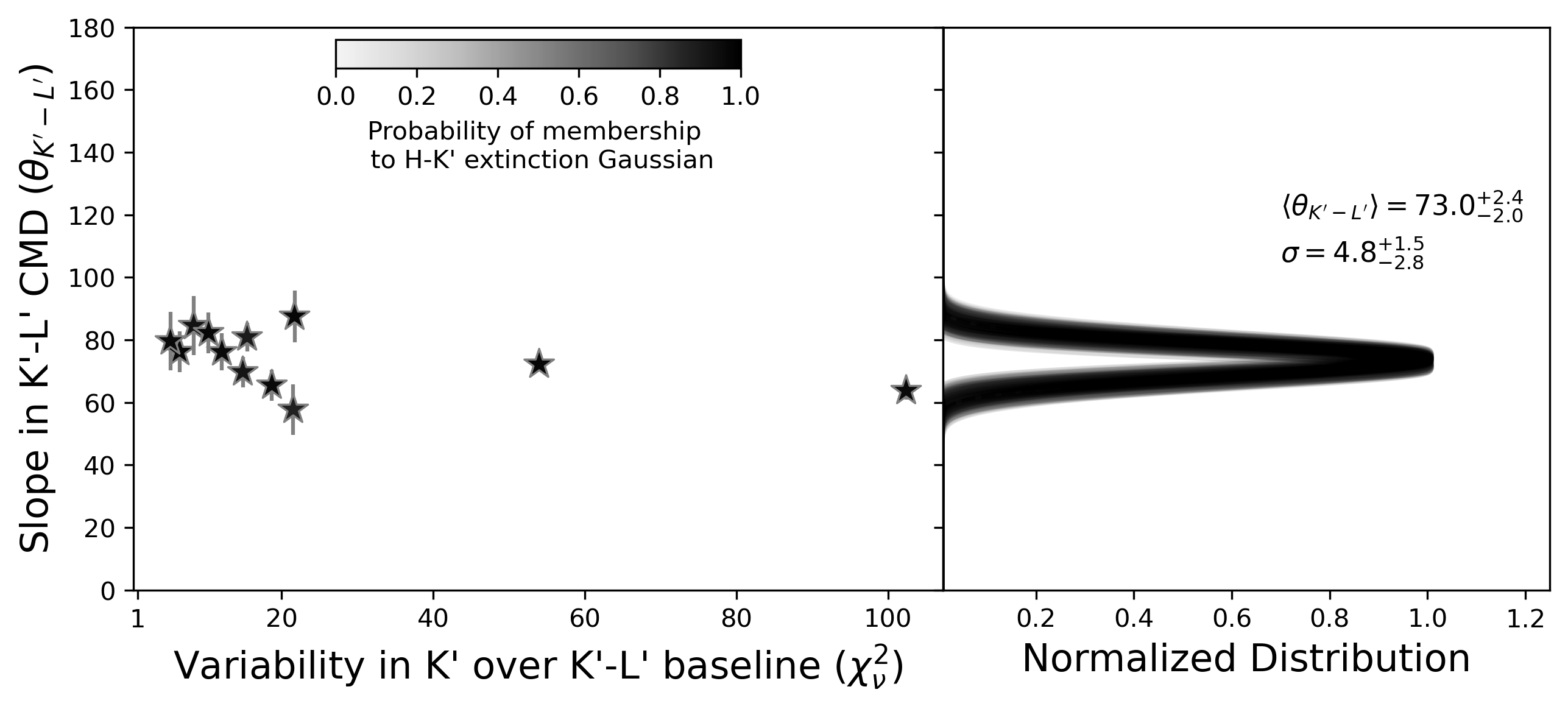}
\caption{(Left) Slopes ($\mathrm{\theta}_{\mathrm{K'-L'}}$) in CMD space against the K' variability ($\mathrm{\chi}^2_{\mathrm{\nu}}$) for the 12 sources that are variable in K' over the K'-L' sampling and are consistent with extinction variability in H-K'. (Right) Results of the one-component error-corrected Gaussian mixture model (ecGMM). The model prefers a mean slope $\langle{\mathrm{\theta}_{\mathrm{K'-L'}}}\rangle$ around $73 \pm 2$ degrees.}
\label{fig:slope_results_kplp}
\end{figure*}

Of these 36 extinction sources identified in H-K', 12 are also found to be variable over their color sampling in K'-L' (these are marked with stars in Figure \ref{fig:slope_results_hkp}). The slopes for these 12 sources, which we refer to as ``K'-L' extinction-sources" are reported in column 8 of Table \ref{table:extinction} and shown in the left panel of Figure \ref{fig:slope_results_kplp}. To model the extinction characteristics at longer wavelengths, we model the "K'-L' extinction sources" with a single Gaussian using ecGMM and find a best fit mean slope of $\langle{\mathrm{\theta}_{\mathrm{K'-L'}}\rangle} = 73 \pm 2$ degrees in the K'/K'-L' CMD (see right panel of Figure \ref{fig:slope_results_kplp}).

\input{extinction.tex}

\section{Discussion} \label{sec:discussion}

\subsection{Color distributions of Galactic Center stars} \label{section:color_GC}
\begin{figure}[ht]
\centering
\includegraphics[width=0.45\textwidth]{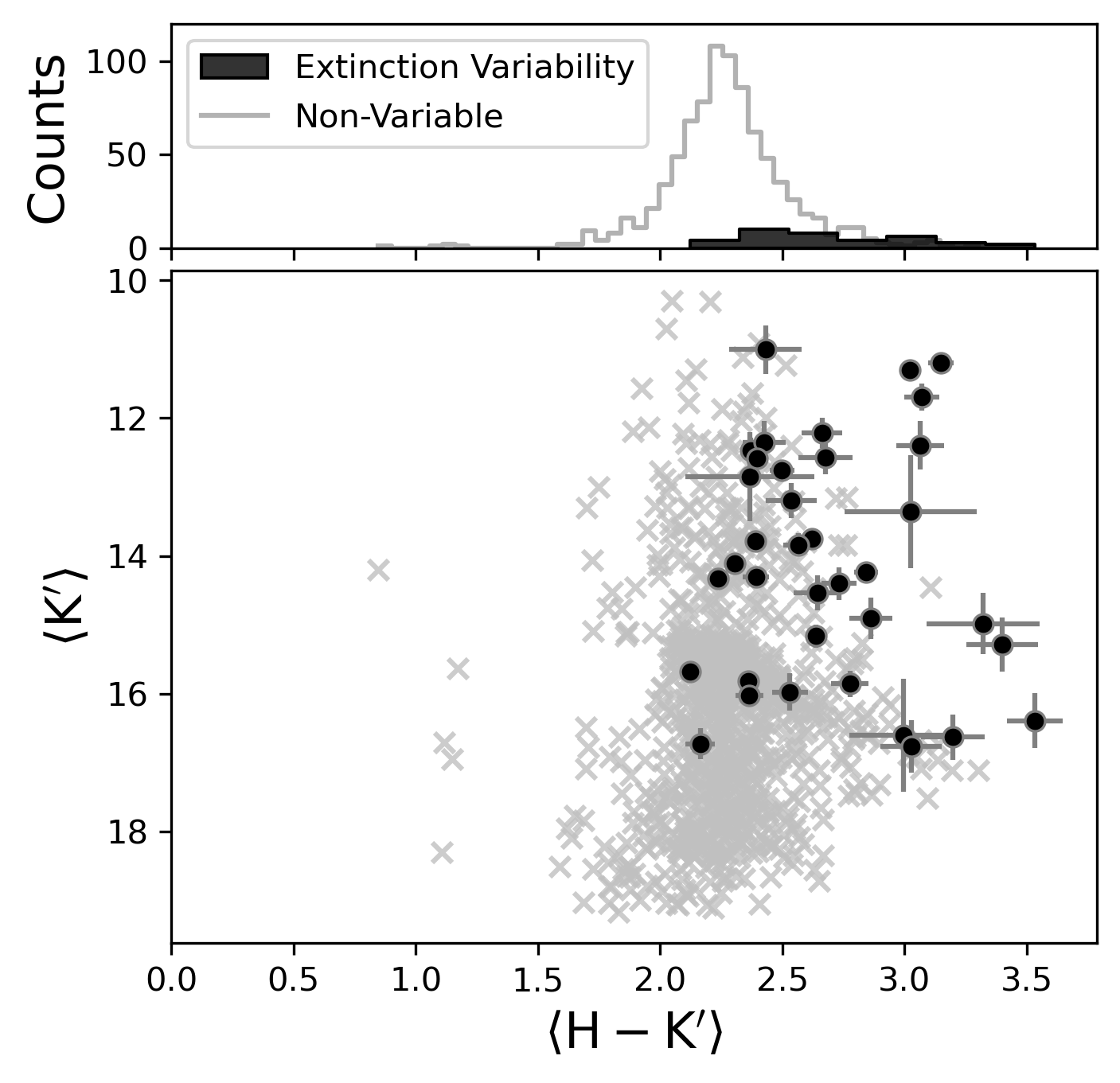}
\caption{(Bottom) Color-magnitude diagram (mean K' vs. mean H-K') of extinction variable sources (marked in black) and non-variables found in H-K' (marked in grey crosses). The error bars on the extinction sources are derived from the observed $\Delta{\mathrm{K'}}$. (Top) Histogram of mean H-K' color for non-variables (in grey) and sources that have variability consistent with extinction (in shaded black).}
\label{fig:CMD_sample}
\end{figure}

Most of the bulk extinction towards the Galactic Center occurs before 8 kpc (e.g. \citealt{NL_2021}). Our experiment, however, is not sensitive to this bulk extinction screen; instead, we probe \textit{changes} in extinction as fast-moving Galactic Center stars pass behind extinguishing structures along the line-of-sight. These structures are likely to be more localized to the GC, as the feature must be angularly small enough to dim the light of a single star. If stars in our experiment are also local to (or behind) the GC, they will tend to be more reddened than the non-variables. As shown in Figure \ref{fig:CMD_sample}, we find that stars with variability dominated by changing extinction \textit{do} tend to be redder: the extinction variability population has a mean H-K' color of 2.7 mags compared to 2.2 for non-variables. The color distributions of these two populations is also significantly different. While extinction sources are far fewer in number, according to a two-sided Kolmogorov–Smirnov test, the extinction variability distribution is significantly different from that of the non-variables, with a p-value much smaller than 0.05 of $2\times10^{-12}$. Extinction sources, then, are unlikely to be foreground to the GC: they tend to be quite red and have a distribution dissimilar to the non-variables.

\subsection{Variability fraction of stars at the Galactic Center}

$26\% \pm 7\%$ of the entire sample's K' variable sources show long-term color variability dominated by extinction effects, which suggests that the true intrinsic variability fraction is lower than estimated from single-wavelength observations. In order to do an equal comparison to \citealt{Gautam_2019}, we only look at stars that are brighter than $\mathrm{K'} = 16$; this reduces the sample to 85 sources. Re-running the ecGMM as described in \ref{sec:ecGMM}, we find that $32\% \pm 10\%$ of the reduced sample shows long-term variability dominated by extinction effects. Taking the variability fraction of $50 \pm 2\%$ over 12 years estimated by \citealt{Gautam_2019}, we can approximate the intrinsic photometric variability of stars in the Galactic Center at around $34\% \pm 10\%$. While still high, this result is more consistent with the variability fraction of stars observed outside the Galactic Center region, which is between $1$ and $6\%$ for late-type clusters (see \citealt{Figuera_2016}, \citealt{Figuera_2016b}) and $10$ to $40\%$ for young populations (see \citealt{Rice_2015}, \citealt{2014_Kourniotis}).

Sources however, can still have underlying photometric variability such as binary stars. Within our sample, there are three known binaries: S2-36, IRS 16SW, and S4-258. Both S2-36 and IRS 16SW have variability that is not consistent with extinction. Their movement in the CMD appears to be fairly vertical, within $2\sigma$ of $90^{\mathrm{\circ}}$ (see Table \ref{table:non-extinction}). This behavior is expected of binaries observed in the near- to mid-infrared: because the near-infrared is close to the Rayleigh-Jean tail, binaries will tend to exhibit no color change with magnitude over phase if variations are temperature induced (see \citealt{Rice_2015}). The last binary in our sample, S4-258, has long-term color variability suggestive of extinction effects. In both H-K' and K'-L', we see large overall dimming in magnitude (around $\mathrm{\Delta{K'}} \sim 2$). This change is consistent with extinction, with a probability of membership to the H-K' extinction Gaussian of $96\%$ (see Table \ref{table:extinction} and Table \ref{sec:linear_fitting_results}). 

This dimming event could be produced by the binary system from dust generation, although unlikely. S4-258 is a candidate short period (around 1.14 days) O/WR binary (see \citealt{Gautam_2019}, \citealt{2020MNRAS.492.2481W}). Given that it has strong wind lines found by \citealt{Pfuhl_2014}, such a system could produce dust. Wolf-Rayet binaries -- especially wide binaries -- can periodically produce dust behind the shocks of colliding winds where the environment is cool and dense enough for dust condensation. The production of dust can increase the amount of extinction, increase flux at longer wavelengths due to heated dust, and create variability in the near-infrared (e.g. \citealt{Williams_2009}, \citealt{2016ApJ...818..117L}). But, we would expect to see variable brightening and dimming on roughly the binary orbital timescale instead of the observed long-term trend (see \citealt{2019_Williams}). More likely than a dust production event, then, is that S4-258 is passing behind one (of many) extinguishing features present at the Galactic Center, especially toward the dusty and high-extinction IRS 13 structure. 

The remainder of the variability -- not consistent with extinction effects or vertical movement in the CMD -- could have a wide variety of origins. While the bulk of non-extinction sources has slopes within 3$\mathrm{\sigma}$ of 90 degrees (i.e., no color change with K' variability), 25 exhibit brightening-reddening effects, having significant slopes in the CMD in the opposite direction that would be caused by extinction. Brightening-reddening could be a sign of changes in accretion onto the star or changes in dust scattering properties, as is the case with observations of young stellar objects harboring circumstellar material (e.g. \citealt{Poppenhaeger_2015}, \citealt{Rice_2015}).

\subsection{Extinction features at the Galactic Center} \label{sec:dipping}

\begin{figure}[ht]
\centering
\includegraphics[width=0.43\textwidth]{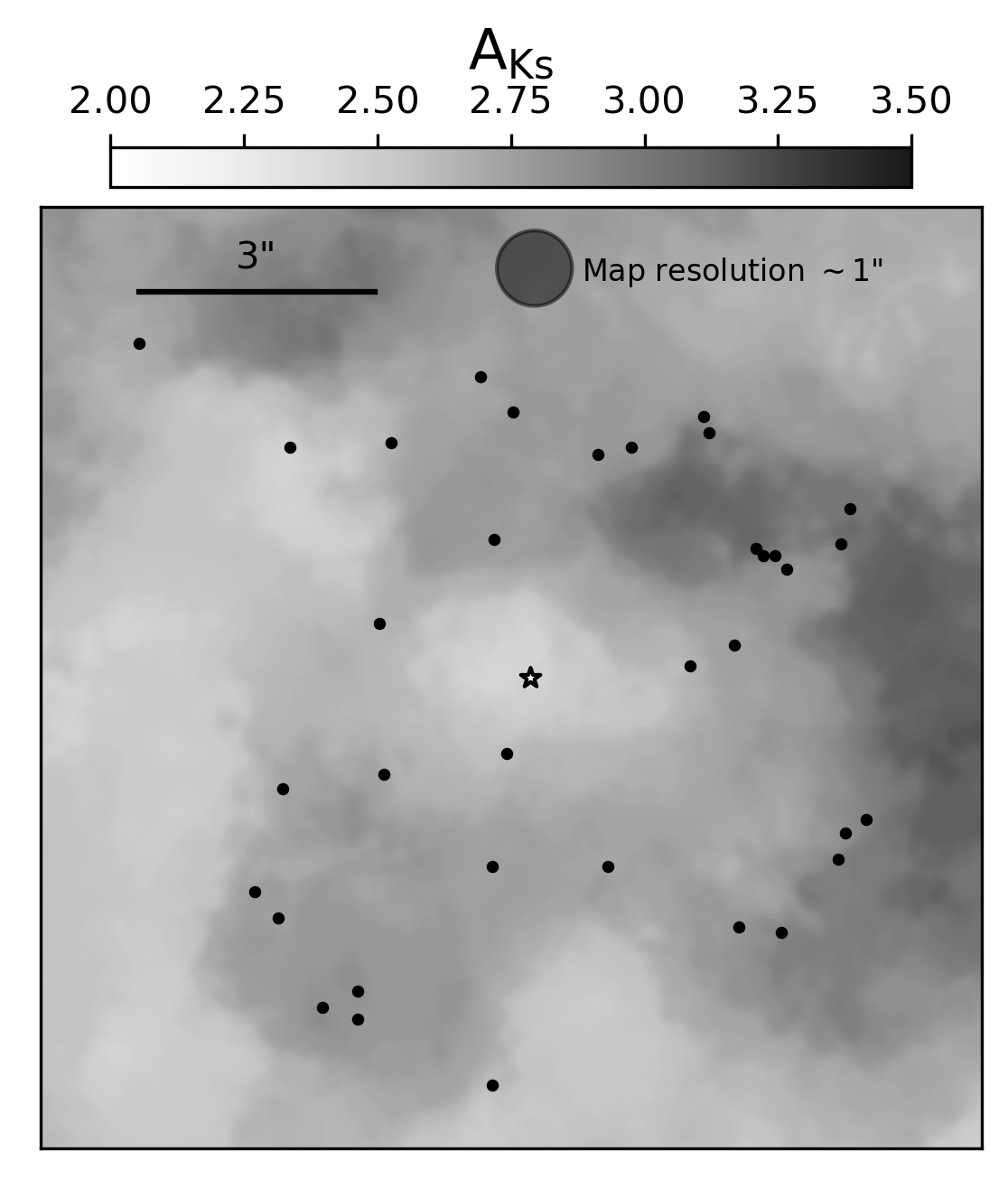}
\caption{Extinction map from \citealt{Schodel2010} which has a resolution of $\sim$1". Extinction sources are marked in black and tend to be in more extinguished regions; SgrA* is marked with a star at the center of the field. Our experiment's resolution -- $\sim$50 mas -- is smaller than the black points.}
\label{fig:ext_map}
\end{figure}
\begin{figure}[ht]
\includegraphics[width=0.43\textwidth]{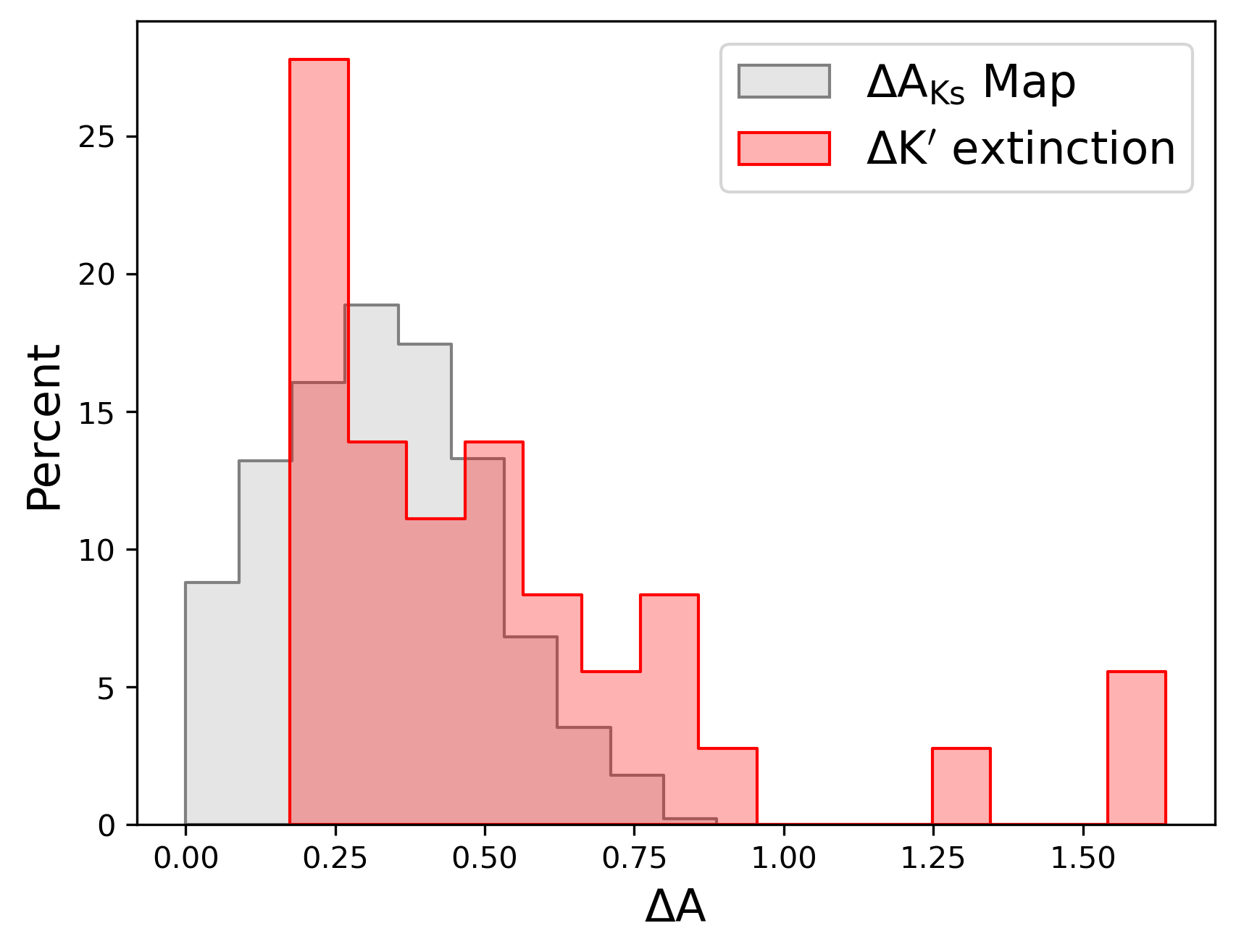}
\caption{Average extinction variability in \citealt{Schodel2010}'s extinction map (in grey) compared to the change in magnitude ($\mathrm{\Delta{K'}}$) for each extinction event (in red). The changes in extinction, $\mathrm{\Delta{K'}}$, that we observe are on average consistent with the extinction variations seen in \citealt{Schodel2010}, although we observe changes that are around a magnitude larger.}
\label{fig:extinction_hist_changes}
\end{figure}

We find extinction variations on 20 times smaller scales compared to previous studies. Figure \ref{fig:ext_map} shows the 36 extinction variability sources superimposed on the \citealt{Schodel2010} extinction map with a resolution of 1". Given a mean proper motion of 5 mas/year and a sampling of 9 years, we observe extinction changes on $\sim50$ mas scales, or 400 AU at the distance of 8 kpc. The typical change in extinction for our sources ($\mathrm{\Delta{K'}} \sim 0.5$ mag, but at most $\sim 1$ mag) is similar to that seen in earlier studies focusing on integrated line-of-sight extinction variations. As shown in Figure \ref{fig:extinction_hist_changes}, we find extinction variations similar to the spatial variations in the \citealt{Schodel2010} map, if not slightly larger. 

Our results agree with previous measurements of the dust localized to the Galactic Center. \citealt{Chatzopoulos2015}, for instance, use the differential extinction of the NSC stellar population in conjunction with their proper motions to derive local extinction variations; they find local variations between 0.15 to 0.8 magnitudes in the central parsec. Furthermore, these results are also consistent with those of \citealt{Paumard_2004}, who find that the \textit{Minispiral} contributes $\mathrm{A_{K}} \sim 0.8$ mags of extinction toward the Northern arm. 

The origin of the extinction variability could be thin filaments, similar to the population observed by \citealt{Mu_i__2007}, with sizes $\leq100$ mas. Within the extinction sample we tend to see long-term dimming (e.g., S4-258, S4-4, S3-167, S3-129, S2-85, see Appendix \ref{sec:kp_lightcurves}). With further observations, such sources may also exhibit a rise in magnitude as they traverse out from behind an extinguishing feature. But, of particular interest are extinction variability sources which have ``dipping" events -- dimming and subsequent brightening associated with changing extinction captured by the observation baseline -- which can be used to derive the scale and density associated with the extinguishing medium. Any source that has a light curve showing an isolated fall and rise that can be fit with a Gaussian is identified as ``dipping''

\begin{figure*}
\centering 
\includegraphics[width=0.98\textwidth]{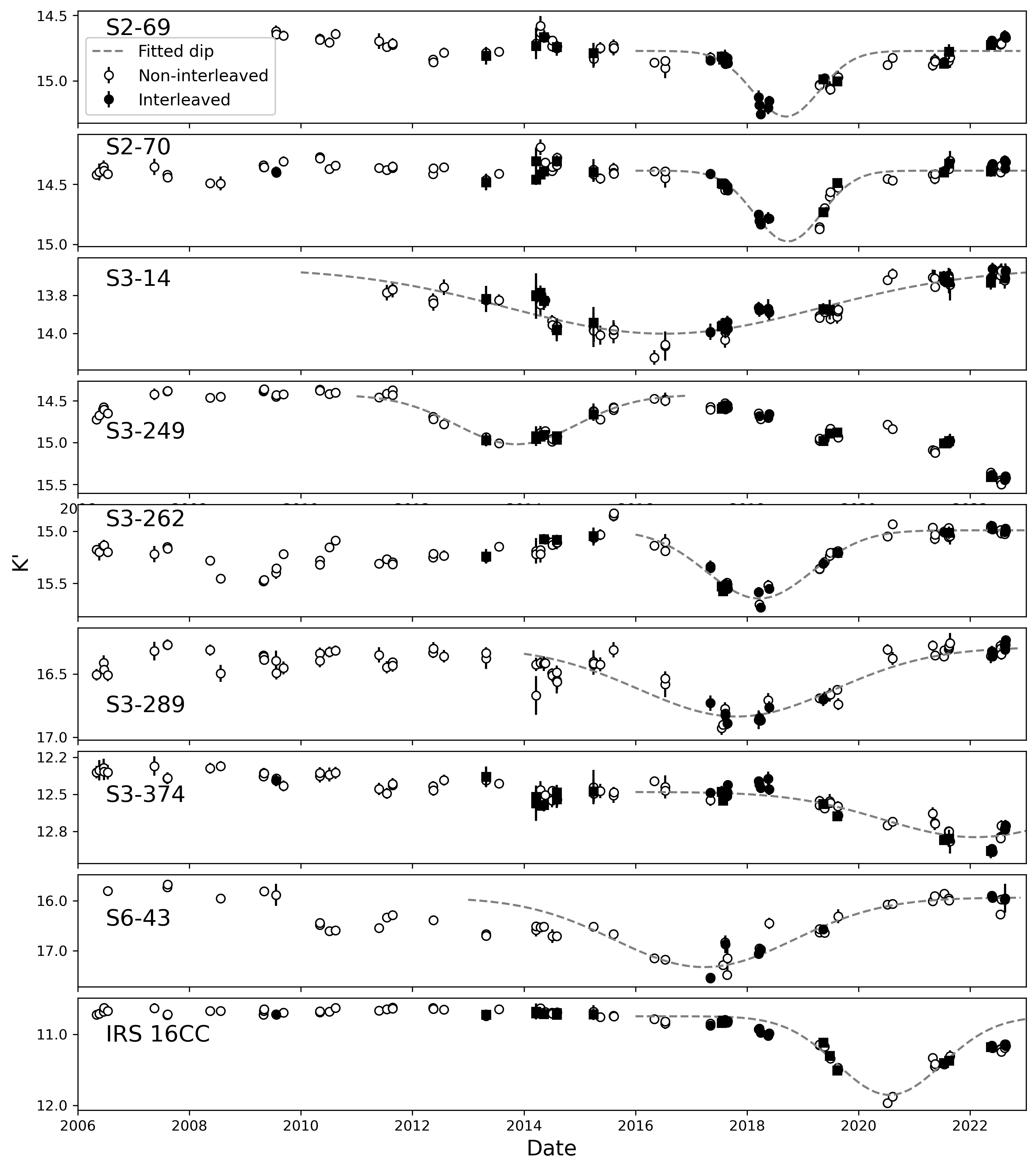}
\caption{All 9 dipping sources' time-series in K'. K' single band data is marked with open circles, H-K' color sampling are marked with black circles, and K'-L' color sampling are marked with black squares. We fit a Gaussian model to each dip in K', which is shown by the dashed grey line.}
\label{fig:dipping_events}
\end{figure*}

In our sample, we find ``dipping'' events associated with the nine following stars: S2-69, IRS 16CC, S2-70, S3-262, S3-14, S3-374, S3-289, S6-43, and S3-249 (see Figure \ref{fig:dipping_events}). The dipping events are approximated by a Gaussian fit to the K' light-curve, where the duration of the dip is estimated as three times the Gaussian width. These results are presented in Table \ref{table:dipping_events} along with each star's proper motion. We find that ``dipping" events captured in our dataset tend to last around 4-8 years and correspond to changes in magnitude ranging from 0.3 mags to 1.4 mags in K'. Factoring in the proper motion of each star -- and assuming a stationary extinguishing source at 8 kpc -- this translates to a typical cross-section around $\sim450$ AU for the extinguishing medium. These filaments, if in the same volume and under the gravitational influence of a SMBH, are likely also moving at hundreds of kilometers per second. Our estimate of size, then, could be a factor of $\sim 2$ larger (or smaller) if we take this into account.

The dips observed are extinction enhancements and can be used to estimate the neutral hydrogen ($\mathrm{H}_2$) column density of the extinguishing filament. The relationship between visual extinction and the column density of $\mathrm{H}_2$, as described in \citealt{Ext_2009}, is approximately, 
\begin{equation}
\frac{\mathrm{N_{H}}}{\mathrm{A_{V}}} \sim 2 \times 10^{21}~ \mathrm{atoms} \   \mathrm{cm^{-2}} \ \mathrm{mag^{-1}}.
\end{equation}
From \citealt{ext_gc_visual}, the visual extinction, $\mathrm{A_{V}}$, towards the Galactic Center is near $40 \pm 2$ mag; in the near-IR towards the central parsec region, \citealt{Fritz_2011} find $\mathrm{A_{K'}} = 2.5 \pm 0.1$, which is $0.06 \times \mathrm{A_{V}}$. The number density of the extinguishing source, then, can be related to the estimated dip depth in K' ($\mathrm{\Delta{K'}}$) with the following equation: 
\begin{equation}
\frac{\Delta{\mathrm{N_{H}}}}{\Delta{\mathrm{K'}}} \sim 1.2 \times 10^{20}~ \mathrm{atoms} \ \mathrm{cm^{-2}} \  \mathrm{mag^{-1}}.
\end{equation}
From the measured $\Delta{\mathrm{K'}}$ from the Gaussian fits and taking into account the derived filament cross sections, the typical filament density is around $3 \times 10^{4} \sim \mathrm{atoms/cm^{3}}$ (see Table \ref{table:dipping_events}). These densities, while large compared to local values, are consistent with the gas densities observed in the Galactic Center region. For instance, \citealt{1993ApJ...402..173J} find a neutral hydrogen gas density ranging from $10^{4}$ to $10^{5} \mathrm{atoms/cm^3}$ in the central two parsecs.

\begin{figure*}[ht]
\centering 
\includegraphics[width=0.95\textwidth]{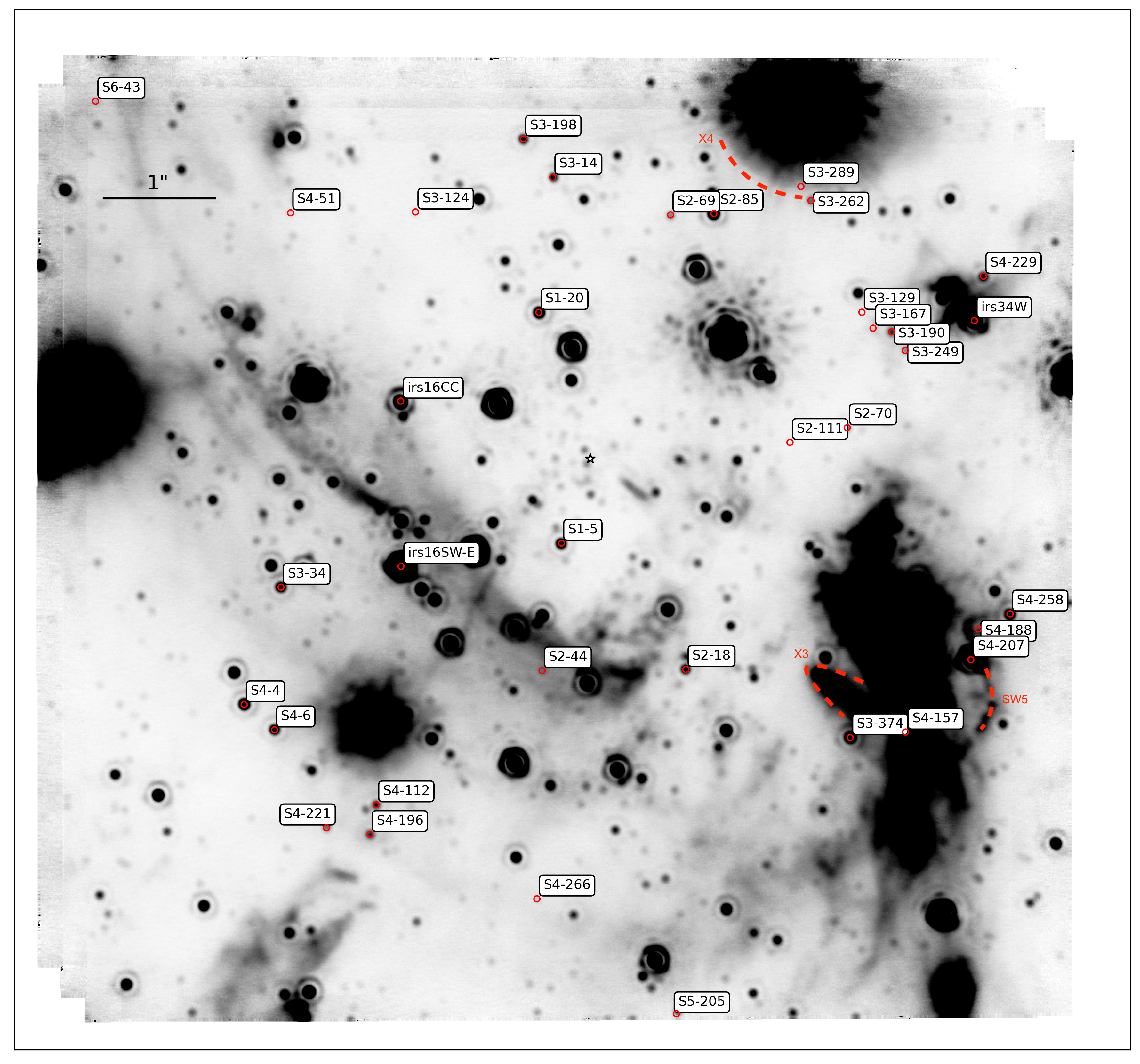}
\caption{Location on the L' image (2021-08-15) for all extinction variability sources. Three known L' features, identified by \citealt{Mu_i__2007}, are also marked in red.}
\label{fig:location_extinction}
\end{figure*}

Some sources are coincident with known L' counterparts: S4-188 and S4-207, for instance, are near the SW5 filament identified in \citealt{Mu_i__2007}. S3-262 and S3-374 are close to X4 and the tail of X3, respectively (see Figure \ref{fig:location_extinction}). In the case of X4, this object is an apparent bow-shock source hypothesized to have originated from an interaction between the IRS 16 cluster stellar winds and the winds of the likely AGB star IRS 3 (see \citealt{2005A&A...433..117V}); another interpretation is that X4 is a dusty shell ejected from IRS 3 and tidally distorted by SgrA*, with an average envelope density between $\mathrm{n_{H}} \sim 6 \times 10^{3}$ and $\sim 3 \times 10^{4}$, depending on the clumpiness of the gas (see \citealt{2017ApJ...837...93Y}). Our result of $\sim 2 \times 10^{4} \mathrm{atoms/cm^{3}}$ for S3-262 is consistent with these densities. Nevertheless, most of the ``dipping'' sources and extinction sources are not coincident with known L' filaments; this is not surprising, however, as the derived filament sizes are below the diffraction limit at L' for NIRC2.

\input{dipping.tex}

\subsection{Near-infrared extinction at the Galactic Center}
The sub-sample of variable stars that are consistent with extinction provides a unique opportunity to study the extinction at the Galactic Center. Our population modeling of the CMD slopes suggests a mean $\langle{\theta_{H-K'}}\rangle = 55 \pm 2$ degrees and a mean $\langle{\theta_{K'-L'}}\rangle = 73 \pm 2$ degrees. From this, we estimate the relative extinction $\mathrm{A_{H}/A_{K'}}$ and $\mathrm{A_{L'}/A_{K'}}$ by the following equations:
\begin{equation}
\begin{split}
\frac{\mathrm{A_{H}}}{\mathrm{A_{K'}}} = 1 + \frac{1}{\mathrm{s}} \\
\frac{\mathrm{A_{L'}}}{\mathrm{A_{K'}}} =  1 - \frac{1}{\mathrm{s}}
\end{split}
\end{equation}
where the slope (s) relates to the angle in the CMD according to Equation \ref{equation:angle_to_slope}. From this, we find $\frac{\mathrm{A_{H}}}{\mathrm{A_{K'}}} = 1.67 \pm 0.05$ and $\frac{\mathrm{A_{L'}}}{\mathrm{A_{K'}}} = 0.69 \pm 0.03$.

The color-dependent effective wavelength ($\mathrm{\lambda}_{\mathrm{eff}}$), however, could result in a biasing of the CMD slope and should be considered when comparing these results to literature values. In our case, where NIR extinction is modeled as a decreasing power law, this will bias the slope to shallower values: as a star gets more red, the extinction values decrease with increasing effective wavelength. As a test case to see the magnitude of this effect we model a young star using SPISEA \citealt{Hosek_2020} assuming the extinction law of \citealt{Schodel2010} and an extinction of $\mathrm{A}_{\mathrm{Ks}} = 2.64$, which is average for the GC field (see \citealt{Schodel2010}); the effective wavelength is calculated following the procedure of \citealt{Tokunaga_2005}. We estimate the effective wavelength changes for each extinction star, assuming the $\mathrm{\Delta}{\mathrm{K'}}$ values shown in Table \ref{table:extinction}. We find that there is a slight change in the effective wavelength, especially for stars with larger $\mathrm{\Delta}{\mathrm{K'}}$. The median effective wavelength changes for each band are the following: $\mathrm{\lambda}_{\mathrm{H,eff}} \sim -0.007$ $\mu$m, $\mathrm{\lambda}_{\mathrm{K',eff}} = -0.005$ $\mu$m, and $\mathrm{\lambda}_{\mathrm{L',eff}} = -0.002$ $\mu$m. As extinction has a diminishing effect, it is not surprising that longer wavelengths (i.e. L’) are less affected. 

To estimate the size of this bias, we use the \citealt{Schodel2010} law and find the change in extinction that would be the result of the $\mathrm{\delta}{\mathrm{\lambda}_{\mathrm{eff}}}$ for each extinction star in the sample. The median changes in extinction just due to effective wavelength changes are the following: $\mathrm{\Delta}{\mathrm{A_{H}}}$ = 0.042 mags, $\mathrm{\Delta}{\mathrm{A_{K’}}}$ = 0.020 mags, and $\mathrm{\Delta}{\mathrm{A_{L’}}}$ = 0.003 mags. Translating this to a slope for each source, we find that the median error on the measured slope due only to changing effective wavelengths in the HK CMD is $\sim$ 0.86 degrees (maximum 1.8 degrees) and in the KL CMD is $\sim$ 0.1 degrees  (maximum 0.2 degrees). This corresponds to a bias of $\frac{\mathrm{A_{H}}}{\mathrm{A_{K'}}} \sim 0.02$ and $\frac{\mathrm{A_{L'}}}{\mathrm{A_{K'}}} \sim 0.001$. While we have assumed an extinction law to arrive at estimate this bias – and a steeper extinction law, especially around H-band, would increase the bias – this effect is not large, especially around L-band, where the difference between our results and literature is most prominent. Nevertheless, we include this estimated bias when comparing our results to literature values. 

Furthermore, when comparing to literature values, we also consider the effective wavelength of our experiment and its estimated error. Again, using SPISEA \citealt{Hosek_2020} with the same setup as above and assuming the \citealt{Schodel2010} extinction law, we estimate the effective wavelength of a 5,000 K, 10,000 K, and 25,000 K star, set at extinctions of $A_{Ks}$ = 1.5, 2, 2.5, and 3. We find the following average effective wavelength and error per band: $\lambda_{H}$ = $1.654 \pm 0.014$ $\mu$m, $\lambda_{K’}$ = $2.125 \pm 0.008$ $\mu$m,  $\lambda_{L’}$ = $3.741 \pm 0.005$ $\mu$m. While we assume an extinction law here, assuming a different power law slope within the errors of \citealt{Schodel2010}, affects the estimates far below the errors assumed above.

\begin{figure}[ht]
\centering
\includegraphics[width=0.47\textwidth]{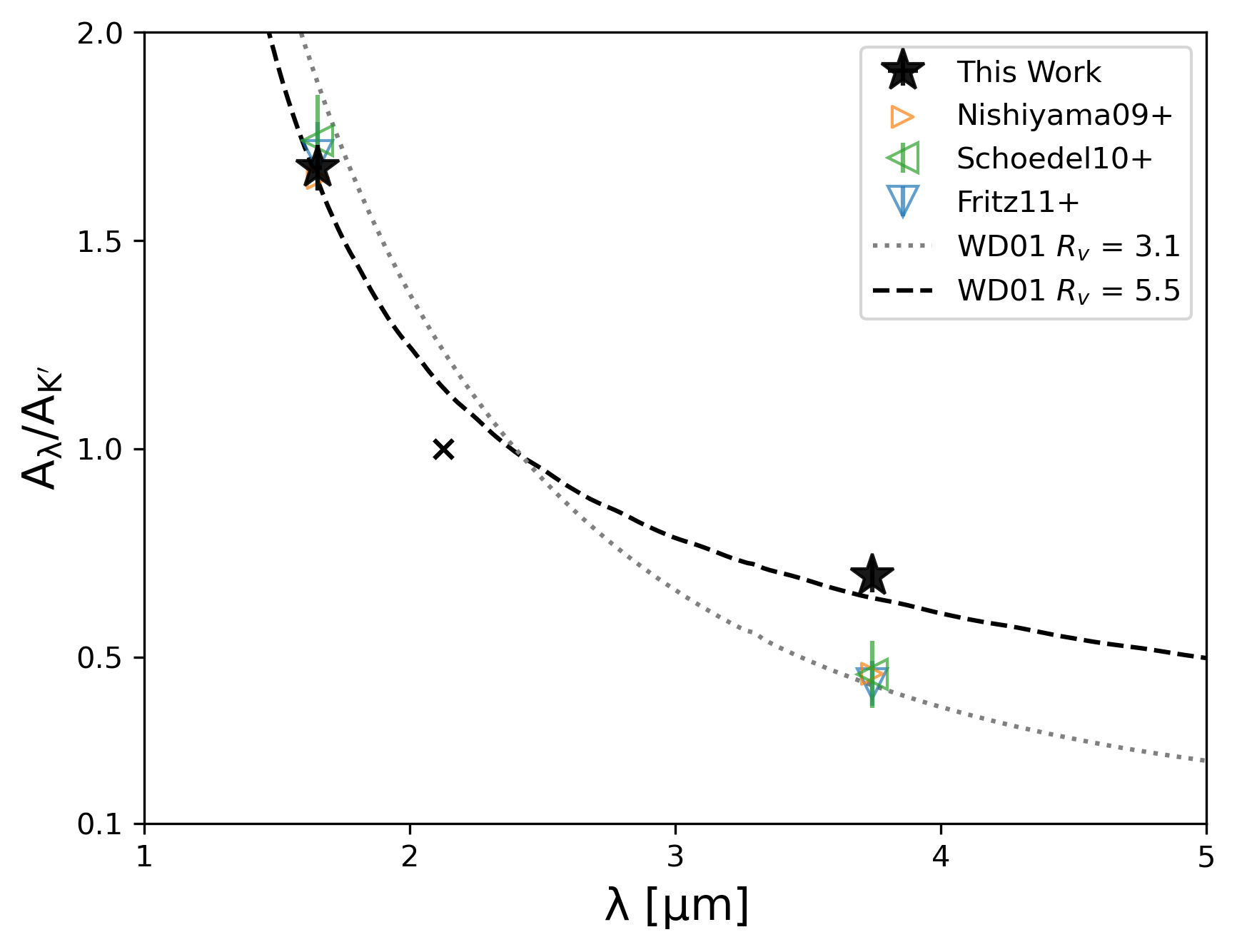}
\caption{Wavelength (in $\mathrm{\mu}$m) compared to the relative extinction (in reference to K') at our experiment's effective wavelengths:  H = $1.654 \mu$m, K' = $2.125 \mu$m, L' = $3.741 \mu$m. Plotted are values found by other studies such as \citealt{Nishiyama_2009}, \citealt{Schodel2010}, and \citealt{Fritz_2011}. Both \citealt{Nishiyama_2009} and \citealt{Schodel2010} are plotted assuming the derived broken power-law slopes and their errors from each paper. For \citealt{Fritz_2011}, the value is calculated assuming the first column of Table 9 for the NIRC2 H, K' and L' filters. We plot the effect of grain size (using $\mathrm{R_{V}}$ as a proxy) on the relative extinction, assuming the model as presented in \citealt{WD_2001}. In general, our results (in black) are consistent for H. We find more extinction at L' with our methods, suggesting a greyer extinction in the mid-IR, possibly due to larger dust grains local to the Galactic Center.}
\label{fig:rel_extinction}
\end{figure}

While $\mathrm{A_{H}/A_{K'}}$ is in agreement with literature measurements (see \citealt{Nishiyama_2009}, \citealt{Schodel2010}, and \citealt{Fritz_2011} along with Figure \ref{fig:rel_extinction}), there is a large discrepancy at $\mathrm{A_{L'}/A_{K'}}$. This discrepancy could be explained by grain populations local to the Galactic Center which are different from those along the line-of-sight. Previous studies of the IR extinction invoke different dust environments to partly understand the excess extinction (or, flattening of the extinction power law) starting at the mid-IR, or $\sim$3$\mathrm{\mu}$m. While this flattening in the mid-IR is not unique to Galactic Center sight-lines, it appears to be characteristic of regions with dense and large grains, such as molecular clouds. \citealt{ext_gc_visual} find that including a population of larger dust grains (dense regions and a larger total-to-selective extinction ratio, $\mathrm{R_{V}}$) along with more processed and diffuse dust (lower $\mathrm{R_{V}}$) reproduces the extinction curve. \citealt{Voshchinnikov_2017} similarly model the extinction with a multi-component model. Their model consists of two components: a foreground cloud of processed grains that is relatively transparent between 4 and 7 $\mathrm{\mu}$m, and a second cloud more localized (within 1 kpc) to the Galactic Center composed of primarily unprocessed and fresh silicate grains. Therefore, we may be witnessing the effect of large grains at the Galactic Center, which would result in the excess $\mathrm{A_{L'}}$ as observed. As a qualitative example, we show the effect on extinction of two different values total-to-selective extinction ratios: $\mathrm{R_{V}} = 5.5$ which is dominated by larger grains than $\mathrm{R_{V}} = 3.1$ (see Figure \ref{fig:rel_extinction}). We assume the model ``B" presented in \citealt{WD_2001} with silicate-graphite grains. In general, that the $\mathrm{A_{L'}/A_{K'}}$ ratio is much more consistent with the $\mathrm{R_{V}} = 5.5$ model compared to the $\mathrm{R_{V}} = 3.1$ model.

\begin{figure}[ht]
\centering
\includegraphics[width=0.47\textwidth]{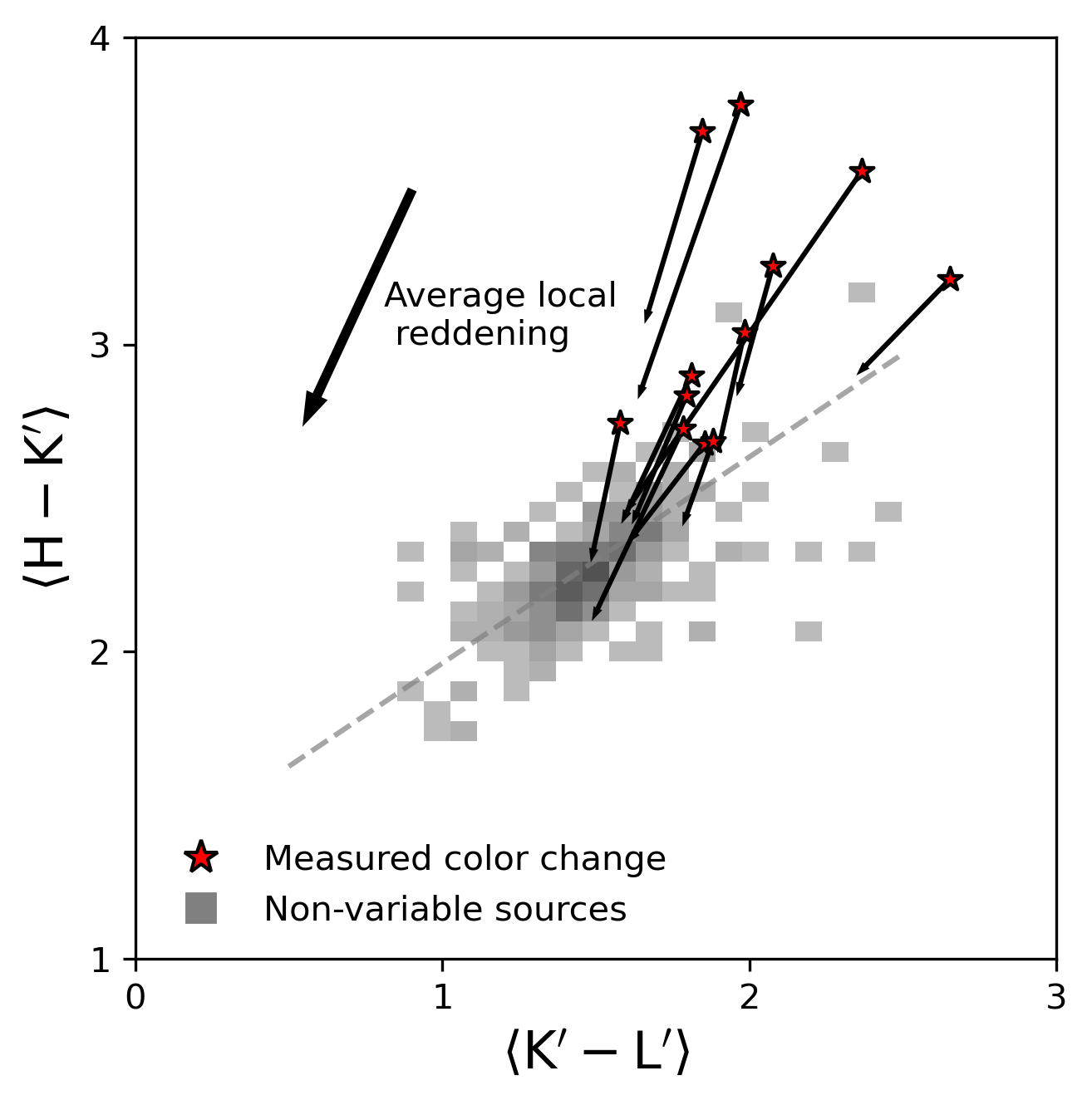}
\caption{Plot of mean H-K' and K'-L' color for sources that are detected in all 3 wavelengths. Non-variable sources are plotted in grey and their L' information is provided in Table \ref{table:non_var_lp}. We plot the movement in color-color of sources where we measure a slope in both the H-K' and K'-L' CMD. The sources which are consistent with extinction follow our average local reddening vector as they change color; the mean colors tend to follow the integrated reddening along the line-of-sight, assuming \cite{Fritz_2011}, which is marked by the dashed grey line.}
\label{fig:local_reddening}
\end{figure}

A population of large grains local to the Galactic Center may not be surprising. While shocks (i.e. from supernova or turbulence) can break up large grains via shattering, smaller grains may struggle to survive at the Galactic Center. The central parsec has a high ambient UV radiation field from $\sim100$ OB and WR stars (e.g. \citealt{1991_Krabbe}, \citealt{2007_Martins}). High UV photons preferentially evaporate and destroy small grains; large grains, then, may preferentially survive and further grow via coagulation. Furthermore, late-type stars -- a large fraction of the Galactic Center stars (see \citealt{Do_2013}) -- are thought to be important producers of dust in the universe, ejecting larger grains (a $\geq 0.1 \mu$m) produced in their atmospheres into the ISM (see \citealt{Hirashita_2013}). Wolf-Rayet (WR) binaries also may be responsible for the large size grains, especially those that are carbon rich (WC). \citealt{Lau_2020}, for instance, find that most efficient dust producing WC stars tend to form relatively large grains. The dust we see, therefore, could also be relatively fresh and biased towards large sizes.

Continuing the long-term tracking of these stars, especially into the mid-infrared (M-band) and shorter wavelengths (J-band), along with future JWST observations would facilitate modeling of dust in our Galaxy's central region. Further multi-wavelength observations will also increase the observational baseline for detecting extinction variability, as these events occur on the order of years.

\section{Summary} \label{sec:summary}
Extinction effects local to the GC are investigated using observations of high proper-motion stars in H, K' and L'. We find that around a third of the bright photometrically variable sample ($32\% \pm 10\%$) is dominated by variations in the line-of-sight extinction rather than intrinsic photometric changes. This lowers the implied fraction of photometric variable stars found by \citealt{Gautam_2019} to $34\% \pm 10\%$, which is more consistent with the properties found elsewhere in the Galaxy. 

The variable extinction sources probe spatial changes in the extinction that are proportional to their proper motions ($\sim$50 mas) and that are smaller than previous works. Of these extinction sources, a small fraction exhibit ``dipping" events: a relatively symmetrical dim and rise in magnitude over years. We interpret these events as due to dense ($\rho \sim 10^4$ atoms/$\mathrm{cm^3}$) filaments or clumps of dust and gas with sizes on the order of hundreds of AU ($\sim$100 mas). 

Finally, we find that while the relative extinction is consistent with previous results at H-band, our measurements at L' disagree with previous measurements along the entire integrated line-of-sight. Because the variable extinction stars are more reddened than the rest of the sample, we infer that the material dominating these changes is \textit{local} to Galactic Center. This leads to the possible conclusion that the difference in the extinction law could be due to larger dust grains local to the Galactic Center environment.

\section{Acknowledgements}
This work is supported by the National Science Foundation Graduate Research Fellowship under Grant No. DGE-2034835, The Gordon E \& Betty I. Moore Foundation under award No. 11458, and the National Science Foundation under Grant No. 1909554. A.M.G. acknowledges support from her Lauren B. Leichtman and Arthur E. Levine Endowed Astronomy Chair.

The authors thank the anonymous referee for helpful comments that improved this work. We thank Francisco Nogueras-Lara, Matt Malkan, and Eric Becklin for helpful discussions. We thank the staff of the Keck Observatory for their help in obtaining the observations. The data presented herein were obtained at the W. M. Keck Observatory from telescope time allocated to the National Aeronautics and Space Administration through the agency's scientific partnership with the California Institute of Technology and the University of California. The Observatory was made possible by the generous financial support of the W. M. Keck Foundation. Finally, we wish to recognize and acknowledge the very significant cultural role and reverence that the summit of Maunakea has always had within the indigenous Hawaiian community. We are most fortunate to have the opportunity to conduct observations from this mountain.
\facility{Keck:II (NIRC2)}
\software{emcee \citep{2013emcee}, SPISEA \citep{Hosek_2020}, Astropy \citep{astropy:2013, astropy:2018, astropy:2022}, numpy \citep{harris2020array}, matplotlib \citep{Hunter:2007}, scipy \citep{2020SciPy-NMeth}, KAI \citep{Lu_Keck-DataReductionPipelines_KAI_v1_0_0_Release_2021}, ecGMM \citep{Hao_2009}, dust\textunderscore{extinction} \citep{Gordon_dust_extinction_2024}}
\clearpage
\appendix
\setcounter{table}{0}
\renewcommand{\thetable}{A\arabic{table}}
\section{Color observation details} \label{appendix:observations}
\input{HK_measurments}
\input{KL_measurments}
\section{Methodology} 
\setcounter{table}{0}
\renewcommand{\thetable}{B\arabic{table}}
\subsection{Line-fitting in the CMD} \label{sec:linear_fitting_results}
If the K' variability is due to changing levels of extinction along the line-of-sight for these high proper motion stars, there should be a linear trend in their color-magnitude diagram (CMD). We describe the linear trend with a slope characterized by an angle $\theta$ from the horizontal. Specifically, we fit a linear function ($\mathrm{y = sx + b}$) in a Cartesian coordinate system in which the x-axis is the color and the y-axis is the K' magnitude. As described in \citealt{Hogg2010DataAR}, the probability of getting a measurement $\mathbf{Z_{i}} = (\mathrm{x}_{i}, \mathrm{y}_{i})$ given a true value $\mathbf{Z} = (\mathrm{x}, \mathrm{y})$ is written as the following:
\begin{equation}
    \mathrm{p}(\mathrm{x}_{i}, \mathrm{y}_{i} | \mathbf{S_{i}}, \mathrm{x}, \mathrm{y}) = 
    \frac{1}{2\pi\sqrt{\mathrm{det}{S_{i}}}}\exp{\left (\frac{-1}{2}([\mathbf{Z_{i} - Z}]^{T}\mathbf{S_{i}}^{-1}[\mathbf{Z_{i}} - \mathbf{Z}])\right)}
\end{equation}
where $\mathrm{S}_{i}$ is the covariance tensor given by,
\begin{equation}
    \mathbf{S_{i}} = 
    \begin{bmatrix}\sigma_{x_{i}}^2 & \sigma_{xy_{i}}\\ \sigma_{y_{i}} & \sigma_{yx_{i}}^2\end{bmatrix} 
\end{equation}
Using these probabilities, this approach minimizes each observation's orthogonal displacement from the best fit line. Each point will have an orthogonal displacement ($\Delta_{i}$) from the linear relationship,
\begin{equation}
    \centering
    \mathrm{\Delta}_{i} = \mathbf{\hat{v}^{T}}\mathbf{Z_{i}} - \mathrm{b}\mathrm{cos}\mathrm{\theta}
\end{equation}
and $\mathbf{\hat{v}}$ is the vector orthogonal to the slope, described by s:
\begin{equation}
    \centering
    \mathbf{\hat{v}} = \frac{1}{\sqrt{1+s^2}} \begin{bmatrix} -s\\ 1\end{bmatrix} = \begin{bmatrix} -\mathrm{sin}{\mathrm{\theta}}\\ \mathrm{cos}{\mathrm{\theta}} \end{bmatrix}
\end{equation}
The projected orthogonal variance is described by $\Sigma_{i}^2$:

\begin{equation}
    \centering 
    \Sigma_{i}^2 = \mathbf{\hat{v}^{T}}\mathbf{S_{i}}\mathbf{\hat{v}}
\end{equation}
The Log-Likelihood for ($\mathrm{\theta}, \mathrm{b\cos{\theta}}$) is written as:
\begin{equation}
\ln \mathcal{L} = K - \sum_{i=1}^{N}\frac{{\Delta_{i}}^2}{{2\Sigma_{i}}^2}
\label{eq:lnlike}
\end{equation}
where K is a constant and N is the number of observations. 

Our reported slope estimates are found by minimizing the Log-Likelihood with the MCMC package \texttt{emcee} (see \citealt{2013emcee}). We use uniform priors on both $\theta$ ($-\pi < \theta < \pi$) and $\mathrm{b\cos(\theta)}$ ($-1000 < \mathrm{b\cos(\theta)} < 1000$).
\subsection{Error-corrected Gaussian mixture modeling} \label{sec:ecGMM}
For error-corrected Gaussian mixture model (ecGMM), we follow \cite{Hao_2009}. In the mixture model, each Gaussian in the mixture of K is described by the following parameters: a weight $\alpha_{\mathrm{K}}$, a width $\sigma_{\mathrm{K}}$, and a mean CMD slope $\langle{\theta}\rangle_{\mathrm{K}}$. The weights of the K Gaussians are normalized such that the sum is 1. For a sample of $\mathrm{N_{stars}}$ with measured CMD slopes $\theta$ and associated errors $\delta$, the parameters that optimize the ecGMM are found by maximizing the following likelihood:
\begin{equation}
    \prod_{j=1}^{N_{stars}} \left(\sum_{i=1}^{K} \frac{\alpha_{i}}{\sqrt{2\pi(\sigma_{i}^2 + \delta_{j}^2)}})\exp{\left[\frac{-1}{2}\frac{(\theta - \langle{\theta}\rangle_{i})^2}{\sigma_{i}^2 + \delta_{j}^2}\right]} \right)
\end{equation}

The exception maximization (EM) method is used to maximize the likelihood and optimize parameters of interest for a mixture with K Gaussians (with the parameters $\mathrm{\alpha_{K}}$, $\mathrm{\sigma_{K}}$, $\langle{\theta}\rangle_{K}$). In order to pick the best number of Gaussians for the mixture and avoid over-fitting in the modeling, we use the Bayesian Information Criteria (BIC) to distinguish between the performance of one, two, and three Gaussian model. The model with the smallest BIC value is favored. 

As EM methods are sensitive to initial parameter guesses and do not provide error estimates for each parameter, we use a resampling method. We randomly pick with replacement a new sample with length $\mathrm{N_{stars}}$ and feed this sample into the ecGMM model to estimate a weight $\mathrm{\alpha_{K}}$, a width $\mathrm{\sigma_{K}}$, and a mean CMD slope $\mathrm{\langle{\theta}\rangle_{K}}$. This process is repeated 1000 times with different samples. The lower error on each parameter is defined as the the difference between the mean and lower quartile while the upper error is defined as the difference between mean and upper quartile for the 1000 samples. 

From our random samples of the ecGMM ($\langle{\theta}\rangle_{K}$, $\sigma_{K}$, $\alpha_{K}$), we calculate the membership probability of each star. For a fit with two Gaussians ($\mathrm{G}_{1}$ and $\mathrm{G}_{2}$) the probability of a star x with a slope $\theta$ and slope error $\delta$ being drawn from $\mathrm{G}_{1}$ is given by:
\begin{equation}
\mathrm{P}(\mathrm{G}_{1}|\mathrm{X=x})=\frac{\alpha_{1}\mathrm{p}_{\mathrm{X}|\mathrm{G}_{1}}(\mathrm{x})}{\alpha_{1}\mathrm{p}_{\mathrm{X}|\mathrm{G}_{1}}(x) + \alpha_{2}\mathrm{p}_{\mathrm{X}|\mathrm{G}_{2}}(\mathrm{x})}
\end{equation}

where p is the associated probability density function, 
\begin{equation}
    \mathrm{p(X=x}(\theta, \delta)|\langle{\theta}\rangle, \sigma) = \\
    \frac{1}{\sqrt{2\pi(\sigma^2 + \delta^2)}}\exp{\left[\frac{-1}{2}\frac{(\theta - \langle{\theta}\rangle)^2}{\sigma^2 + \delta^2}\right]}
\end{equation}
\clearpage
\subsection{Likely extinction CMD: H-K'}
\begin{figure}[ht]
\centering
\includegraphics[width=0.75\textwidth]{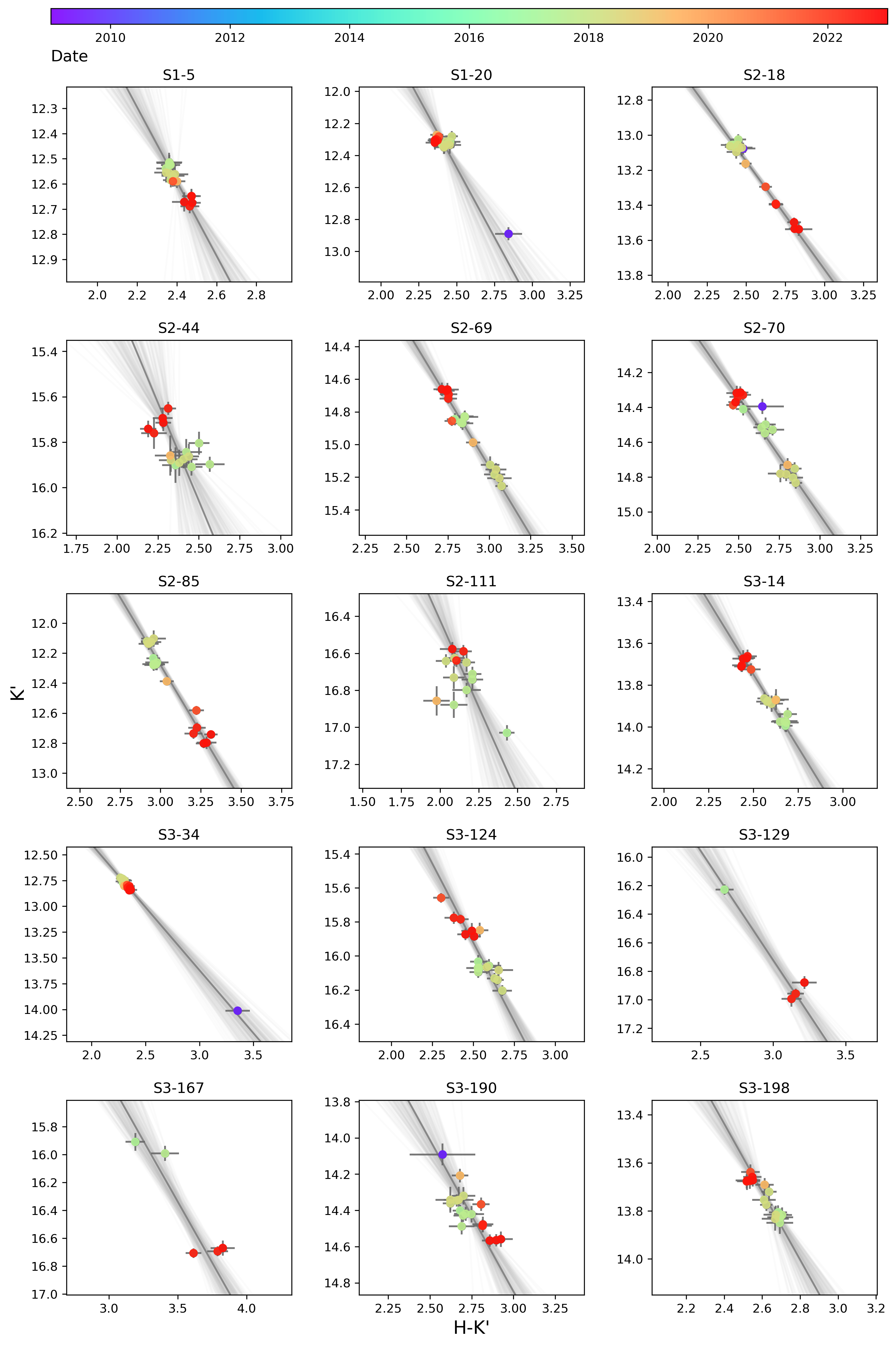}
\caption{K'/H-K' CMD for extinction variability stars. The over-plotted grey lines are 100 randomly sampled points from the MCMC chains (1).}
\end{figure}
\begin{figure}[ht]
\centering
\includegraphics[width=0.75\textwidth]{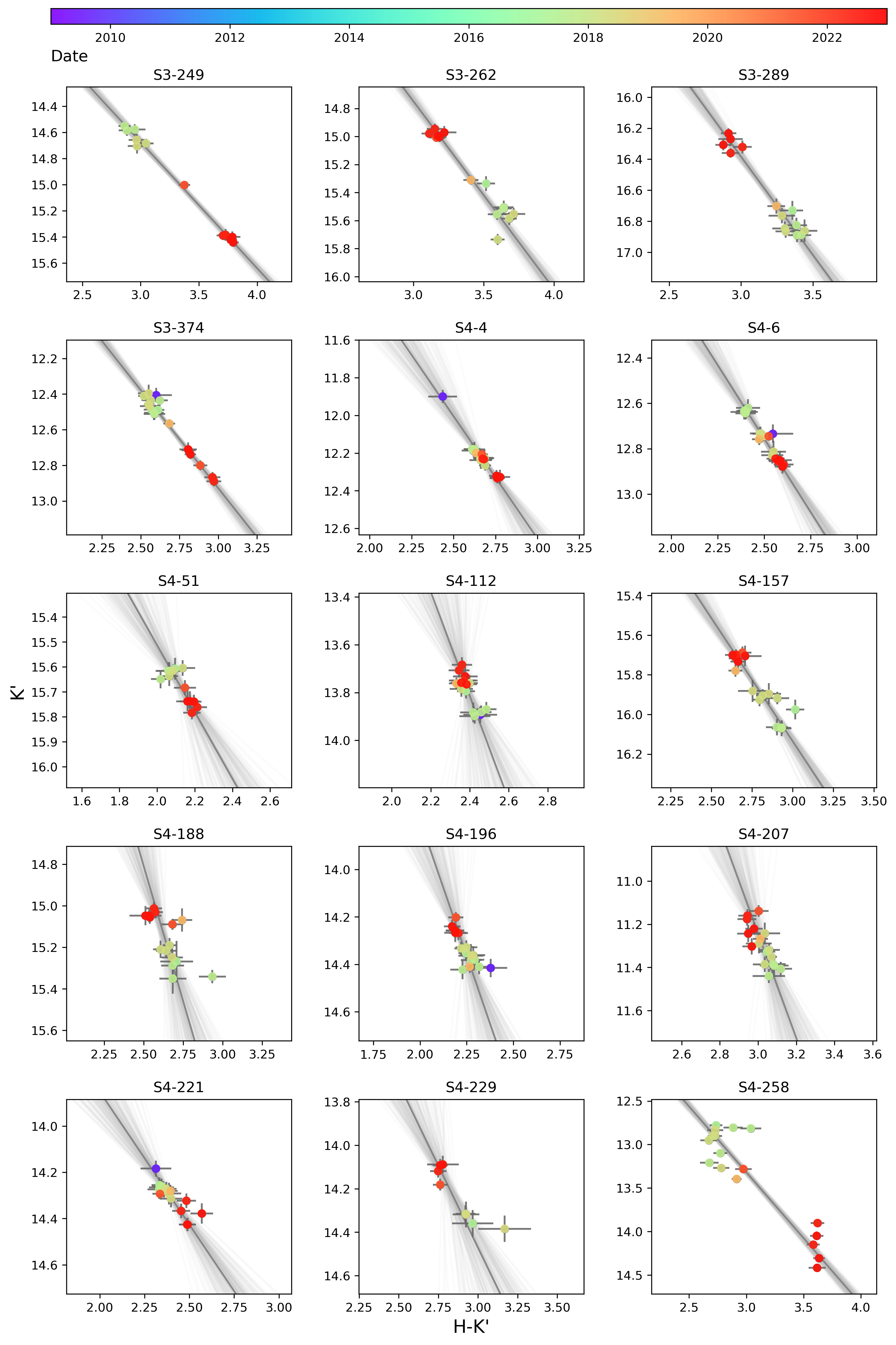}
\caption{K'/H-K' CMD for extinction variability stars. The over-plotted grey lines are 100 randomly sampled points from the MCMC chains (2).}
\end{figure}
\begin{figure}[ht]
\centering
\includegraphics[width=0.75\textwidth]{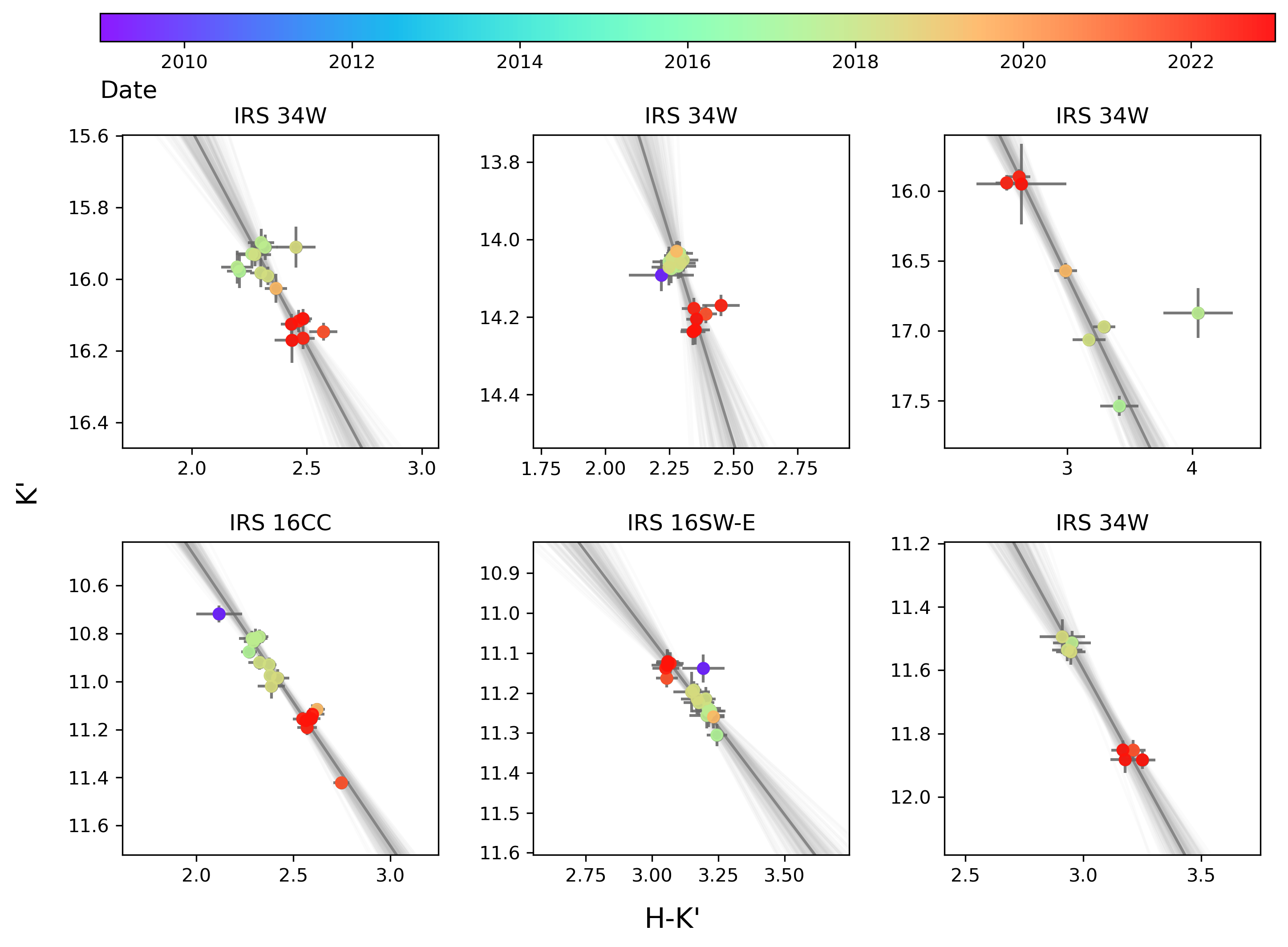}
\caption{K'/H-K' CMD for extinction variability stars. The over-plotted grey lines are 100 randomly sampled points from the MCMC chains (3).}
\end{figure}
\clearpage
\subsection{Likely extinction CMD: K'-L'}
\begin{figure}[ht]
\centering
\includegraphics[width=0.75\textwidth]{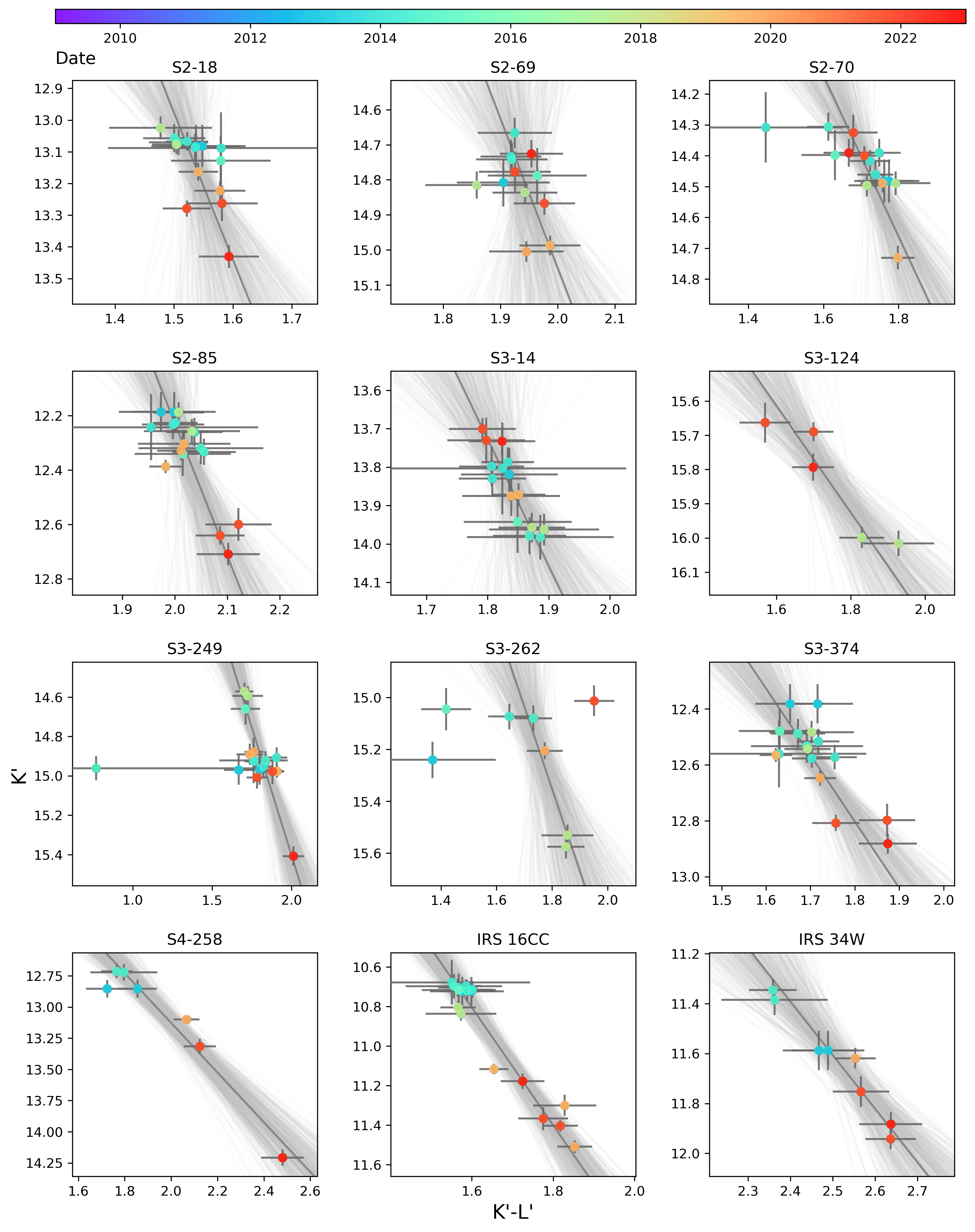}
\caption{K'/K'-L' CMD for extinction variability stars. The over-plotted grey lines are 100 randomly sampled points from the MCMC chains.}
\end{figure}

\clearpage
\subsection{Stars with variability not consistent with changing extinction} \label{appendix:non-extinction_results}
\input{non-extinction.tex}
\subsection{Other CMD}
\figsetstart
\figsetnum{1}
\figsettitle{Other CMD}

\figsetgrpstart
\figsetgrpnum{figurenumber.1}
\figsetgrptitle{other CMD HK 1}
\figsetplot{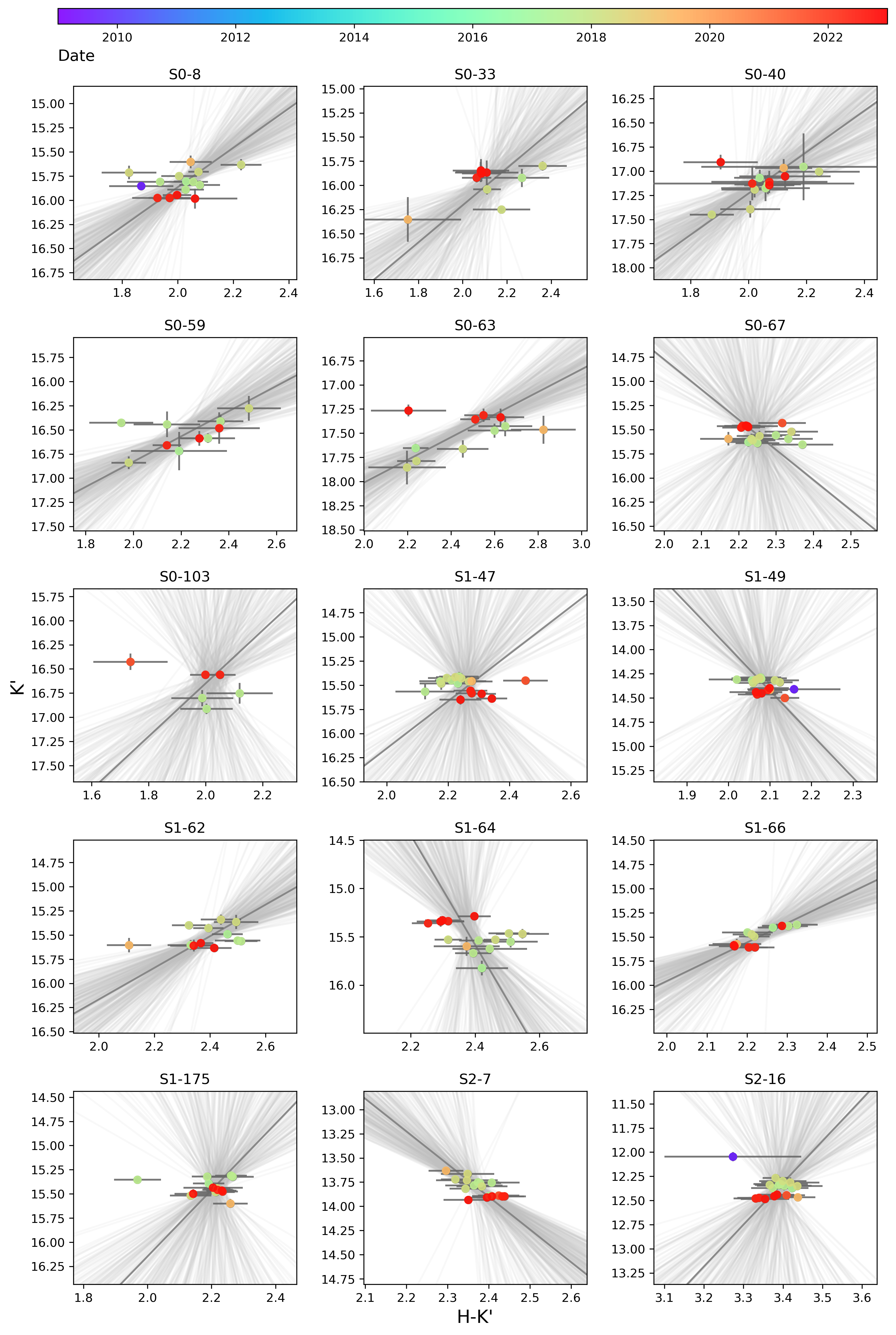}
\figsetgrpnote{K'/H-K' CMD for non-extinction stars. The over-plotted grey lines are 100 randomly sampled points from the MCMC chains. (1)}
\figsetgrpend

\figsetgrpstart
\figsetgrpnum{figurenumber.2}
\figsetgrptitle{other CMD HK 2}
\figsetplot{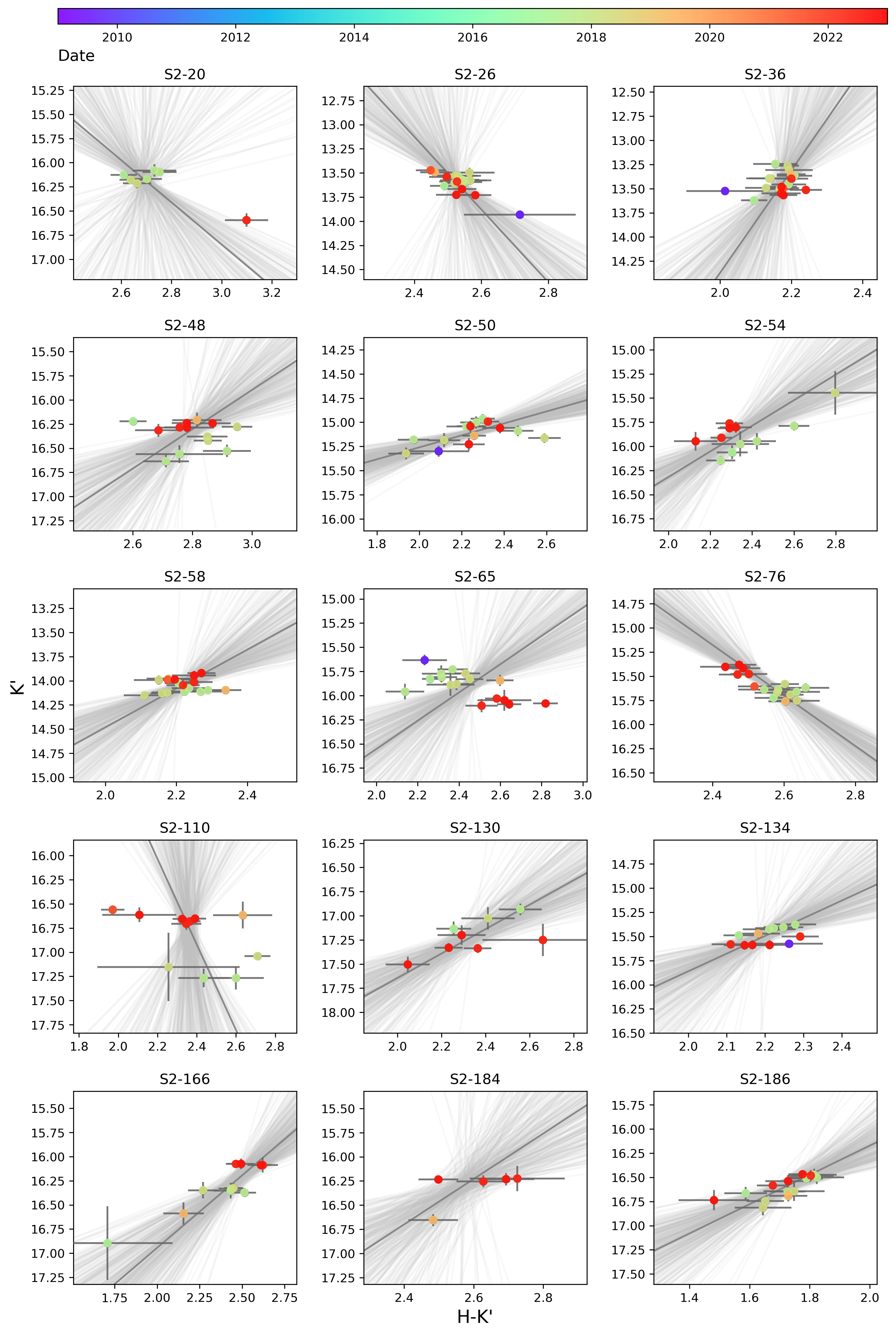}
\figsetgrpnote{K'/H-K' CMD for non-extinction stars. The over-plotted grey lines are 100 randomly sampled points from the MCMC chains. (2)}
\figsetgrpend

\figsetgrpstart
\figsetgrpnum{figurenumber.3}
\figsetgrptitle{other CMD HK 3}
\figsetplot{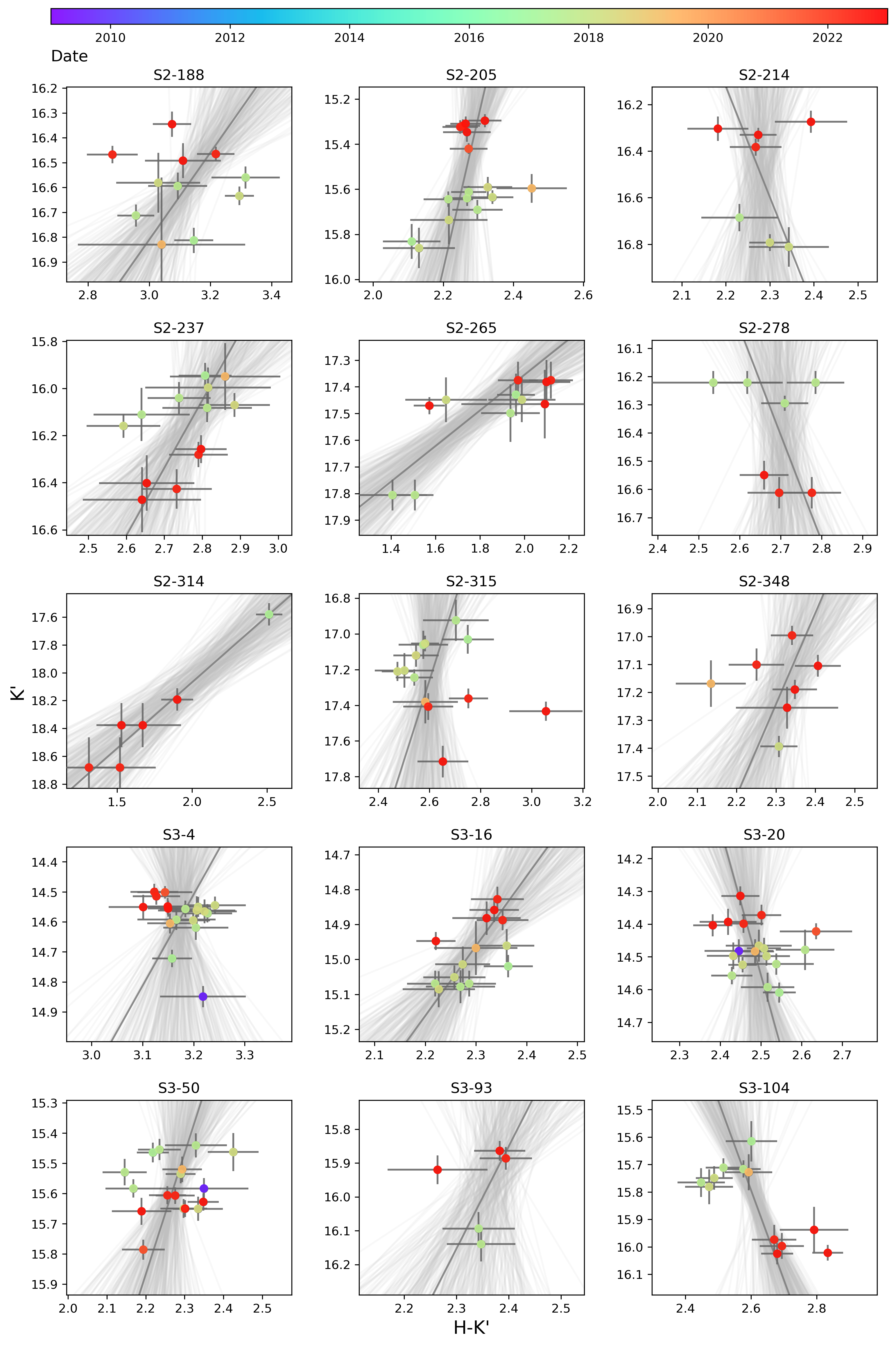}
\figsetgrpnote{K'/H-K' CMD for non-extinction stars. The over-plotted grey lines are 100 randomly sampled points from the MCMC chains. (3)}
\figsetgrpend

\figsetgrpstart
\figsetgrpnum{figurenumber.4}
\figsetgrptitle{other CMD HK 4}
\figsetplot{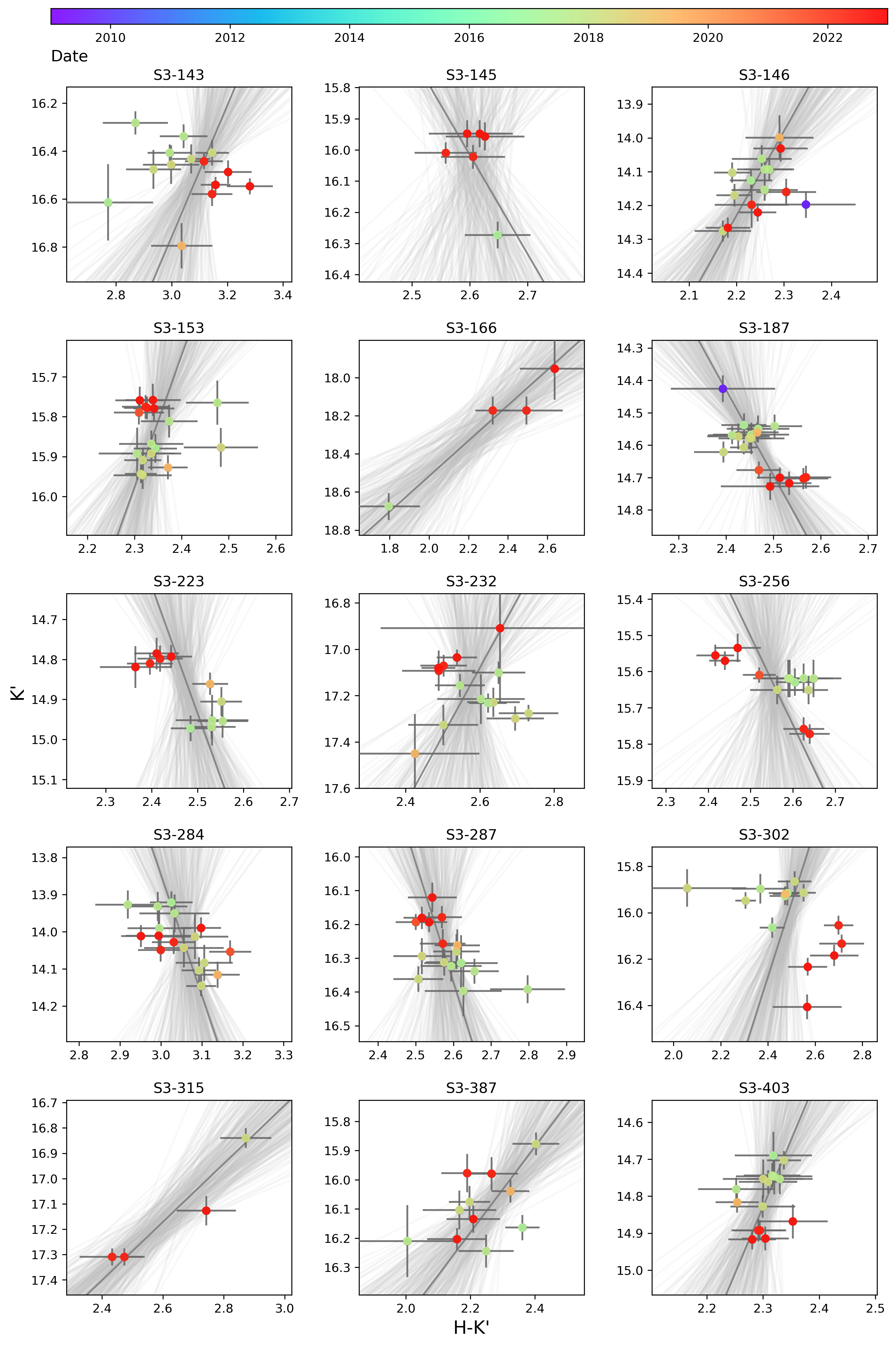}
\figsetgrpnote{K'/H-K' CMD for non-extinction stars. The over-plotted grey lines are 100 randomly sampled points from the MCMC chains. (4)}
\figsetgrpend

\figsetgrpstart
\figsetgrpnum{figurenumber.5}
\figsetgrptitle{other CMD HK 5}
\figsetplot{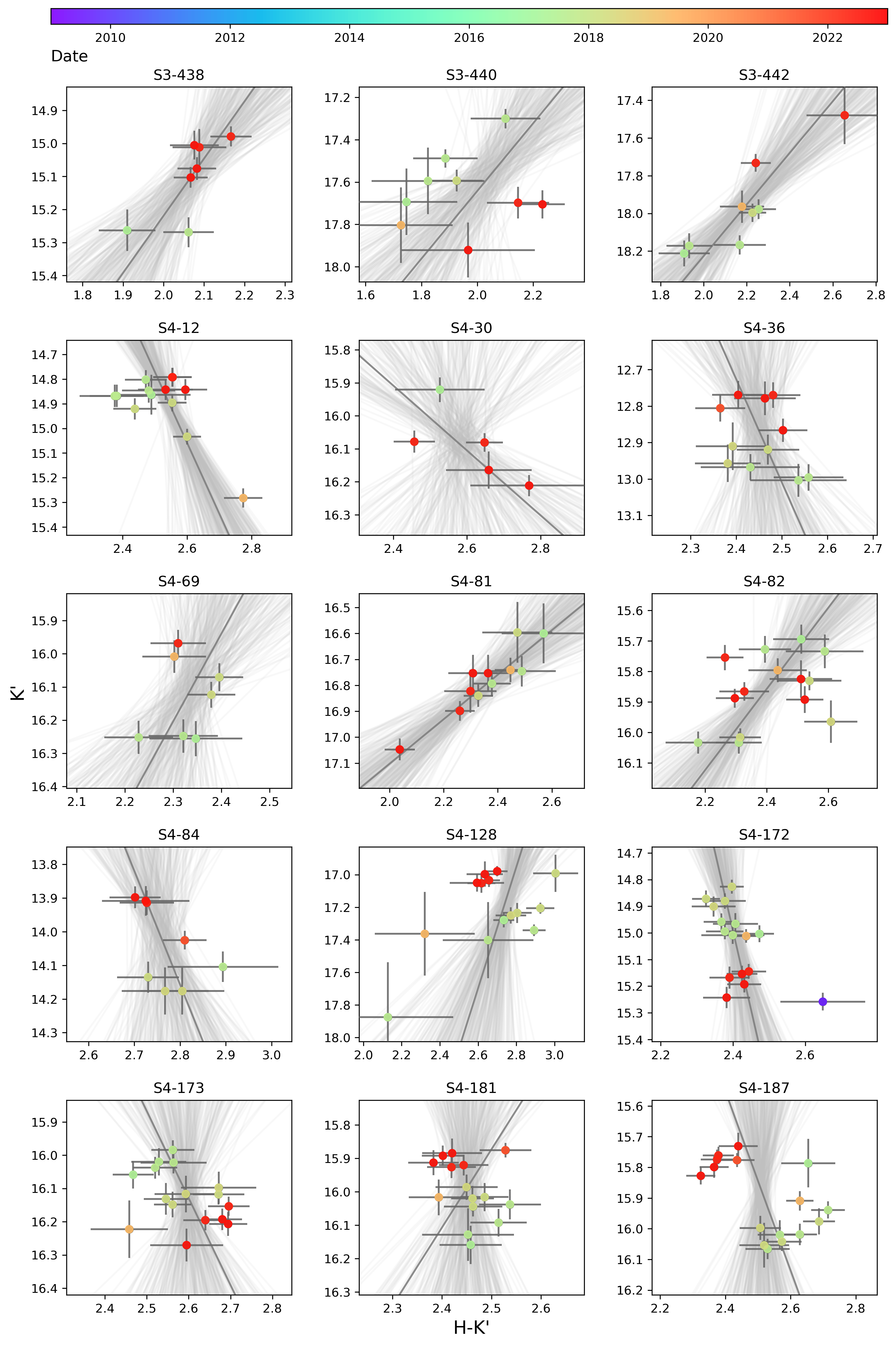}
\figsetgrpnote{K'/H-K' CMD for non-extinction stars. The over-plotted grey lines are 100 randomly sampled points from the MCMC chains. (5)}
\figsetgrpend

\figsetgrpstart
\figsetgrpnum{figurenumber.6}
\figsetgrptitle{other CMD HK 6}
\figsetplot{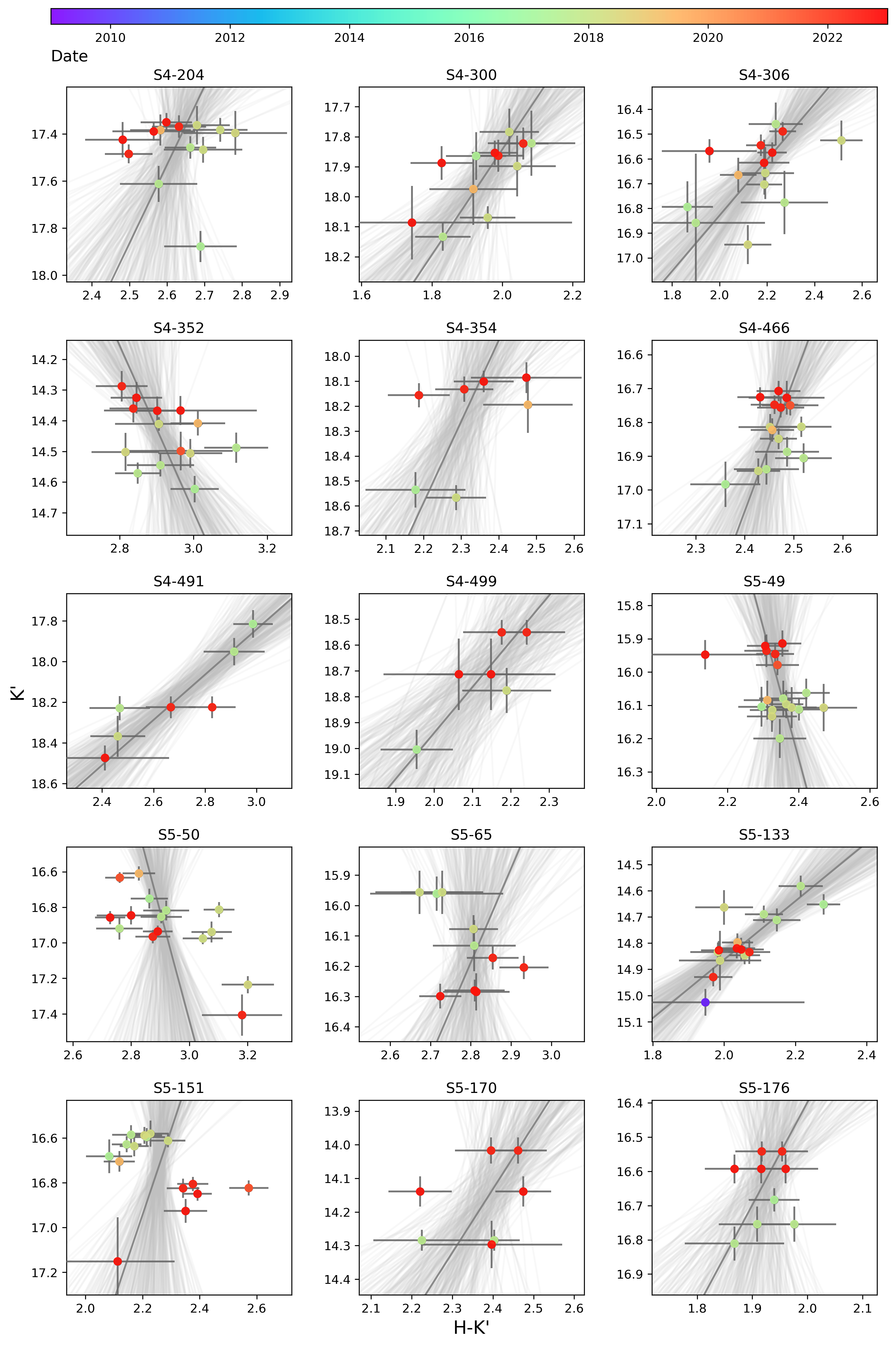}
\figsetgrpnote{K'/H-K' CMD for non-extinction stars. The over-plotted grey lines are 100 randomly sampled points from the MCMC chains. (6)}
\figsetgrpend

\figsetgrpstart
\figsetgrpnum{figurenumber.7}
\figsetgrptitle{other CMD HK 7}
\figsetplot{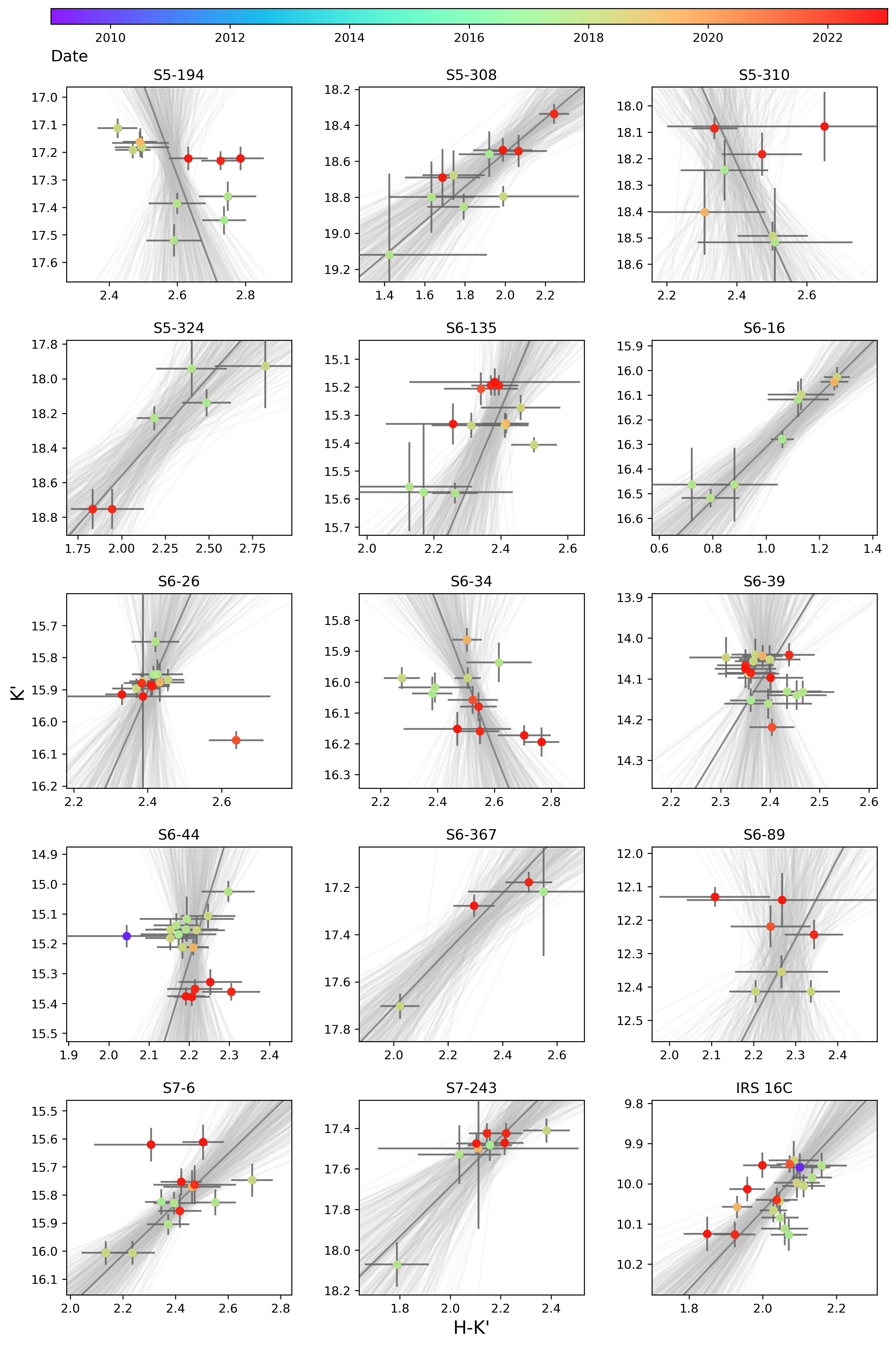}
\figsetgrpnote{K'/H-K' CMD for non-extinction stars. The over-plotted grey lines are 100 randomly sampled points from the MCMC chains. (7)}
\figsetgrpend

\figsetgrpstart
\figsetgrpnum{figurenumber.8}
\figsetgrptitle{other CMD HK 8}
\figsetplot{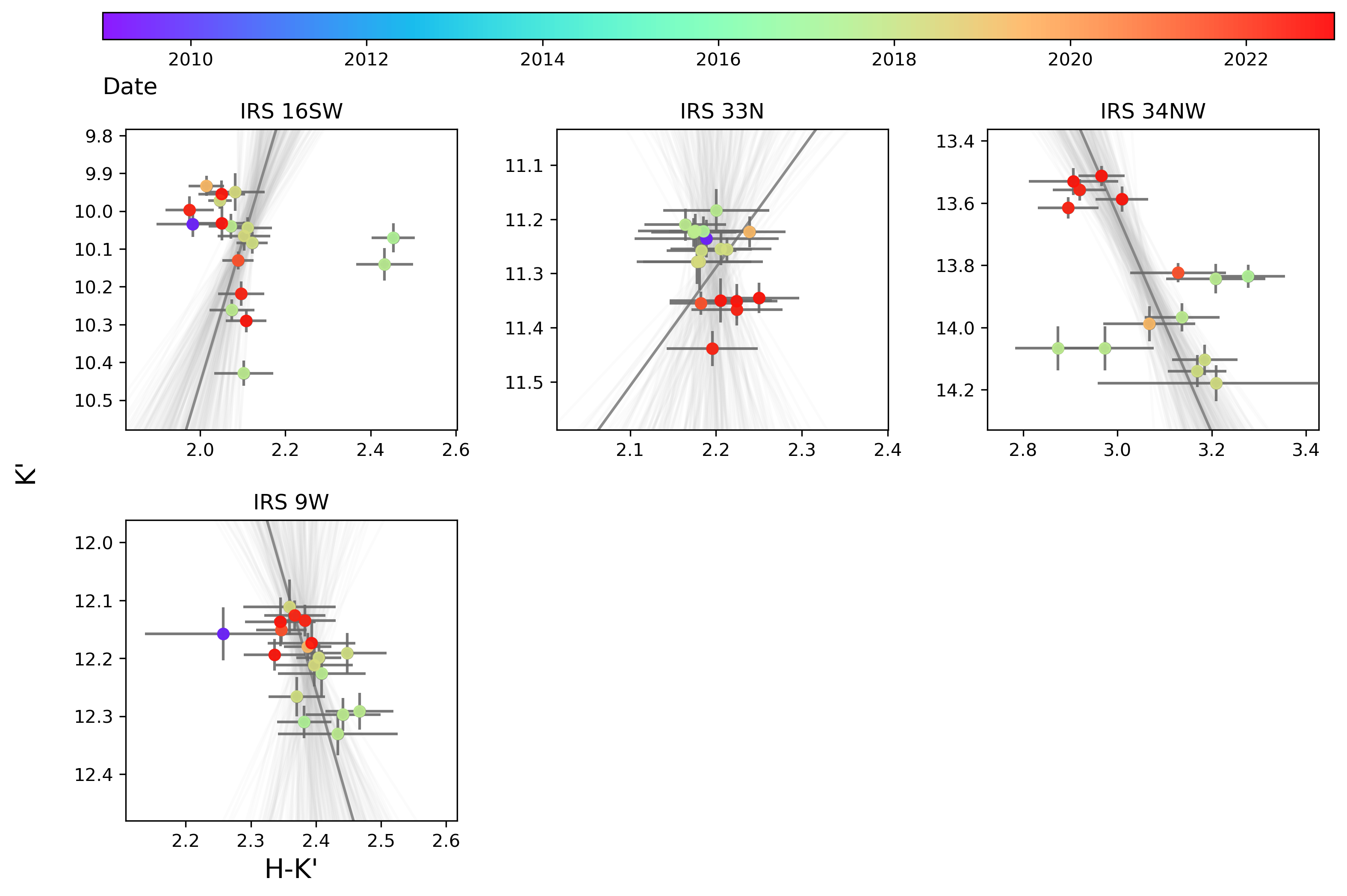}
\figsetgrpnote{K'/H-K' CMD for non-extinction stars. The over-plotted grey lines are 100 randomly sampled points from the MCMC chains. (8)}
\figsetgrpend
\figsetend

\begin{figure}[ht]
\centering
\includegraphics[width=0.7\textwidth]{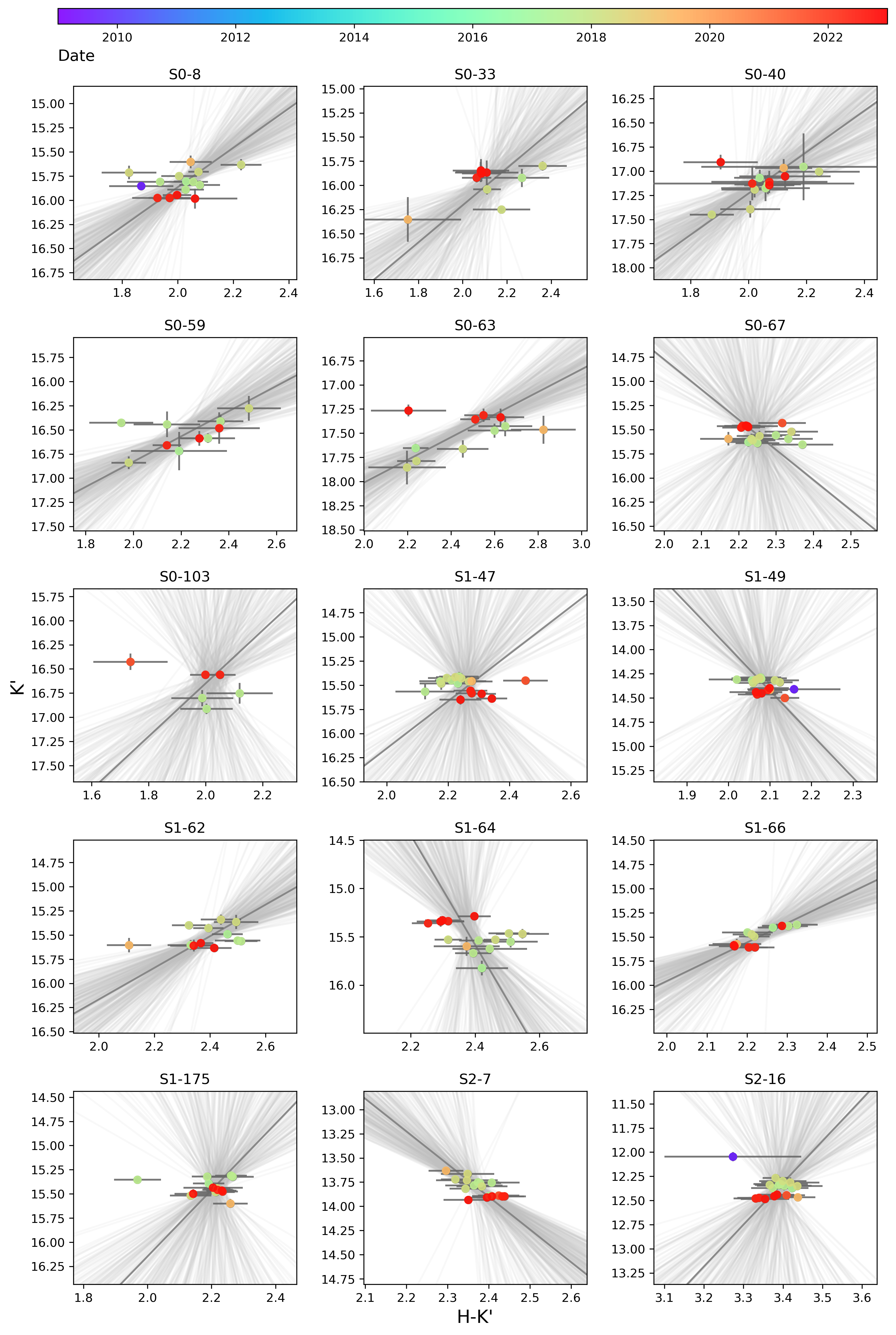}
\caption{K'/H-K' CMD for non-extinction stars and the best fit linear model (in dark grey). The over-plotted grey lines are 100 randomly sampled points from the MCMC chains (1).}
\end{figure}
\begin{figure}[ht]
\centering
\includegraphics[width=0.7\textwidth]{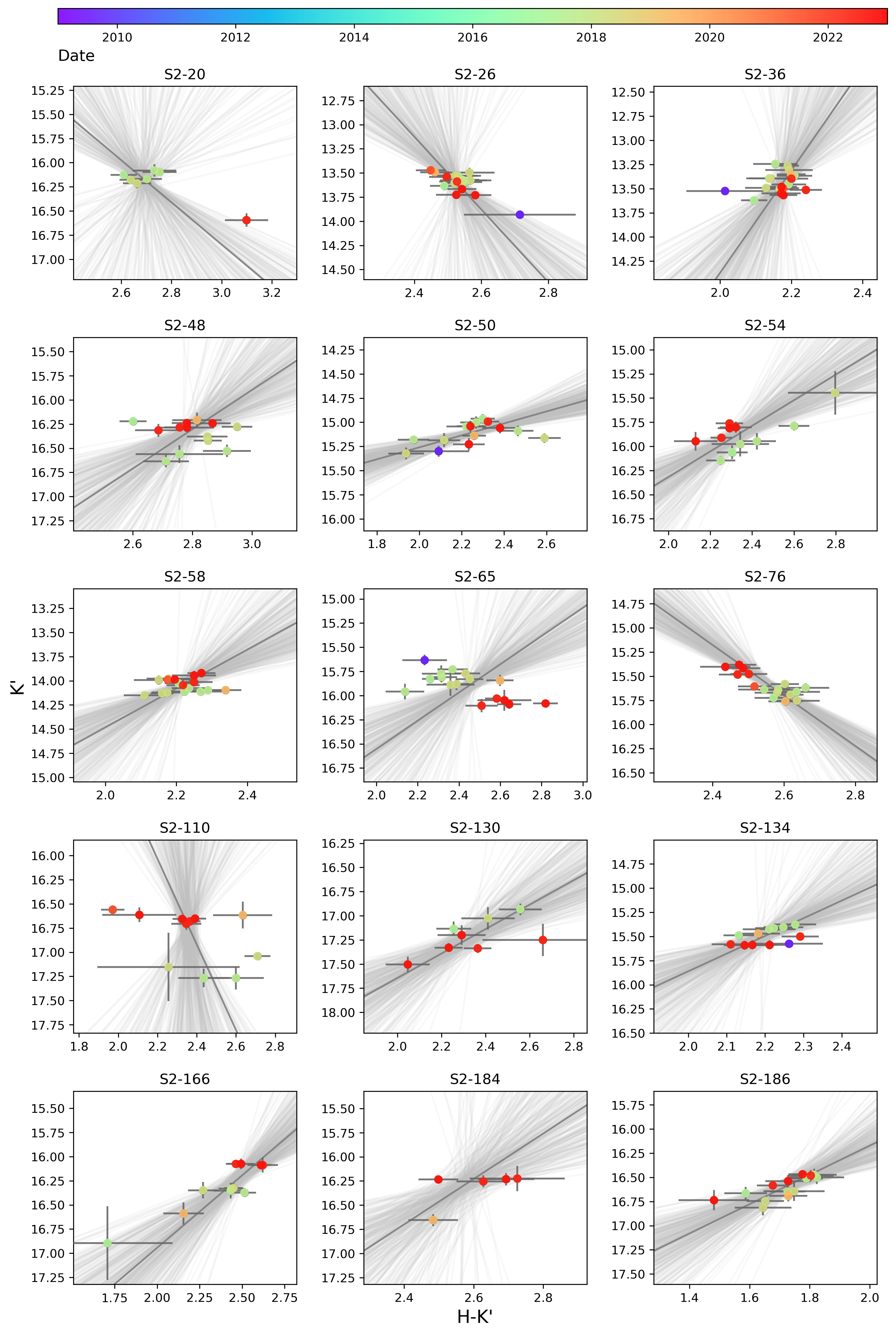}
\caption{K'/H-K' CMD for non-extinction stars and the best fit linear model (in dark grey). The over-plotted grey lines are 100 randomly sampled points from the MCMC chains (2).}
\end{figure}
\begin{figure}[ht]
\centering
\includegraphics[width=0.7\textwidth]{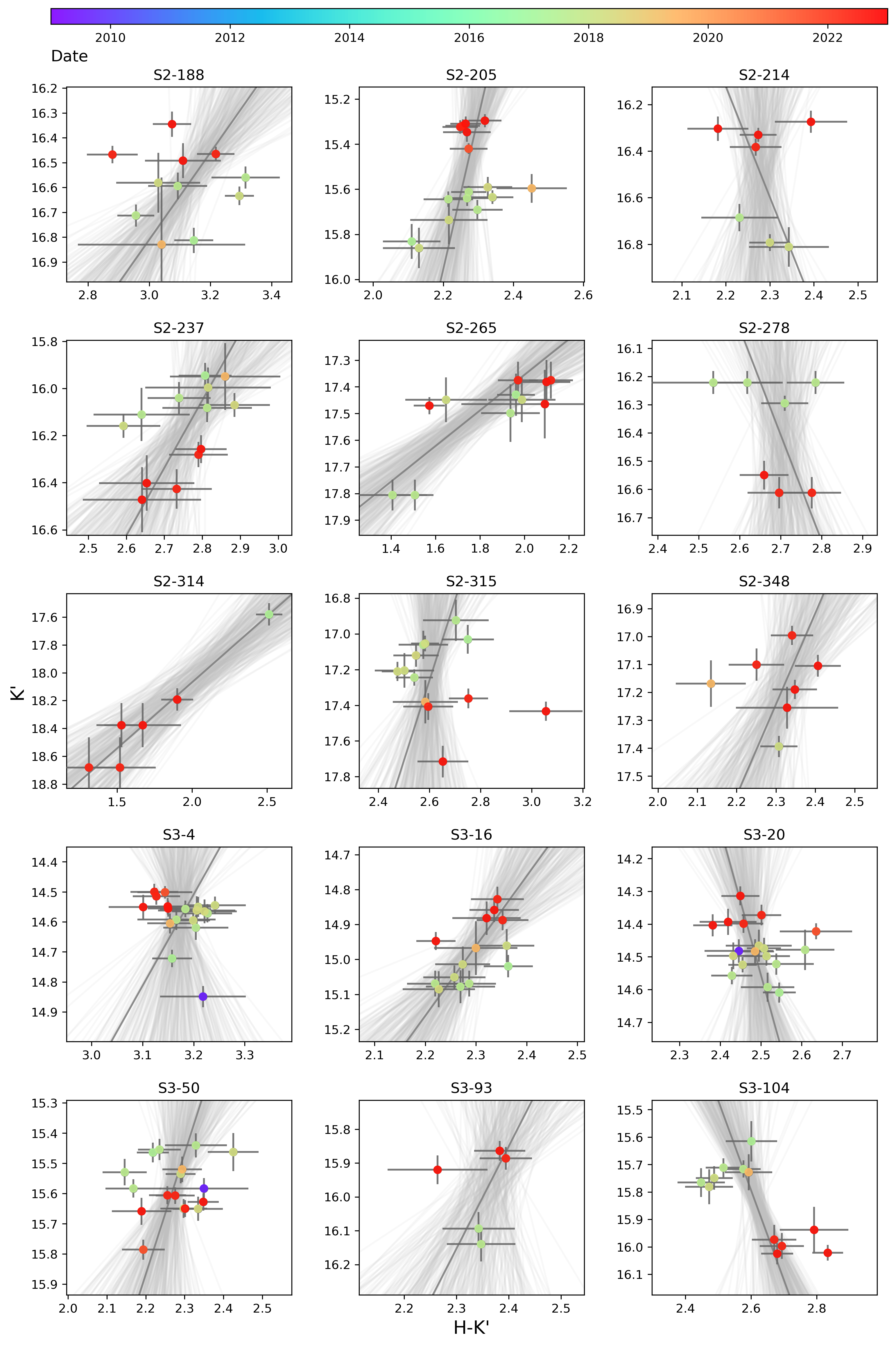}
\caption{K'/H-K' CMD for non-extinction stars and the best fit linear model (in dark grey). The over-plotted grey lines are 100 randomly sampled points from the MCMC chains (3).}
\end{figure}
\begin{figure}[ht]
\centering
\includegraphics[width=0.7\textwidth]{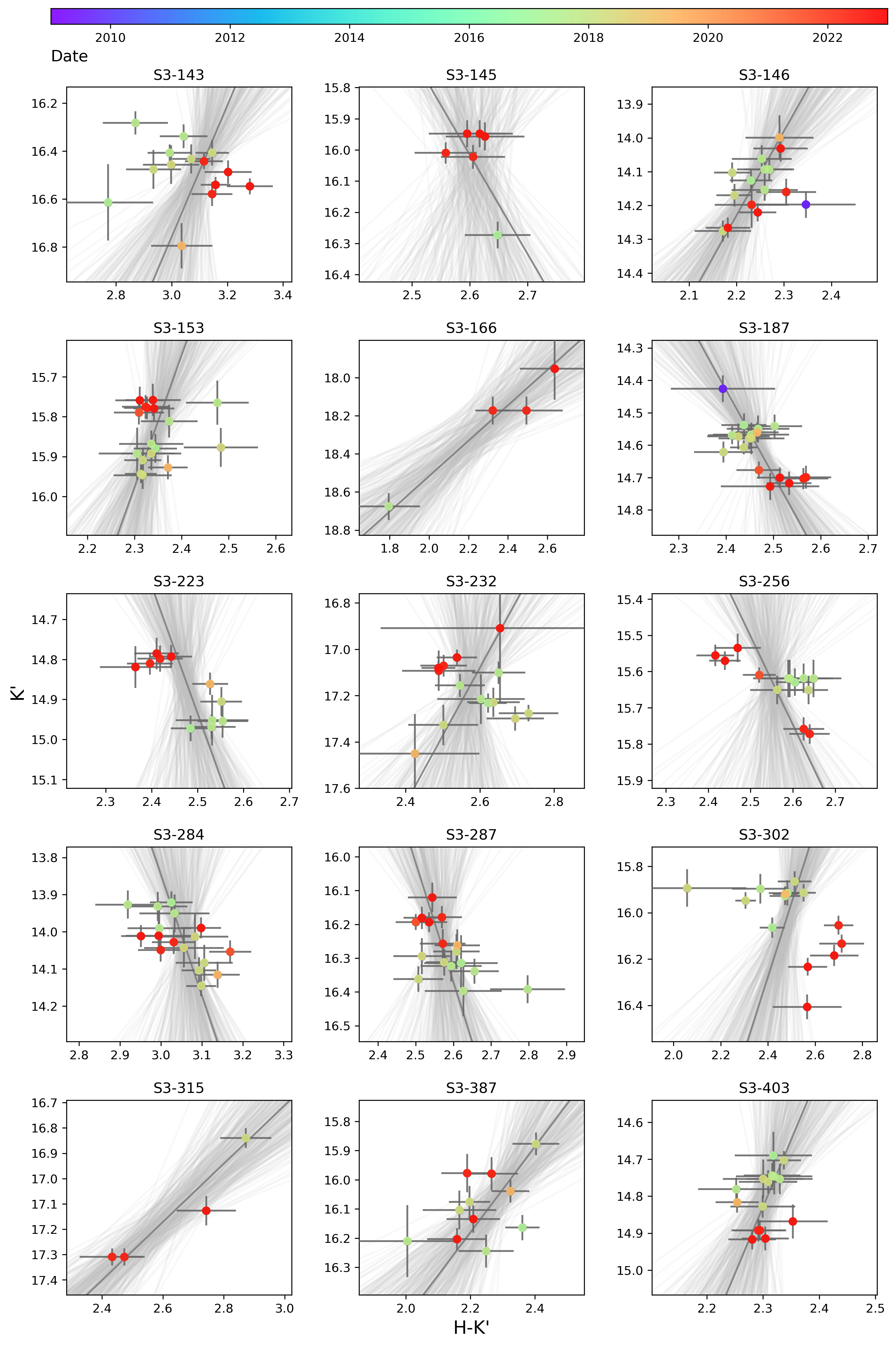}
\caption{K'/H-K' CMD for non-extinction stars and the best fit linear model (in dark grey). The over-plotted grey lines are 100 randomly sampled points from the MCMC chains (4).}
\end{figure}
\begin{figure}[ht]
\centering
\includegraphics[width=0.7\textwidth]{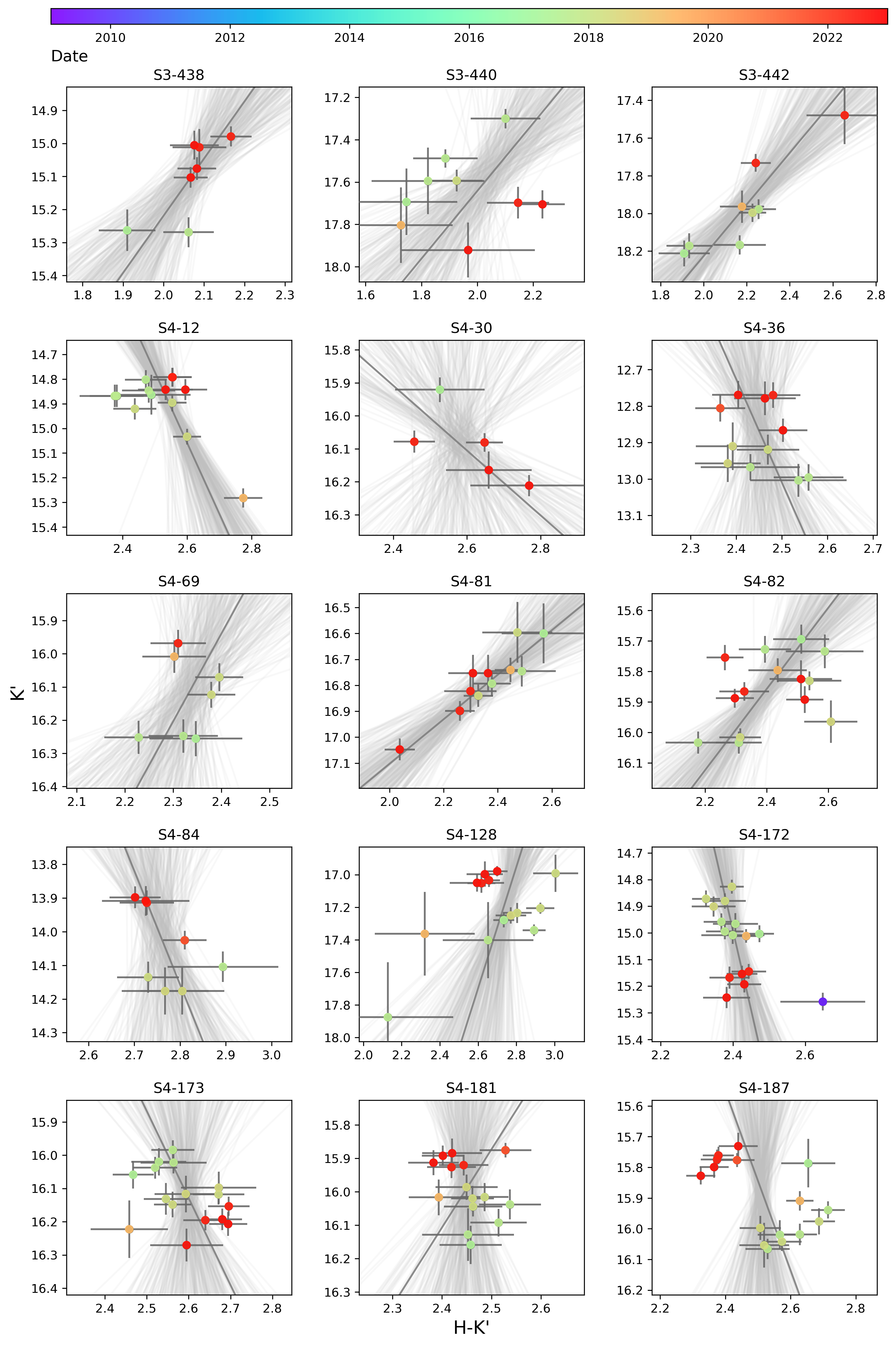}
\caption{K'/H-K' CMD for non-extinction stars and the best fit linear model (in dark grey). The over-plotted grey lines are 100 randomly sampled points from the MCMC chains (5).}
\end{figure}
\begin{figure}[ht]
\centering
\includegraphics[width=0.7\textwidth]{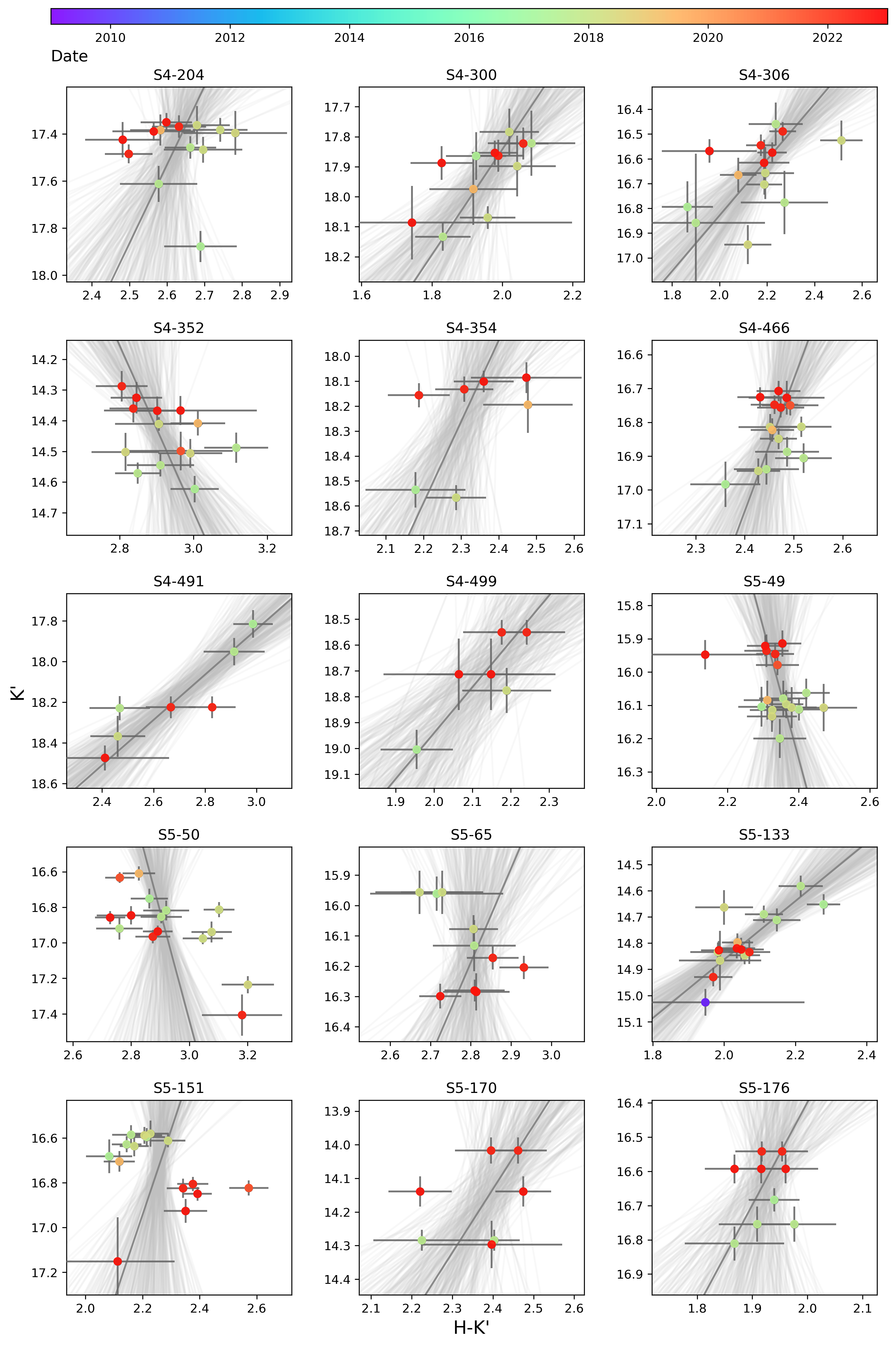}
\caption{K'/H-K' CMD for non-extinction stars and the best fit linear model (in dark grey). The over-plotted grey lines are 100 randomly sampled points from the MCMC chains (6).}
\end{figure}
\begin{figure}[ht]
\centering
\includegraphics[width=0.7\textwidth]{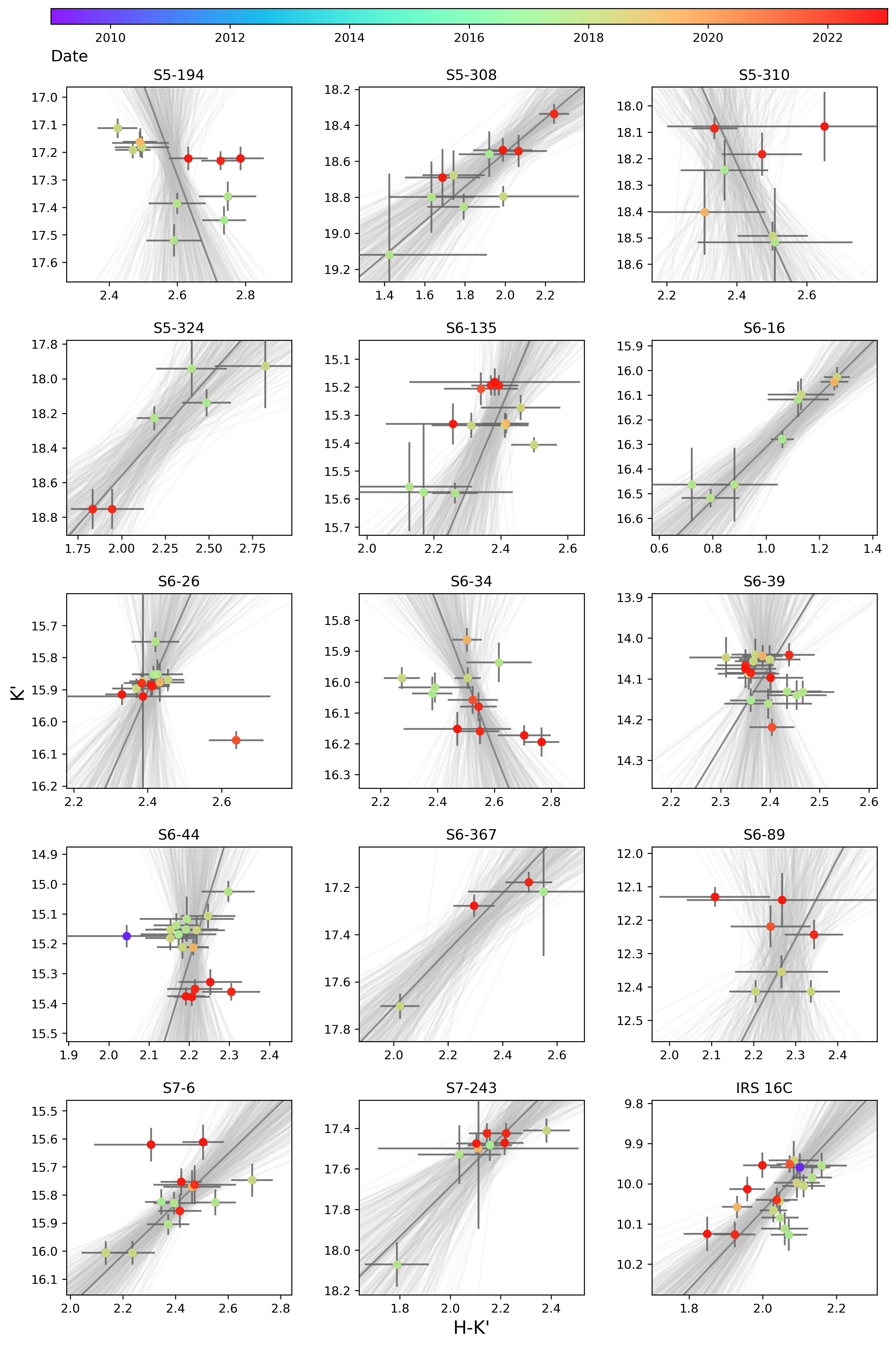}
\caption{K'/H-K' CMD for non-extinction stars and the best fit linear model (in dark grey). The over-plotted grey lines are 100 randomly sampled points from the MCMC chains (7).}
\end{figure}
\begin{figure}[ht]
\centering
\includegraphics[width=0.7\textwidth]{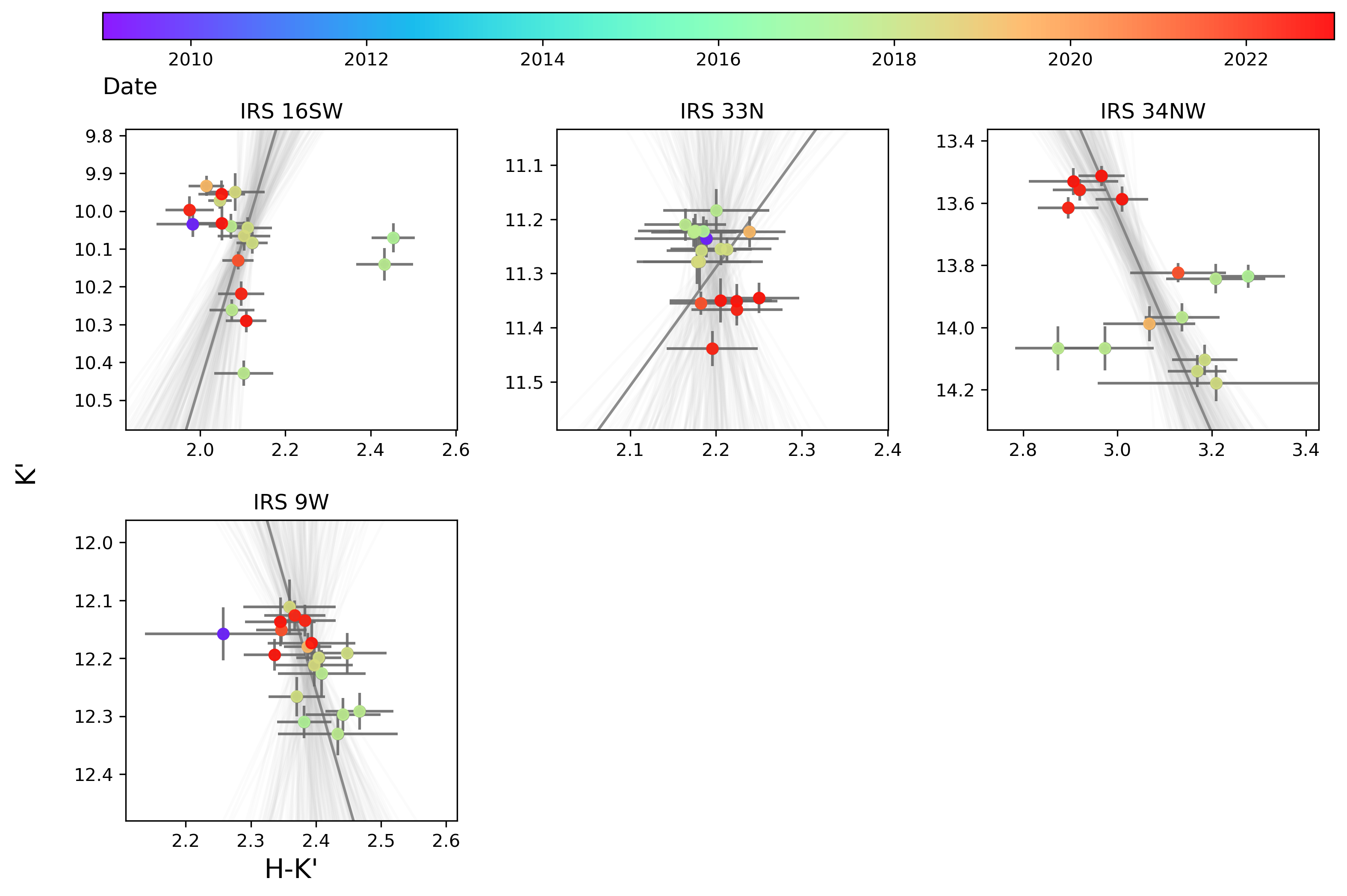}
\caption{K'/H-K' CMD for non-extinction stars and the best fit linear model (in dark grey). The over-plotted grey lines are 100 randomly sampled points from the MCMC chains (8).}
\end{figure}

\clearpage
\section{Likely extinction sources K' lightcurves} 
\label{sec:kp_lightcurves}
\figsetstart
\figsetnum{2}
\figsettitle{Likely extinction sources K' lightcurves}

\figsetgrpstart
\figsetgrpnum{figurenumber.1}
\figsetgrptitle{K' lightcurves for likely extinction sources}
\figsetplot{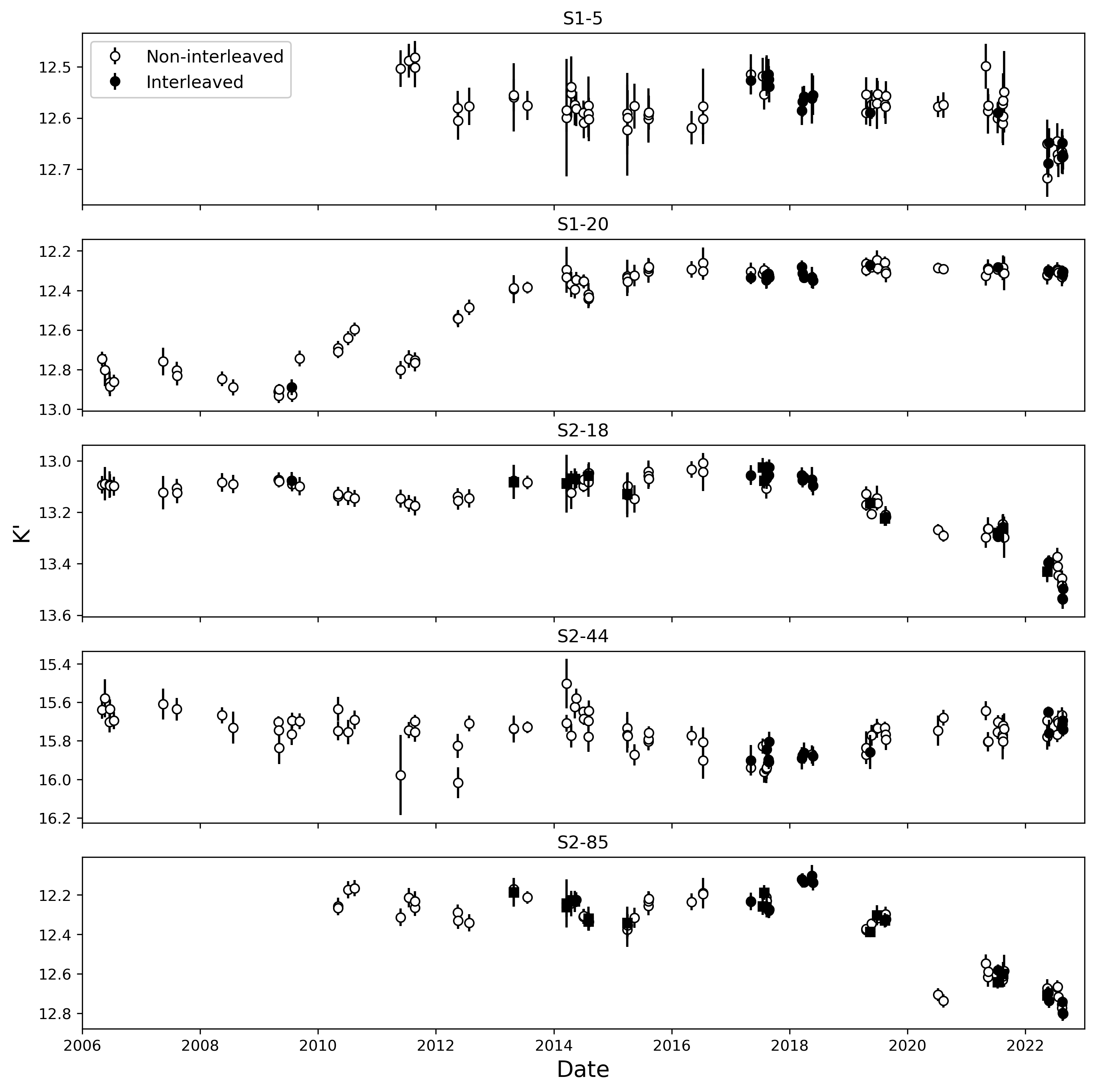}
\figsetgrpnote{K' lightcurve for likely extinction sources. Unfilled points are non-interleaved measurements, while black filled points are interleaved (squares for K'/L' data and circles for K'/H data). (1)}
\figsetgrpend

\figsetgrpstart
\figsetgrpnum{figurenumber.2}
\figsetgrptitle{K' lightcurves for likely extinction sources}
\figsetplot{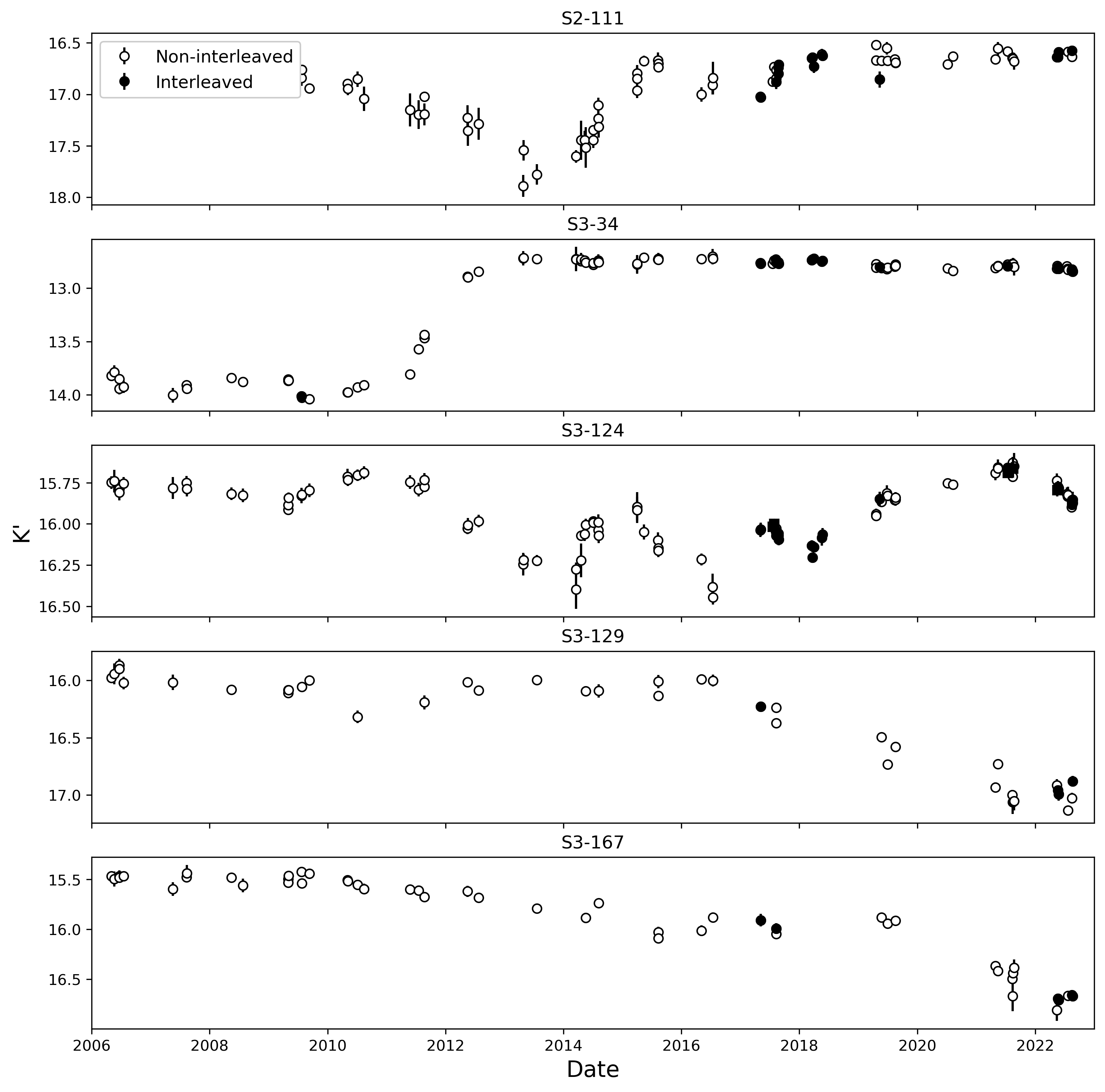}
\figsetgrpnote{K' lightcurve for likely extinction sources. Unfilled points are non-interleaved measurements, while black filled points are interleaved (squares for K'/L' data and circles for K'/H data). (2)}
\figsetgrpend

\figsetgrpstart
\figsetgrpnum{figurenumber.3}
\figsetgrptitle{K' lightcurves for likely extinction sources}
\figsetplot{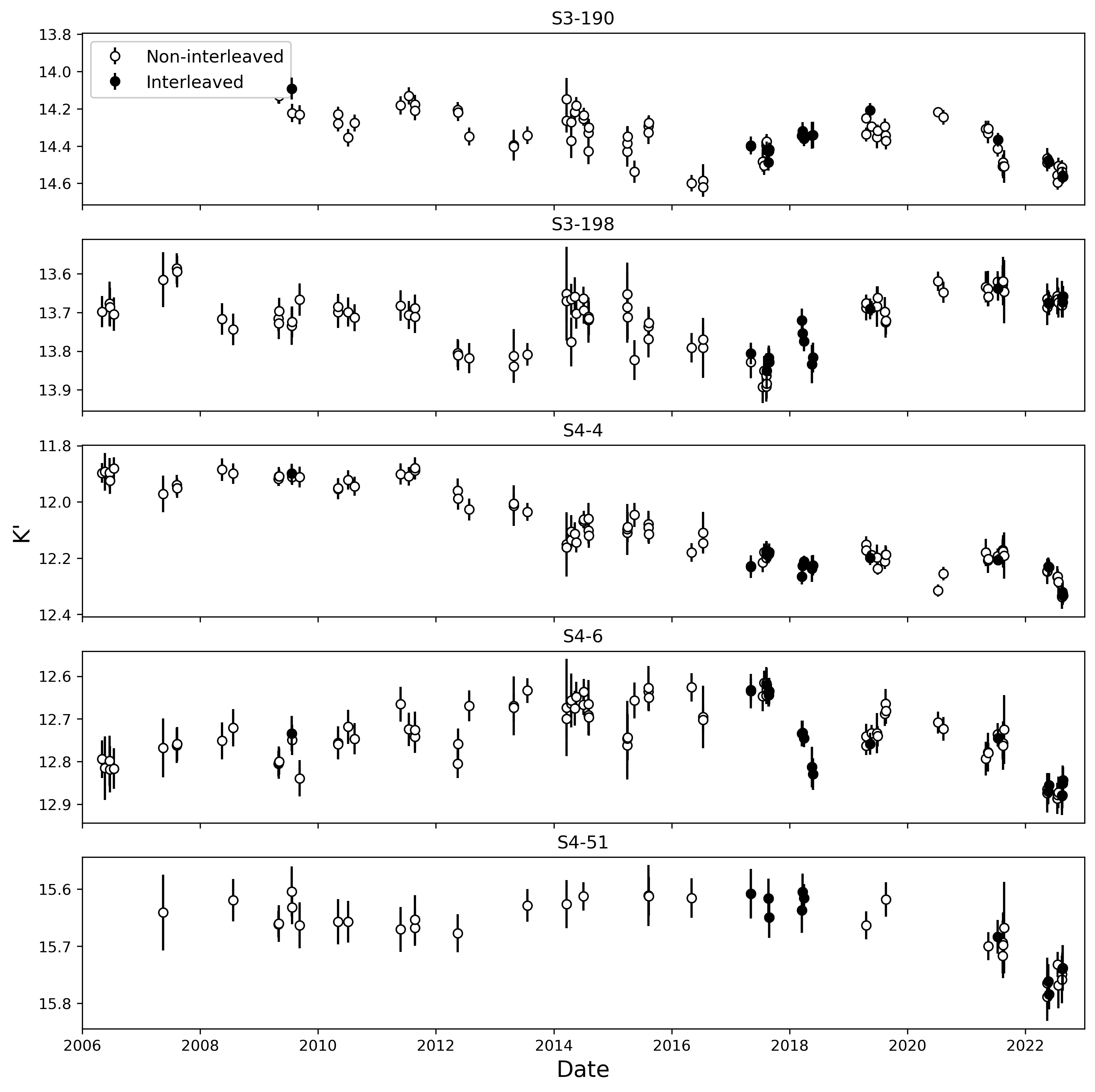}
\figsetgrpnote{K' lightcurve for likely extinction sources. Unfilled points are non-interleaved measurements, while black filled points are interleaved (squares for K'/L' data and circles for K'/H data). (3)}
\figsetgrpend

\figsetgrpstart
\figsetgrpnum{figurenumber.4}
\figsetgrptitle{K' lightcurves for likely extinction sources}
\figsetplot{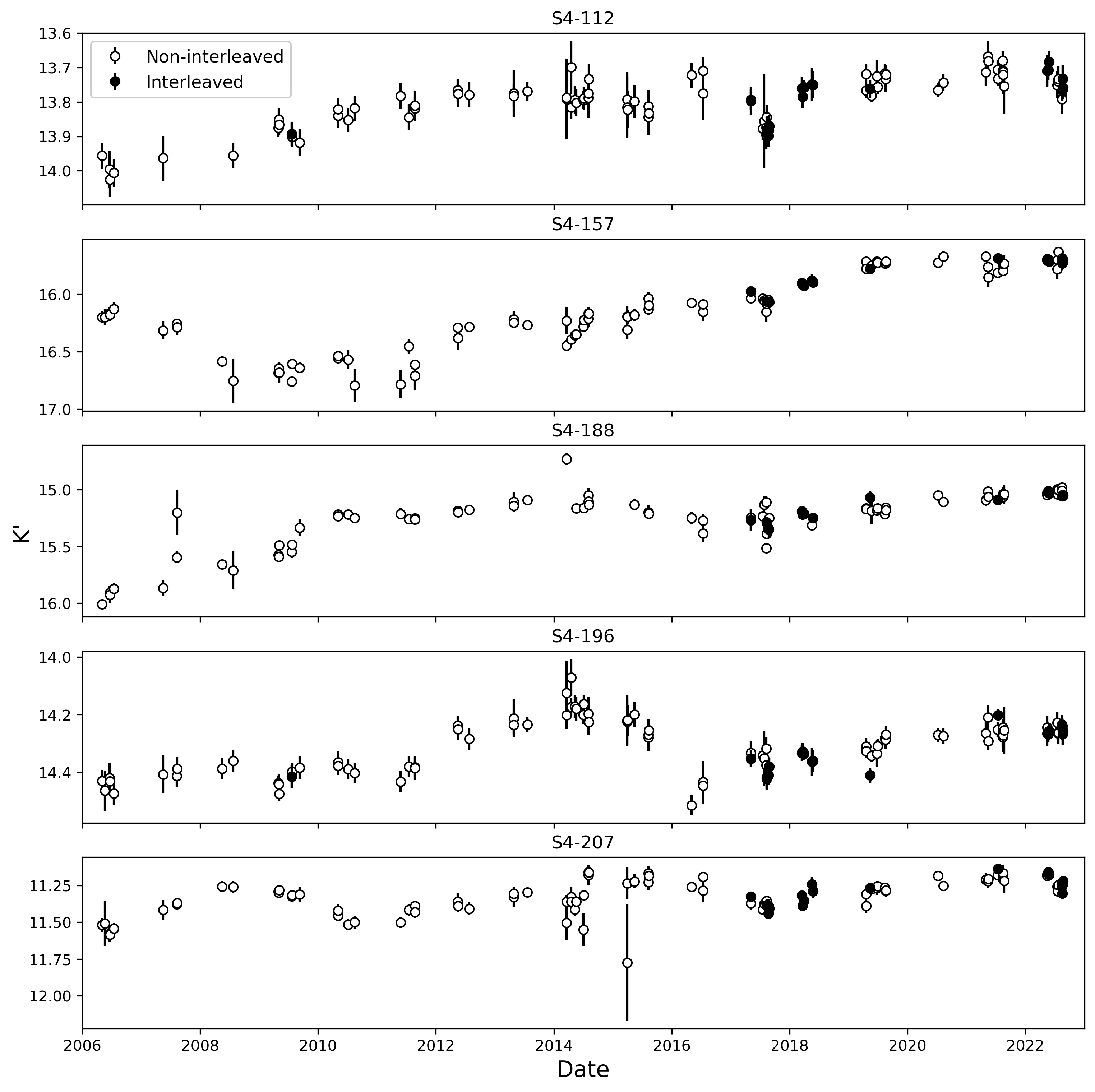}
\figsetgrpnote{K' lightcurve for likely extinction sources. Unfilled points are non-interleaved measurements, while black filled points are interleaved (squares for K'/L' data and circles for K'/H data). (4)}
\figsetgrpend

\figsetgrpstart
\figsetgrpnum{figurenumber.5}
\figsetgrptitle{K' lightcurves for likely extinction sources}
\figsetplot{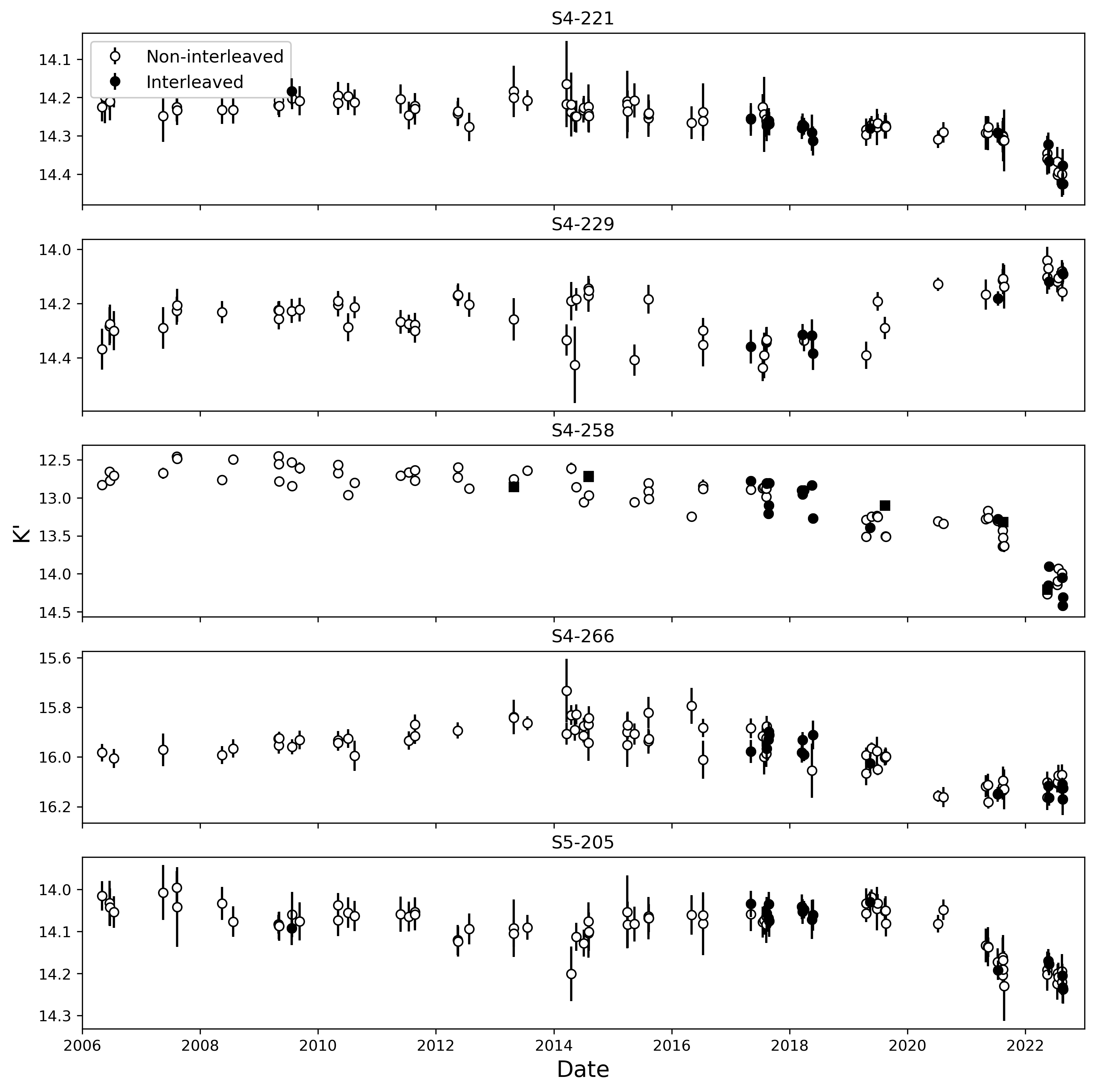}
\figsetgrpnote{K' lightcurve for likely extinction sources. Unfilled points are non-interleaved measurements, while black filled points are interleaved (squares for K'/L' data and circles for K'/H data). (5)}
\figsetgrpend

\figsetgrpstart
\figsetgrpnum{figurenumber.6}
\figsetgrptitle{K' lightcurves for likely extinction sources}
\figsetplot{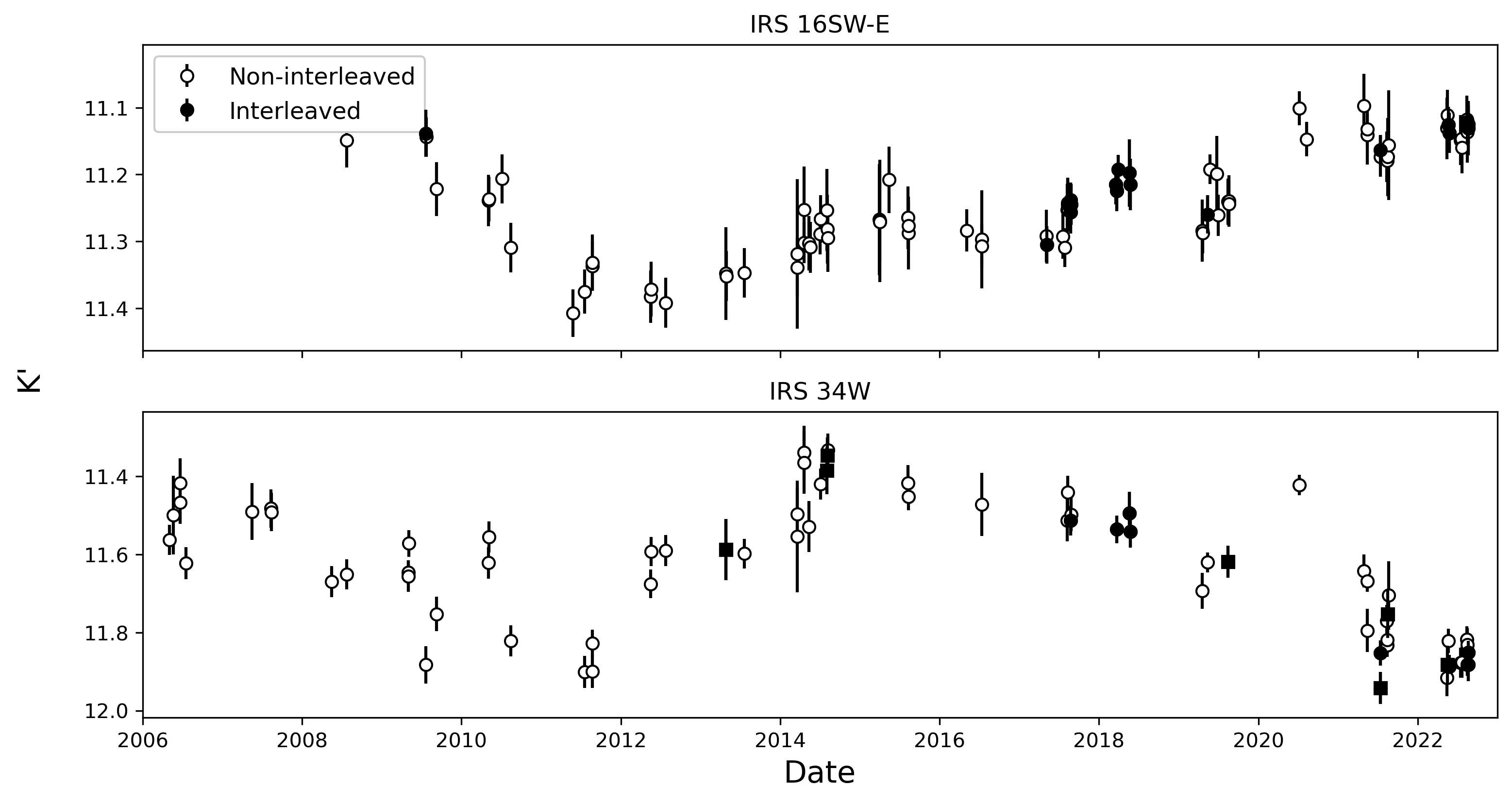}
\figsetgrpnote{K' lightcurve for likely extinction sources. Unfilled points are non-interleaved measurements, while black filled points are interleaved (squares for K'/L' data and circles for K'/H data). (6)}
\figsetgrpend

\begin{figure}[ht]
    \centering
    \includegraphics[width=0.99\textwidth]{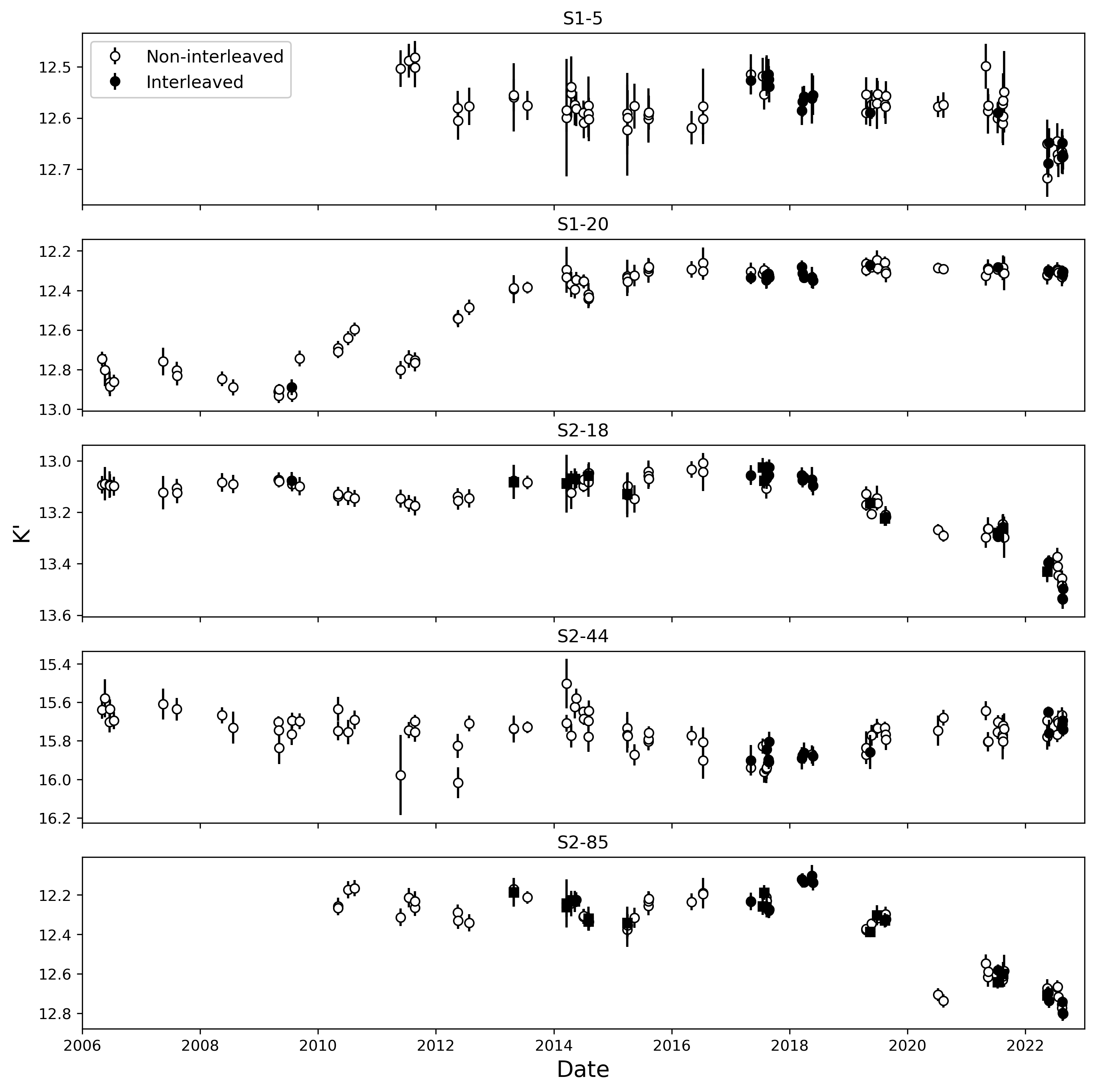}
    \caption{K' lightcurve for likely extinction sources. Unfilled points are non-interleaved measurements, while black filled points are interleaved (squares for K'/L' data and circles for K'/H data) (1).}
\end{figure}
\begin{figure}[ht]
    \centering
    \includegraphics[width=0.99\textwidth]{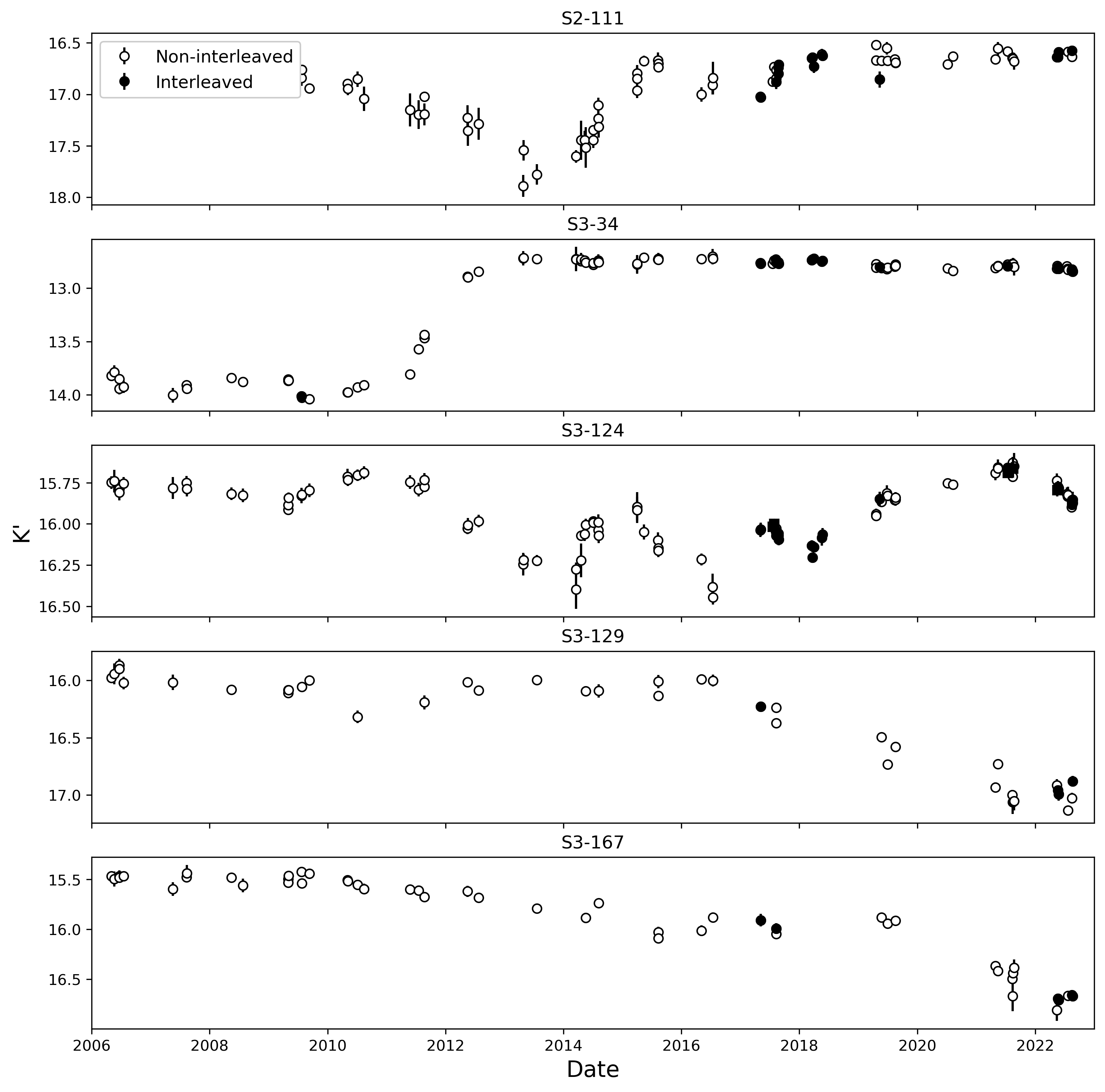}
    \caption{K' lightcurve for likely extinction sources. Unfilled points are non-interleaved measurements, while black filled points are interleaved (squares for K'/L' data and circles for K'/H data) (2.)}
\end{figure}
\begin{figure}[ht]
    \centering
    \includegraphics[width=0.99\textwidth]{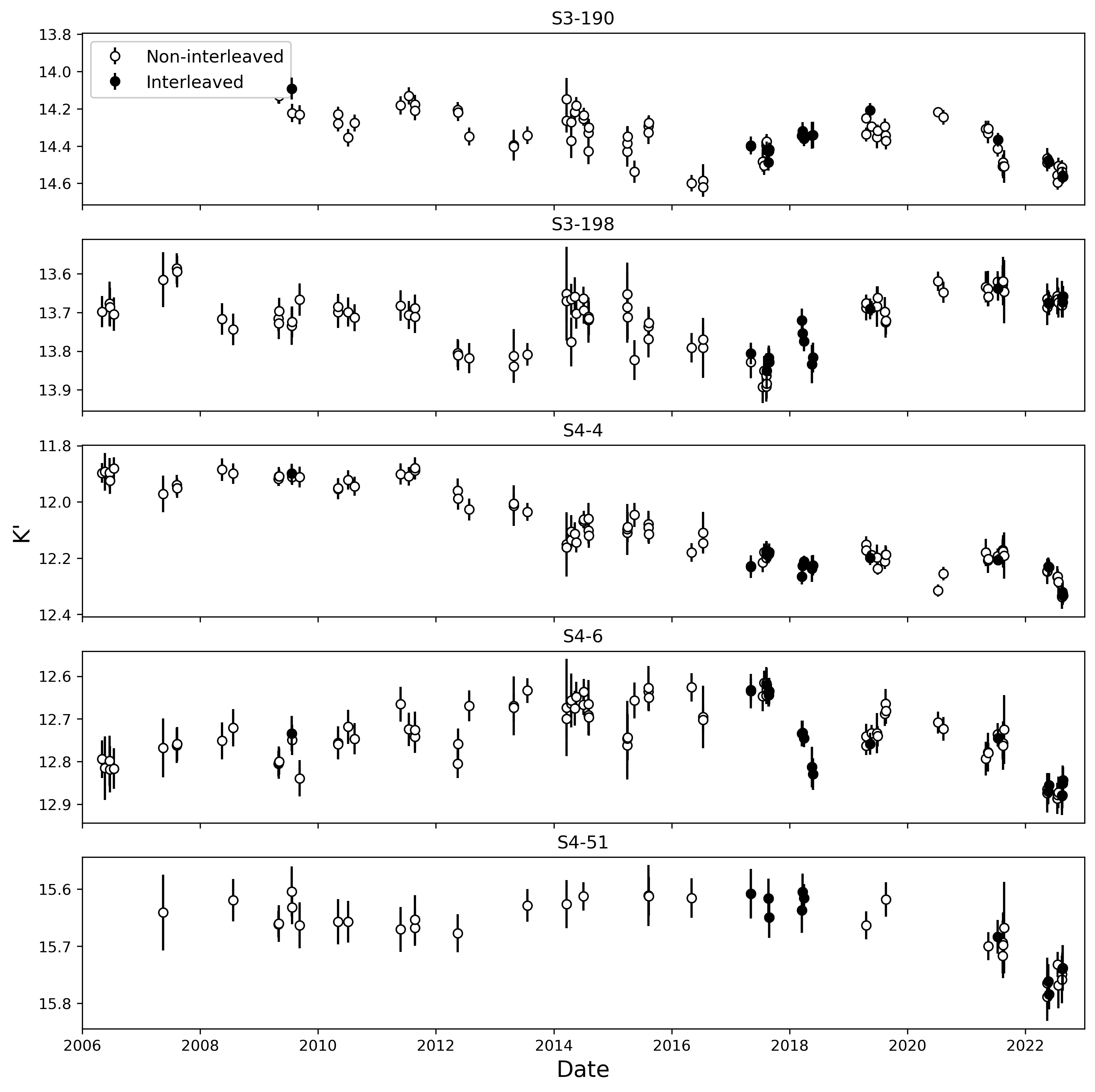}
    \caption{K' lightcurve for likely extinction sources. Unfilled points are non-interleaved measurements, while black filled points are interleaved (squares for K'/L' data and circles for K'/H data) (3).}
\end{figure}
\begin{figure}[ht]
    \centering
    \includegraphics[width=0.99\textwidth]{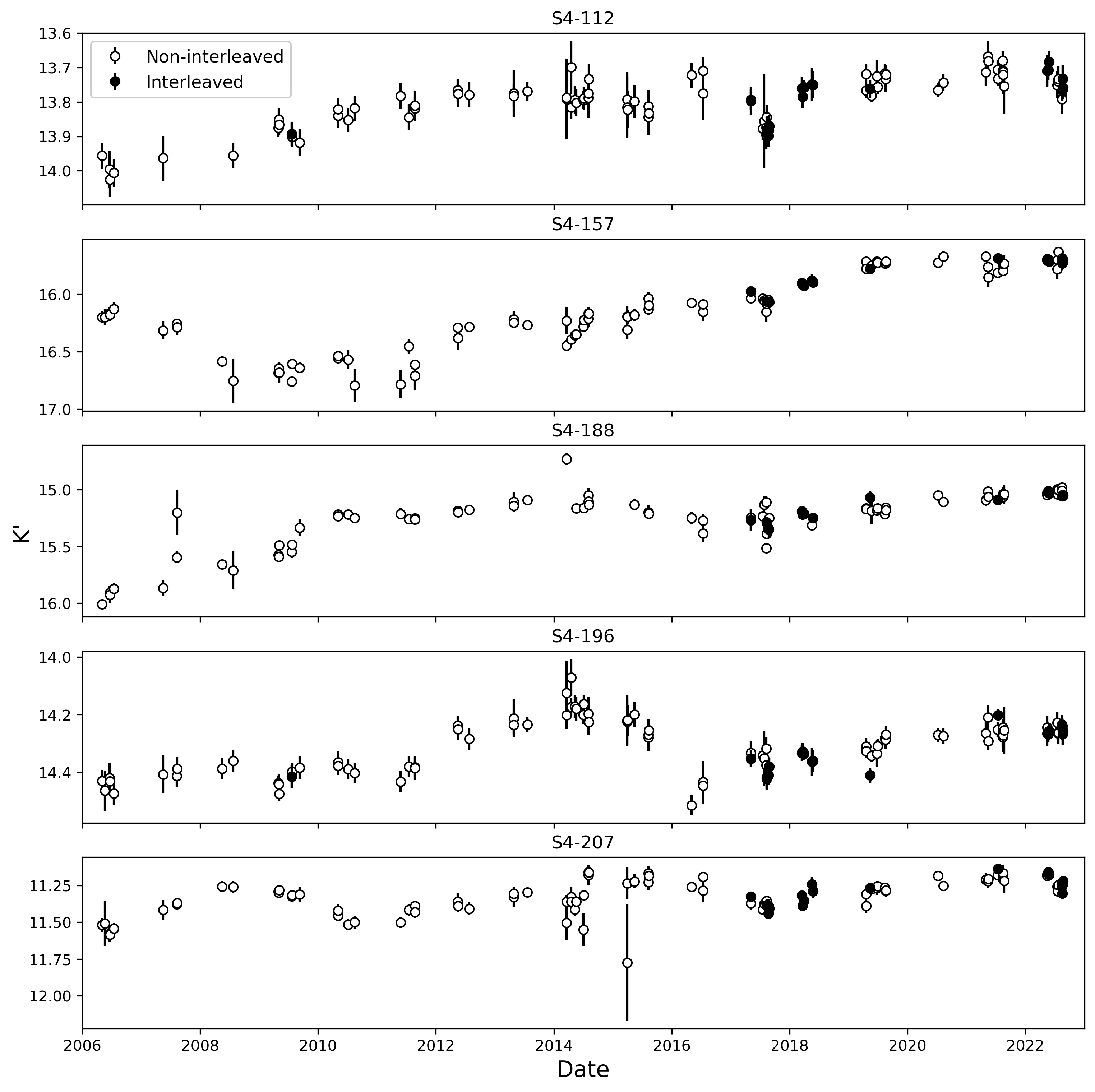}
    \caption{K' lightcurve for likely extinction sources. Unfilled points are non-interleaved measurements, while black filled points are interleaved (squares for K'/L' data and circles for K'/H data) (4)}
\end{figure}
\begin{figure}[ht]
    \centering
    \includegraphics[width=0.99\textwidth]{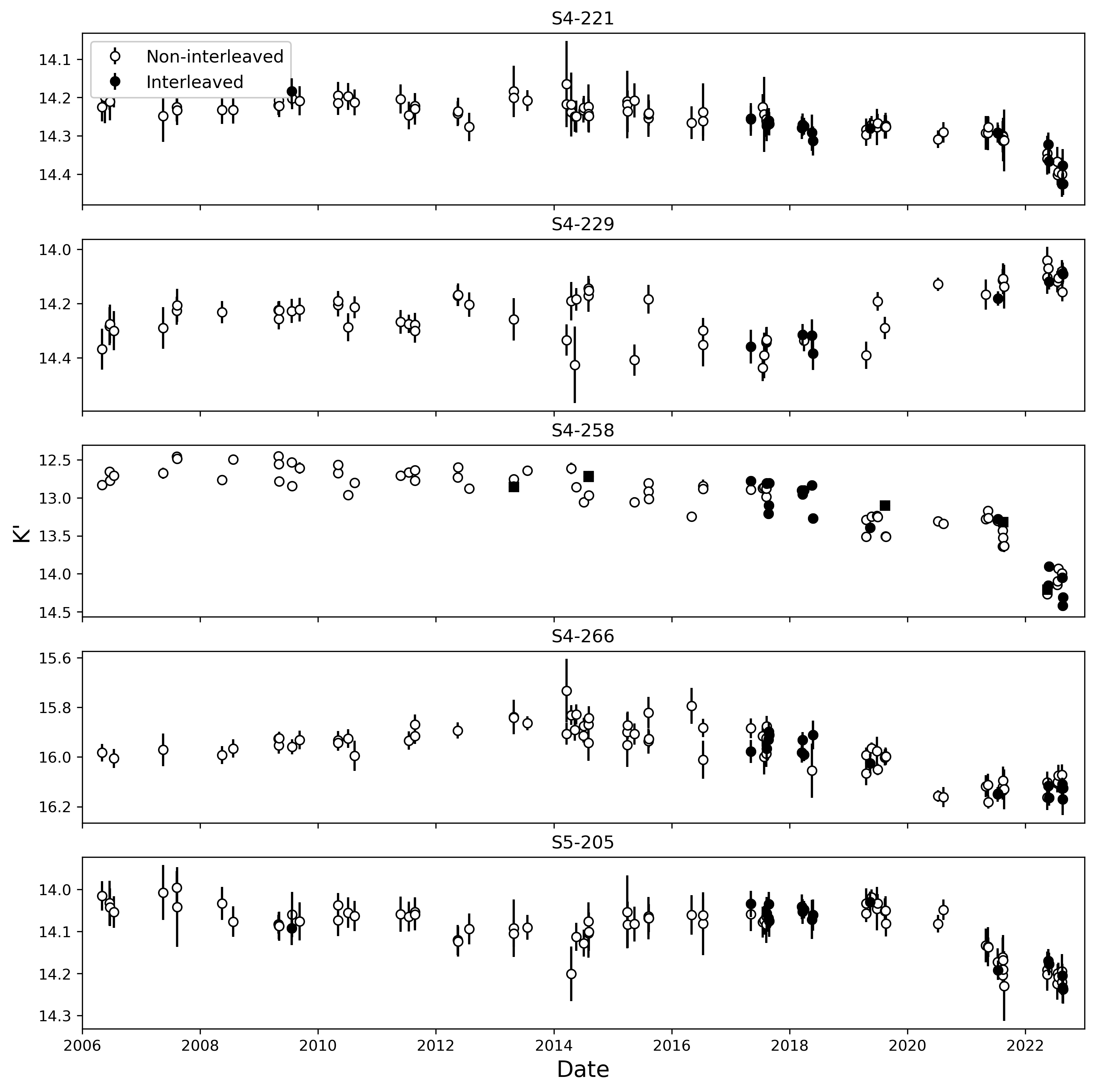}
    \caption{K' lightcurve for likely extinction sources. Unfilled points are non-interleaved measurements, while black filled points are interleaved (squares for K'/L' data and circles for K'/H data) (5).}
\end{figure}
\begin{figure}[ht]
    \centering
    \includegraphics[width=0.99\textwidth]{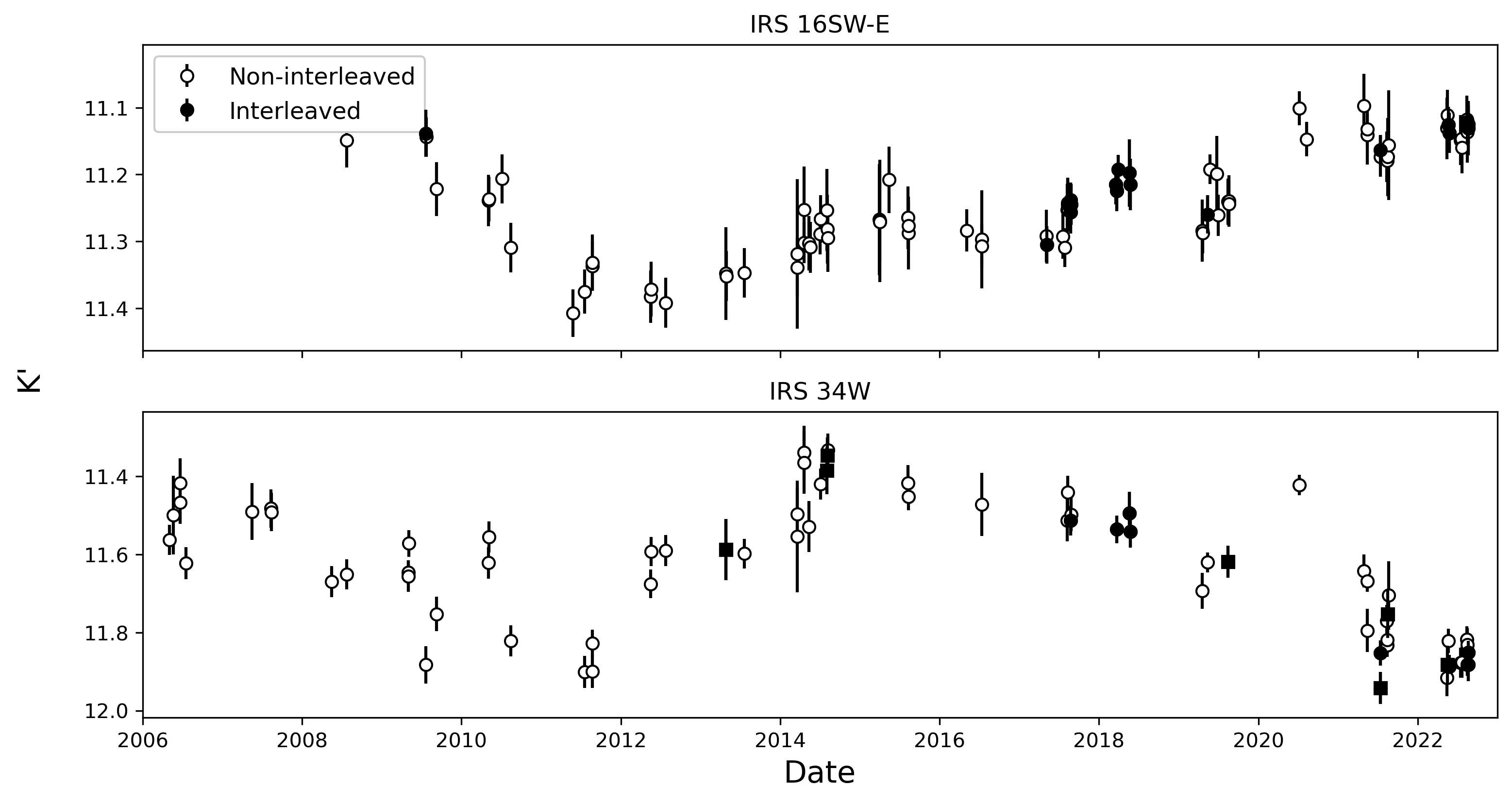}
    \caption{K' lightcurve for likely extinction sources. Unfilled points are non-interleaved measurements, while black filled points are interleaved (squares for K'/L' data and circles for K'/H data) (6).}
\end{figure}

\clearpage
\section{L' magnitudes and colors for non-variable K' sources}
\setcounter{table}{0}
\renewcommand{\thetable}{D\arabic{table}}
\input{lp_phot_non_var.tex}

\clearpage
\bibliography{bib}
\end{document}

%% file: lp_phot.tex
\begin{deluxetable}{lcc}
\tablecaption{Characteristics of L' photometric calibrators.}
\label{table:lp_photometry}
\tablehead{Star & Mean L' mag & L' $\chi^2_{\mathrm{\nu}}$}
\startdata
S3-22 & $9.5 \pm 0.03$ & 0.44 \\
S1-17 &  $10.6 \pm 0.03$ & 0.56 \\
S4-6 &  $11.2 \pm 0.03$ & 0.43 \\
S4-3 &  $11.5 \pm 0.03$ & 1.42 \\
S3-370 &  $12.2 \pm 0.03$ & 0.63 \\
S3-88 &  $12.7 \pm 0.04$ & 0.87 \\
\enddata
\end{deluxetable}

%% file: extinction.tex
\startlongtable
\begin{longrotatetable}
\begin{deluxetable*}{lcccccccccc}
\tablecaption{Characteristics of the variability sample that is consistent with changing extinction. Columns 2 through 5 are related to the K' variability. Column 2 states the mean K' magnitude for each source, $\langle{\mathrm{K'}}\rangle$, and column 3 describes the change in K' magnitude, $\Delta{\mathrm{K'}}$. Column 4 is the reduced chi-squared for variability metric, as outlined in Section \ref{sec:sample} and Column 5 is the degrees-of-freedom for the color measurements (one less than the number of epochs observed). Columns 6 and 9 are the modeled slopes in color-magnitude ($\theta_{\mathrm{H-K'}}$ and $\theta_{\mathrm{K'-L'}}$); columns 7 and 10 are the expected mean color ($\langle{\mathrm{H-K'}}\rangle$ and $\langle{\mathrm{K'-L'}}\rangle$) derived from the slope fits. Column 8, $\mathrm{P}(\mathrm{extinction})_{\mathrm{H-K'}}$, is the probability of membership to the H-K' extinction Gaussian. Finally, column 11 highlights the 9 sources we identify as ``dipping'' (see Section \ref{sec:dipping}).}
\label{table:extinction}
%\\
%\hline
%\hline
\tablehead{Star & $\langle{\mathrm{K'}}\rangle$ & $\Delta{\mathrm{K'}}$ & K' variability ($\chi_{\nu}^2$) & $\mathrm{d.o.f.}_{H-K'}$ & $\theta_{\mathrm{H-K'}}$ & $\langle{\mathrm{H-K'}}\rangle$ & $\mathrm{P}(\mathrm{extinction})_{\mathrm{H-K'}}$ & $\mathrm{\theta}_{\mathrm{K'-L'}}$ & $\langle{\mathrm{K'-L'}}\rangle$ & Dipping? \\
 & mag & mag & & & degrees & mag &  & degrees & mag &  \\}
%\hline
%\hline
%\endfirsthead

%\toprule
%    Star & $K'_{mean}$ & $\Delta{K'}$ & K' variability ($\chi_{\nu}^2$) & $\mathrm{d.o.f.}_{H-K'}$ & $\theta_{H-K'}$ & $\langle{H-K'}\rangle$ & $P(\mathrm{extinction})_{H-K'}$ & $\theta_{K'-L'}$ & $\langle{K'-L'}\rangle$ & Dipping? \\ 
%\hline
%\hline
%\endhead
%\multicolumn{11}{r}{{Continued on next page}} \\ 
%\hline
%\endfoot
%\endlastfoot
\startdata
S1-5      &   12.6 &     0.2 &        4.2 &       16 &   $55.9\pm6.7$ &     2.4 &        0.9 &           --- &     --- &     --- \\
S1-20     &   12.3 &     0.6 &       19.5 &       17 &   $60.1\pm5.9$ &     2.4 &        0.9 &           --- &     --- &     --- \\
S2-18     &   13.2 &     0.5 &       46.5 &       17 &   $51.0\pm1.5$ &     2.5 &        1.0 &  $82.3\pm6.6$ &     --- &     --- \\
S2-44     &   15.8 &     0.3 &        4.8 &       14 &  $60.6\pm11.5$ &     2.4 &        0.8 &           --- &     --- &     --- \\
S2-69     &   14.9 &     0.6 &       34.9 &       16 &   $59.2\pm2.2$ &     2.9 &        0.9 &  $84.5\pm9.5$ &     --- &     yes \\
S2-70     &   14.5 &     0.5 &       31.2 &       17 &   $53.5\pm1.9$ &     2.6 &        1.0 &  $76.2\pm6.5$ &     --- &     yes \\
S2-85     &   12.4 &     0.7 &       61.4 &       16 &   $61.0\pm1.4$ &     3.1 &        0.9 &  $80.9\pm4.5$ &     --- &     --- \\
S2-111    &   16.7 &     0.5 &       10.9 &       12 &   $62.0\pm6.4$ &     2.2 &        0.8 &           --- &     --- &     --- \\
S3-14     &   13.8 &     0.3 &       13.8 &       16 &   $54.4\pm2.8$ &     2.6 &        1.0 &  $79.6\pm9.3$ &     --- &     yes \\
S3-34     &   12.8 &     1.3 &      183.6 &       17 &   $50.7\pm2.6$ &     2.4 &        1.0 &           --- &     --- &     --- \\
S3-124    &   16.0 &     0.5 &       22.2 &       16 &   $61.7\pm2.9$ &     2.5 &        0.9 &  $57.7\pm8.1$ &     --- &     --- \\
S3-129    &   16.8 &     0.8 &       60.9 &        3 &   $56.9\pm3.6$ &     3.0 &        0.9 &           --- &     --- &     --- \\
S3-167    &   16.4 &     0.8 &       40.0 &        4 &   $60.3\pm3.6$ &     3.5 &        0.9 &           --- &     --- &     --- \\
S3-190    &   14.4 &     0.5 &        7.3 &       17 &   $59.1\pm5.6$ &     2.7 &        0.9 &           --- &     --- &     --- \\
S3-198    &   13.7 &     0.2 &        5.0 &       16 &   $54.8\pm5.3$ &     2.6 &        0.9 &           --- &     --- &     --- \\
S3-249    &   15.0 &     0.9 &       92.4 &       11 &   $44.0\pm1.0$ &     3.3 &        0.9 &  $69.7\pm4.9$ &     --- &     yes \\
S3-262    &   15.3 &     0.8 &       58.1 &       13 &   $53.3\pm1.7$ &     3.4 &        1.0 &  $87.5\pm8.2$ &     --- &     yes \\
S3-289    &   16.6 &     0.7 &       39.4 &       13 &   $51.7\pm2.0$ &     3.2 &        1.0 &           --- &     --- &     yes \\
S3-374    &   12.6 &     0.5 &       32.9 &       17 &   $47.8\pm1.5$ &     2.7 &        1.0 &  $76.1\pm6.0$ &     --- &     yes \\
S4-4      &   12.2 &     0.4 &        6.6 &       17 &   $52.4\pm5.1$ &     2.7 &        1.0 &           --- &     --- &     --- \\
S4-6      &   12.8 &     0.3 &        8.2 &       17 &   $52.3\pm3.7$ &     2.5 &        1.0 &           --- &     --- &     --- \\
S4-51     &   15.7 &     0.2 &        4.7 &       12 &   $53.2\pm7.9$ &     2.1 &        0.9 &           --- &     --- &     --- \\
S4-112    &   13.8 &     0.2 &        4.5 &       16 &  $66.1\pm10.2$ &     2.4 &        0.7 &           --- &     --- &     --- \\
S4-157    &   15.9 &     0.4 &       17.9 &       15 &   $51.1\pm2.7$ &     2.8 &        1.0 &           --- &     --- &     --- \\
S4-188    &   15.2 &     0.3 &        9.8 &       14 &   $69.2\pm7.0$ &     2.6 &        0.5 &           --- &     --- &     --- \\
S4-196    &   14.3 &     0.2 &        5.5 &       17 &   $66.6\pm7.3$ &     2.2 &        0.7 &           --- &     --- &     --- \\
S4-207    &   11.3 &     0.3 &        8.3 &       16 &   $67.7\pm6.8$ &     3.0 &        0.6 &           --- &     --- &     --- \\
S4-221    &   14.3 &     0.2 &        4.3 &       17 &   $49.0\pm5.9$ &     2.4 &        1.0 &           --- &     --- &     --- \\
S4-229    &   14.2 &     0.3 &        8.4 &        7 &   $56.7\pm8.4$ &     2.8 &        0.9 &           --- &     --- &     --- \\
S4-258    &   13.4 &     1.6 &      265.0 &       16 &   $56.6\pm0.8$ &     3.0 &        1.0 &  $63.7\pm2.8$ &     --- &     --- \\
S4-266    &   16.0 &     0.3 &        8.2 &       15 &   $50.1\pm4.9$ &     2.4 &        1.0 &           --- &     --- &     --- \\
S5-205    &   14.1 &     0.2 &        6.5 &       17 &   $65.0\pm6.3$ &     2.3 &        0.7 &           --- &     --- &     --- \\
S6-43     &   16.6 &     1.6 &       66.2 &        7 &   $61.6\pm2.5$ &     3.0 &        0.9 &           --- &     --- &     yes \\
IRS 16CC   &   11.0 &     0.7 &       44.5 &       17 &   $50.1\pm1.9$ &     2.4 &        1.0 &  $72.2\pm1.8$ &     --- &     yes \\
IRS 16SW-E &   11.2 &     0.2 &        3.7 &       17 &   $41.3\pm4.0$ &     3.1 &        0.9 &           --- &     --- &     --- \\
IRS 34W    &   11.7 &     0.4 &       22.6 &        7 &   $53.7\pm3.4$ &     3.1 &        1.0 &  $65.4\pm5.0$ &     --- &     --- \\
\enddata
\end{deluxetable*}
\end{longrotatetable}

%% file: dipping.tex
\begin{deluxetable*}{lccccc}
\tablecaption{Measured ``dipping'' events consistent with extinction.}
\label{table:dipping_events}
\tablehead{Star  &  Proper motion  & K' dip depth & Dip length & Implied size & Implied density \\ 
&  mas/year &  mag &  years & AU & atoms/$\mathrm{cm^3}$ \\}
\startdata
S2-69 & 5.4 & $0.50 \pm 0.02$ & $3.60 \pm 0.14$ & $158 \pm 6$ & $2.5 \times 10^{4}$ \\
S2-70 & 6.5 & $0.59 \pm 0.03$ & $3.60 \pm 0.13$ & $187 \pm 7$ & $2.5 \times 10^{4}$ \\
S3-14 & 14.8 & $0.34 \pm 0.02$ & $16.8 \pm 1.26$ & $1867 \pm 139$ & $1.5 \times 10^{3}$ \\
S3-249 & 8.8 & $0.60 \pm 0.02$ & $6.46 \pm 0.27$ & $454 \pm 19$ & $1.0 \times 10^{4}$ \\
S3-262 & 4.5 & $0.66 \pm 0.02$ & $5.65 \pm 0.20$ & $203 \pm 7$ & $2.6 \times 10^{4}$ \\
S3-289 & 3.9 & $0.54 \pm 0.02$ & $10.54 \pm 0.53$ & $332 \pm 17$ & $1.3 \times 10^{4}$ \\
S3-374 & 4.4 & $0.30 \pm 0.01$ & $9.96 \pm 0.76$ & $352 \pm 27$ & $6.9 \times 10^{3}$ \\
S6-43 & 2.6 & $1.39 \pm 0.02$ & $9.65 \pm 0.28$ & $200 \pm 6$ & $5.6 \times 10^{4}$ \\
IRS 16CC & 6.6 & $1.10 \pm 0.02$ & $5.66 \pm 0.09$ & $300 \pm 5$ & $2.9 \times 10^{4}$ \\
\enddata
\end{deluxetable*}

%% file: HK_measurments.tex
\startlongtable
\begin{deluxetable}{cccccc}
\tablecaption{H observations used in this work, with nearest K' match used to infer color information.}
\label{table:HK_measurements}
\tablehead{H observation & K' observation & Med. H Strehl & Med. H FWHM & Med. K' Strehl & Med. K' FWHM \\
UTC & UTC & & mas & & mas \\}
\startdata
2009-7-22 & 2009-7-22 & 0.12 & 68.67 & 0.23 & 73.41\\
2017-5-7 & 2017-5-5 & 0.23 & 50.54 & 0.38 & 58.38 \\
2017-8-13 & 2017-8-11 & 0.21 & 54.02  & 0.43 & 52.64\\
2017-8-23 & 2017-8-23 & 0.11 & 72.89 & 0.32 & 64.15 \\
2017-8-24 & 2017-8-24 & 0.18 & 59.15 & 0.37 & 60.68 \\
2017-8-26 & 2017-8-26 & 0.18 & 56.76 & 0.35 & 58.93 \\
2018-3-17 & 2018-3-17 & 0.28 & 53.52 & 0.48 & 55.83 \\
2018-3-22 & 2018-3-22 & 0.13 & 73.45 & 0.26 & 74.45 \\
2018-3-30 & 2018-3-30 & 0.19 & 58.70 & 0.37 & 61.03 \\
2018-5-19 & 2018-5-19 & 0.11 & 83.43 & 0.25 & 78.42 \\
2018-5-24 & 2018-5-24 & 0.12 & 76.45 & 0.23 & 77.99 \\
2019-5-13 & 2019-5-13 & 0.22 & 56.46 & 0.38 & 60.11 \\
2021-7-14 & 2021-7-14 & 0.10 & 83.92 & 0.22 & 82.95 \\
2022-5-21 & 2022-5-21 & 0.26 & 52.85& 0.44 & 55.10 \\
2022-5-25 & 2022-5-25 & 0.22 & 55.75 & 0.39 & 59.63 \\
2022-8-16 & 2022-8-16 & 0.21 & 57.91 & 0.33 & 65.20 \\
2022-8-19 & 2022-8-19 & 0.24 & 53.90 &0.43& 56.61 \\
2022-8-20 & 2022-8-20 & 0.19 & 61.80 & 0.34 &64.23 \\
\enddata
\end{deluxetable}

%% file: KL_measurments.tex
\startlongtable
\begin{deluxetable}{cccccc}
\tablecaption{L' observations used in this work, with nearest K' match used to infer color information.}
\label{table:KL_measurements}
\tablehead{L' observation & K' observation & Med. L' Strehl & Med. L' FWHM & Med. K' Strehl & Med. K' FWHM \\
UTC & UTC & & mas & & mas \\}
\startdata
2013-4-24* & 2013-4-26 & 0.42 & 90.83 & 0.27 & 64.81 \\
2013-4-25* & 2013-4-26 & 0.35  & 100.81 & 0.27 & 90.83 \\
2014-3-19* & 2014-3-19 & 0.29  & 116.6 & 0.13 & 104.06 \\
2014-3-20 & 2014-3-20 & 0.40  & 92.74 & 0.28 &67.48 \\
2014-4-18* & 2014-4-18 & 0.37 & 99.03 & 0.22 & 78.14 \\
2014-5-11 & 2014-5-11 & 0.40 & 91.03 & 0.26 & 68.32 \\
2014-8-3* & 2014-8-3 & 0.31 & 114.15 & 0.17 & 92.53 \\
2014-8-4 & 2014-8-4 & 0.42 & 92.44 & 0.29 & 64.25 \\
2015-3-31* & 2015-3-31 & 0.39 & 96.19 & 0.28& 68.22 \\
2017-7-16 & 2017-7-18 & 0.67 & 84.15 & 0.28 & 64.46 \\
2017-7-26* & 2017-7-27 & 0.47 & 89.72 & 0.20 & 78.21 \\
2019-5-14* & 2019-5-13 &0.46& 90.98 & 0.38& 60.11 \\
2019-6-19* & 2019-6-25 & 0.32& 101.34 & 0.24 & 75.44 \\
2019-8-14 & 2019-8-14 & 0.51& 87.78 & 0.30 & 67.40 \\
2021-7-13 & 2021-7-13 & 0.44 & 94.12 & 0.30 & 67.73 \\
2021-8-15 & 2021-8-15 & 0.61 & 81.47 & 0.47 & 50.19 \\
2022-5-15* & 2022-5-15 & 0.51 & 85.25 & 0.27 & 69.13 \\
\enddata
\tablecomments{Newly reported L' observations are marked with an asterisk.}
\end{deluxetable}

%% file: non-extinction.tex
\startlongtable
\begin{deluxetable*}{lccccccc}
\tablecaption{Characteristics of the variability sample that is not consistent with changing extinction. Columns 2 through 5 are related to the K' variability. Column 2 states the mean K' magnitude for each source, $\langle{\mathrm{K'}}\rangle$ and column 3 describes the change in K' magnitude, $\Delta{\mathrm{K'}}$. Column 4 is the reduced chi-squared for variability metric, as outlined in Section \ref{sec:sample} and Column 5 is the degrees-of-freedom for the color measurements (one less than the number of epochs observed). Columns 6 is the modeled slope in color-magnitude ($\theta_{\mathrm{H-K'}}$); column 7 is the expected mean color $\langle{\mathrm{H-K'}}\rangle$ derived from the slope fits. Column 8, $\mathrm{P}(\mathrm{extinction})_{\mathrm{H-K'}}$, is the probability of membership to the H-K' extinction Gaussian.}
\label{table:non-extinction}
\tablehead{Star & $\langle{\mathrm{K'}}\rangle$ & $\Delta{\mathrm{K'}}$ & K' variability ($\chi_{\nu}^2$) & $\mathrm{d.o.f.}_{H-K'}$ & $\theta_{\mathrm{H-K'}}$ & $\langle{\mathrm{H-K'}}\rangle$ & $\mathrm{P}(\mathrm{extinction})_{\mathrm{H-K'}}$ \\
 & mag & mag & & & degrees & mag &  \\}
\startdata
S0-8    &   15.8 &     0.4 &        7.1 &       14.0 &   $116.1\pm9.1$ &     2.0 &        0.0 \\
S0-33   &   16.0 &     0.6 &        9.7 &        9.0 &  $116.9\pm13.2$ &     2.1 &        0.0 \\
S0-40   &   17.1 &     0.5 &        6.9 &       13.0 &  $114.9\pm10.8$ &     2.1 &        0.0 \\
S0-59   &   16.5 &     0.6 &        6.0 &        9.0 &  $127.4\pm10.0$ &     2.2 &        0.0 \\
S0-63   &   17.5 &     0.6 &        9.3 &       10.0 &   $130.5\pm7.8$ &     2.4 &        0.0 \\
S0-67   &   15.5 &     0.2 &        4.8 &       15.0 &   $87.9\pm11.5$ &     2.2 &        0.0 \\
S0-103  &   16.7 &     0.5 &       10.1 &        5.0 &   $97.4\pm14.4$ &     2.0 &        0.0 \\
S1-47   &   15.5 &     0.2 &        4.1 &       16.0 &   $92.3\pm12.4$ &     2.3 &        0.0 \\
S1-49   &   14.4 &     0.2 &        6.1 &       17.0 &    $86.9\pm9.1$ &     2.1 &        0.0 \\
S1-62   &   15.5 &     0.3 &        4.6 &       11.0 &  $121.4\pm11.0$ &     2.4 &        0.0 \\
S1-64   &   15.5 &     0.5 &       11.0 &       14.0 &    $82.5\pm6.7$ &     2.4 &        0.0 \\
S1-66   &   15.5 &     0.2 &        9.0 &       11.0 &   $116.6\pm8.5$ &     2.2 &        0.0 \\
S1-175  &   15.4 &     0.3 &        5.3 &       14.0 &   $96.0\pm11.5$ &     2.2 &        0.0 \\
S2-7    &   13.8 &     0.3 &        8.7 &       16.0 &    $73.8\pm6.2$ &     2.4 &        0.2 \\
S2-16   &   12.4 &     0.4 &        6.6 &       17.0 &    $95.8\pm9.5$ &     3.4 &        0.0 \\
S2-20   &   16.2 &     0.5 &        9.9 &        6.0 &   $78.4\pm19.5$ &     2.7 &        0.3 \\
S2-26   &   13.6 &     0.5 &       10.0 &       17.0 &    $76.7\pm8.2$ &     2.5 &        0.2 \\
S2-36   &   13.4 &     0.4 &       12.6 &       17.0 &   $100.1\pm5.9$ &     2.2 &        0.0 \\
S2-48   &   16.4 &     0.4 &        7.2 &       12.0 &  $116.1\pm11.3$ &     2.8 &        0.0 \\
S2-50   &   15.1 &     0.4 &        4.7 &       13.0 &   $148.2\pm6.9$ &     2.2 &        0.0 \\
S2-54   &   15.9 &     0.7 &        6.5 &       10.0 &  $127.1\pm10.1$ &     2.3 &        0.0 \\
S2-58   &   14.0 &     0.2 &        5.0 &       16.0 &  $116.3\pm10.4$ &     2.2 &        0.0 \\
S2-65   &   15.9 &     0.5 &        8.8 &       15.0 &  $124.2\pm14.0$ &     2.5 &        0.0 \\
S2-76   &   15.6 &     0.4 &       11.9 &       15.0 &    $69.1\pm4.7$ &     2.6 &        0.4 \\
S2-110  &   16.8 &     0.7 &       20.6 &       10.0 &    $89.1\pm8.4$ &     2.4 &        0.0 \\
S2-130  &   17.2 &     0.6 &        6.4 &        7.0 &  $128.3\pm10.4$ &     2.3 &        0.0 \\
S2-134  &   15.5 &     0.2 &        6.2 &       11.0 &  $118.5\pm10.8$ &     2.2 &        0.0 \\
S2-166  &   16.3 &     0.8 &        8.8 &       10.0 &   $123.7\pm8.9$ &     2.4 &        0.0 \\
S2-184  &   16.3 &     0.4 &       12.8 &        4.0 &  $112.6\pm12.2$ &     2.6 &        0.0 \\
S2-186  &   16.6 &     0.3 &        5.3 &       13.0 &   $123.4\pm9.7$ &     1.7 &        0.0 \\
S2-188  &   16.6 &     0.5 &        8.2 &       10.0 &  $119.6\pm11.3$ &     3.1 &        0.0 \\
S2-205  &   15.6 &     0.6 &       21.4 &       15.0 &    $97.8\pm5.1$ &     2.3 &        0.0 \\
S2-214  &   16.5 &     0.5 &       30.1 &        6.0 &    $89.7\pm6.2$ &     2.3 &        0.0 \\
S2-237  &   16.2 &     0.5 &        4.9 &       12.0 &  $108.5\pm10.7$ &     2.8 &        0.0 \\
S2-265  &   17.5 &     0.4 &        7.5 &       10.0 &   $146.3\pm6.9$ &     1.8 &        0.0 \\
S2-278  &   16.4 &     0.4 &       17.4 &        6.0 &    $88.2\pm9.1$ &     2.7 &        0.0 \\
S2-314  &   18.3 &     1.1 &       10.4 &        5.0 &   $136.5\pm7.1$ &     1.7 &        0.0 \\
S2-315  &   17.2 &     0.8 &        8.5 &       12.0 &    $93.0\pm8.3$ &     2.6 &        0.0 \\
S2-348  &   17.2 &     0.4 &       11.5 &        6.0 &  $105.3\pm10.8$ &     2.3 &        0.0 \\
S3-4    &   14.6 &     0.3 &        8.0 &       17.0 &   $91.8\pm10.3$ &     3.2 &        0.0 \\
S3-16   &   15.0 &     0.3 &        5.5 &       13.0 &  $116.5\pm10.3$ &     2.3 &        0.0 \\
S3-20   &   14.5 &     0.3 &        6.1 &       17.0 &    $83.1\pm9.2$ &     2.5 &        0.0 \\
S3-50   &   15.6 &     0.3 &        6.7 &       16.0 &   $100.5\pm9.7$ &     2.3 &        0.0 \\
S3-93   &   16.0 &     0.3 &        9.7 &        4.0 &  $101.3\pm13.5$ &     2.4 &        0.0 \\
S3-104  &   15.8 &     0.4 &       11.6 &       11.0 &    $73.5\pm6.4$ &     2.6 &        0.2 \\
S3-143  &   16.5 &     0.5 &        4.3 &       13.0 &  $107.6\pm13.3$ &     3.1 &        0.0 \\
S3-145  &   16.0 &     0.3 &        9.8 &        5.0 &   $87.9\pm11.9$ &     2.6 &        0.0 \\
S3-146  &   14.1 &     0.3 &        4.8 &       14.0 &  $111.3\pm10.5$ &     2.2 &        0.0 \\
S3-153  &   15.8 &     0.2 &        4.7 &       16.0 &  $105.3\pm10.5$ &     2.3 &        0.0 \\
S3-166  &   18.2 &     0.7 &       13.5 &        3.0 &  $137.1\pm10.4$ &     2.3 &        0.0 \\
S3-187  &   14.6 &     0.3 &        4.9 &       17.0 &    $70.3\pm9.4$ &     2.5 &        0.4 \\
S3-223  &   14.9 &     0.2 &        5.1 &       10.0 &   $80.0\pm12.7$ &     2.5 &        0.2 \\
S3-232  &   17.2 &     0.5 &        4.3 &       13.0 &   $95.7\pm12.8$ &     2.6 &        0.0 \\
S3-256  &   15.6 &     0.2 &        5.6 &       12.0 &   $84.0\pm16.7$ &     2.6 &        0.2 \\
S3-284  &   14.0 &     0.2 &        3.9 &       16.0 &   $83.9\pm13.1$ &     3.1 &        0.1 \\
S3-287  &   16.3 &     0.3 &        4.4 &       15.0 &   $81.4\pm11.8$ &     2.6 &        0.1 \\
S3-302  &   16.0 &     0.5 &       13.5 &       13.0 &   $106.2\pm9.9$ &     2.5 &        0.0 \\
S3-315  &   17.1 &     0.5 &       28.1 &        3.0 &   $131.0\pm8.0$ &     2.6 &        0.0 \\
S3-387  &   16.1 &     0.4 &        5.6 &       10.0 &  $124.2\pm12.2$ &     2.3 &        0.0 \\
S3-403  &   14.8 &     0.2 &        5.6 &       14.0 &   $103.9\pm9.7$ &     2.3 &        0.0 \\
S3-438  &   15.1 &     0.3 &        7.5 &        6.0 &  $120.0\pm11.9$ &     2.1 &        0.0 \\
S3-440  &   17.6 &     0.6 &        6.4 &        8.0 &  $121.6\pm14.2$ &     2.0 &        0.0 \\
S3-442  &   18.0 &     0.7 &        9.0 &        7.0 &   $126.1\pm9.3$ &     2.2 &        0.0 \\
S4-12   &   14.9 &     0.5 &       14.5 &       12.0 &    $71.0\pm5.9$ &     2.5 &        0.4 \\
S4-30   &   16.1 &     0.3 &        8.7 &        4.0 &   $71.7\pm33.1$ &     2.6 &        0.3 \\
S4-36   &   12.9 &     0.2 &        5.4 &       10.0 &   $80.7\pm14.4$ &     2.5 &        0.2 \\
S4-69   &   16.1 &     0.3 &        6.7 &        6.0 &  $107.7\pm13.7$ &     2.3 &        0.0 \\
S4-81   &   16.8 &     0.5 &        4.9 &       10.0 &   $139.3\pm8.3$ &     2.4 &        0.0 \\
S4-82   &   15.9 &     0.3 &        8.5 &       13.0 &  $126.9\pm10.2$ &     2.4 &        0.0 \\
S4-84   &   14.0 &     0.3 &        6.9 &        7.0 &   $82.0\pm12.2$ &     2.8 &        0.1 \\
S4-128  &   17.2 &     0.9 &        8.1 &       13.0 &   $103.7\pm9.6$ &     2.7 &        0.0 \\
S4-172  &   15.0 &     0.4 &       19.0 &       15.0 &    $81.1\pm5.1$ &     2.4 &        0.0 \\
S4-173  &   16.1 &     0.3 &        4.9 &       15.0 &   $86.7\pm14.3$ &     2.6 &        0.1 \\
S4-181  &   16.0 &     0.3 &        5.2 &       14.0 &   $91.0\pm11.2$ &     2.4 &        0.0 \\
S4-187  &   15.9 &     0.3 &       12.1 &       15.0 &   $89.5\pm10.2$ &     2.5 &        0.0 \\
S4-204  &   17.5 &     0.5 &        7.0 &       12.0 &  $102.9\pm11.8$ &     2.6 &        0.0 \\
S4-300  &   17.9 &     0.3 &        4.6 &       11.0 &  $119.5\pm11.8$ &     2.0 &        0.0 \\
S4-306  &   16.7 &     0.5 &        4.2 &       13.0 &   $131.6\pm9.9$ &     2.2 &        0.0 \\
S4-352  &   14.4 &     0.3 &        5.4 &       13.0 &   $71.1\pm10.7$ &     2.9 &        0.4 \\
S4-354  &   18.3 &     0.5 &       16.0 &        6.0 &  $104.7\pm10.9$ &     2.3 &        0.0 \\
S4-466  &   16.8 &     0.3 &        5.3 &       14.0 &   $102.1\pm9.8$ &     2.5 &        0.0 \\
S4-491  &   18.2 &     0.7 &       10.1 &        6.0 &   $131.9\pm7.4$ &     2.7 &        0.0 \\
S4-499  &   18.7 &     0.5 &        7.8 &        5.0 &  $119.2\pm13.4$ &     2.1 &        0.0 \\
S5-49   &   16.0 &     0.3 &        4.4 &       16.0 &   $83.6\pm10.3$ &     2.3 &        0.1 \\
S5-50   &   16.9 &     0.8 &       16.1 &       14.0 &    $86.2\pm6.8$ &     2.9 &        0.0 \\
S5-65   &   16.1 &     0.3 &        6.1 &        9.0 &   $96.2\pm12.6$ &     2.8 &        0.0 \\
S5-133  &   14.8 &     0.4 &        7.1 &       14.0 &   $131.9\pm6.9$ &     2.1 &        0.0 \\
S5-151  &   16.7 &     0.6 &        8.9 &       14.0 &   $96.2\pm10.7$ &     2.3 &        0.0 \\
S5-170  &   14.2 &     0.3 &        9.7 &        6.0 &  $116.4\pm14.8$ &     2.4 &        0.0 \\
S5-176  &   16.7 &     0.3 &        6.4 &        8.0 &  $100.7\pm13.7$ &     1.9 &        0.0 \\
S5-194  &   17.3 &     0.4 &        8.4 &       11.0 &   $80.1\pm12.1$ &     2.6 &        0.2 \\
S5-308  &   18.7 &     0.8 &        5.3 &        9.0 &   $136.4\pm9.6$ &     1.9 &        0.0 \\
S5-310  &   18.3 &     0.4 &        7.2 &        6.0 &   $79.3\pm13.9$ &     2.4 &        0.2 \\
S5-324  &   18.3 &     0.8 &        9.5 &        5.0 &  $131.1\pm10.2$ &     2.2 &        0.0 \\
S6-16   &   16.3 &     0.5 &       18.9 &        7.0 &   $133.1\pm7.2$ &     1.1 &        0.0 \\
S6-26   &   15.9 &     0.3 &        5.2 &       12.0 &  $103.2\pm15.5$ &     2.4 &        0.0 \\
S6-34   &   16.1 &     0.3 &        6.0 &       11.0 &   $72.5\pm14.7$ &     2.5 &        0.4 \\
S6-39   &   14.1 &     0.2 &        3.8 &       16.0 &   $92.2\pm14.2$ &     2.4 &        0.0 \\
S6-44   &   15.2 &     0.4 &        9.5 &       16.0 &    $92.2\pm8.1$ &     2.2 &        0.0 \\
S6-89   &   12.3 &     0.3 &       10.2 &        6.0 &   $98.0\pm16.1$ &     2.3 &        0.0 \\
S6-135  &   15.3 &     0.4 &        8.9 &       12.0 &  $108.7\pm10.5$ &     2.4 &        0.0 \\
S6-367  &   17.3 &     0.5 &       23.0 &        3.0 &   $129.8\pm9.5$ &     2.3 &        0.0 \\
S7-6    &   15.8 &     0.4 &        5.7 &       12.0 &   $137.9\pm8.3$ &     2.4 &        0.0 \\
S7-243  &   17.5 &     0.7 &        6.7 &        8.0 &  $129.2\pm11.6$ &     2.1 &        0.0 \\
IRS 16C  &   10.0 &     0.2 &        4.0 &       17.0 &   $135.7\pm9.1$ &     2.0 &        0.0 \\
IRS 16SW &   10.1 &     0.5 &       19.7 &       17.0 &   $104.7\pm6.7$ &     2.1 &        0.0 \\
IRS 33N  &   11.3 &     0.3 &        5.5 &       17.0 &    $90.2\pm9.8$ &     2.2 &        0.0 \\
IRS 34NW &   13.9 &     0.7 &       29.1 &       14.0 &    $74.0\pm4.1$ &     3.1 &        0.1 \\
IRS 9W   &   12.2 &     0.2 &        4.8 &       17.0 &   $83.5\pm10.7$ &     2.4 &        0.1 \\
\enddata
\end{deluxetable*}

%% file: lp_phot_non_var.tex
\startlongtable
\begin{deluxetable}{lccc}
\tablecaption{L' photometry information and colors for all non-variable stars. Variable stars are selected according to the reduced chi-squared metric discussed in Section \ref{sec:sample}.}
\label{table:non_var_lp}
\tablehead{Star & Mean L' & Mean H-K' & Mean K'-L' \\
 & mag & mag & mag \\}
\startdata
S0-1    &  $13.76\pm0.08$ &   $2.1\pm0.06$ &  $1.07\pm0.09$ \\
S0-11   &  $14.11\pm0.05$ &  $2.09\pm0.06$ &  $1.09\pm0.08$ \\
S0-12   &  $12.85\pm0.05$ &  $2.29\pm0.05$ &  $1.54\pm0.07$ \\
S0-13   &   $12.0\pm0.07$ &  $2.18\pm0.05$ &   $1.37\pm0.1$ \\
S0-14   &  $12.35\pm0.04$ &  $2.05\pm0.05$ &  $1.34\pm0.06$ \\
S0-15   &  $12.29\pm0.06$ &  $2.23\pm0.05$ &  $1.41\pm0.07$ \\
S0-17   &  $14.95\pm0.08$ &  $2.17\pm0.09$ &  $1.24\pm0.12$ \\
S0-18   &  $13.82\pm0.06$ &   $2.2\pm0.06$ &  $1.33\pm0.07$ \\
S0-2    &  $12.89\pm0.05$ &  $2.02\pm0.06$ &   $1.2\pm0.07$ \\
S0-20   &  $14.19\pm0.07$ &  $2.01\pm0.08$ &  $1.67\pm0.09$ \\
S0-24   &  $14.46\pm0.07$ &  $2.17\pm0.09$ &   $1.18\pm0.1$ \\
S0-3    &  $13.46\pm0.06$ &   $2.1\pm0.06$ &  $1.23\pm0.08$ \\
S0-35   &   $13.9\pm0.04$ &   $2.3\pm0.06$ &  $1.42\pm0.06$ \\
S0-6    &  $12.57\pm0.05$ &  $2.25\pm0.06$ &  $1.48\pm0.07$ \\
S0-62   &  $14.18\pm0.05$ &  $2.21\pm0.05$ &  $1.24\pm0.08$ \\
S0-7    &   $13.9\pm0.06$ &  $2.13\pm0.07$ &  $1.24\pm0.08$ \\
S0-9    &  $13.16\pm0.05$ &  $2.01\pm0.05$ &  $1.12\pm0.07$ \\
S1-1    &  $11.76\pm0.04$ &  $2.03\pm0.05$ &  $1.34\pm0.06$ \\
S1-10   &   $13.8\pm0.05$ &  $1.78\pm0.05$ &  $0.98\pm0.07$ \\
S1-12   &  $12.28\pm0.04$ &  $2.04\pm0.05$ &  $1.25\pm0.07$ \\
S1-13   &  $12.45\pm0.05$ &  $2.34\pm0.05$ &  $1.61\pm0.08$ \\
S1-14   &  $11.47\pm0.05$ &   $2.0\pm0.05$ &  $1.32\pm0.07$ \\
S1-17   &   $10.60\pm0.03$ &  $2.43\pm0.05$ &  $1.52\pm0.06$ \\
S1-19   &  $12.26\pm0.04$ &  $2.09\pm0.05$ &  $1.31\pm0.07$ \\
S1-2    &  $13.62\pm0.06$ &  $2.05\pm0.06$ &  $1.18\pm0.08$ \\
S1-21   &  $11.91\pm0.05$ &  $2.02\pm0.05$ &  $1.38\pm0.07$ \\
S1-23   &  $10.06\pm0.05$ &  $2.38\pm0.05$ &  $1.62\pm0.07$ \\
S1-24   &   $9.97\pm0.05$ &  $2.12\pm0.06$ &  $1.46\pm0.08$ \\
S1-25   &  $11.95\pm0.03$ &  $2.27\pm0.05$ &  $1.45\pm0.06$ \\
S1-26   &   $14.0\pm0.06$ &  $2.25\pm0.06$ &  $1.44\pm0.08$ \\
S1-29   &  $13.35\pm0.06$ &  $2.31\pm0.07$ &  $1.96\pm0.08$ \\
S1-3    &  $10.78\pm0.05$ &  $2.09\pm0.06$ &  $1.41\pm0.07$ \\
S1-31   &   $14.2\pm0.08$ &  $2.31\pm0.08$ &  $1.51\pm0.11$ \\
S1-33   &  $12.87\pm0.06$ &  $2.09\pm0.05$ &  $2.19\pm0.08$ \\
S1-34   &  $12.01\pm0.04$ &  $1.74\pm0.05$ &  $0.97\pm0.06$ \\
S1-39   &   $14.2\pm0.06$ &  $2.26\pm0.06$ &  $1.06\pm0.08$ \\
S1-4    &  $11.17\pm0.05$ &   $2.2\pm0.05$ &  $1.39\pm0.07$ \\
S1-43   &  $14.76\pm0.07$ &  $2.16\pm0.09$ &  $1.28\pm0.09$ \\
S1-48   &  $13.55\pm0.07$ &  $2.23\pm0.06$ &  $1.72\pm0.09$ \\
S1-51   &  $14.15\pm0.09$ &  $2.31\pm0.05$ &   $0.85\pm0.1$ \\
S1-53   &  $13.51\pm0.05$ &  $2.22\pm0.06$ &  $1.69\pm0.07$ \\
S1-55   &   $14.17\pm0.1$ &  $2.31\pm0.09$ &  $1.34\pm0.11$ \\
S1-6    &  $13.94\pm0.06$ &  $2.32\pm0.07$ &  $1.52\pm0.07$ \\
S1-67   &   $14.21\pm0.1$ &  $2.32\pm0.06$ &  $1.33\pm0.12$ \\
S1-68   &  $11.85\pm0.04$ &  $2.27\pm0.05$ &  $1.46\pm0.06$ \\
S1-8    &  $12.94\pm0.04$ &  $2.22\pm0.05$ &  $1.29\pm0.06$ \\
S2-11   &  $10.33\pm0.05$ &  $2.13\pm0.05$ &  $1.44\pm0.07$ \\
S2-12   &  $13.69\pm0.06$ &  $2.21\pm0.08$ &  $1.48\pm0.09$ \\
S2-17   &   $9.23\pm0.04$ &  $2.04\pm0.05$ &  $1.46\pm0.07$ \\
S2-19   &  $11.22\pm0.05$ &  $2.11\pm0.06$ &  $1.51\pm0.07$ \\
S2-198  &  $13.55\pm0.06$ &   $2.3\pm0.07$ &  $2.04\pm0.09$ \\
S2-2    &  $12.98\pm0.04$ &  $1.73\pm0.06$ &  $1.08\pm0.06$ \\
S2-200  &  $14.44\pm0.05$ &   $2.3\pm0.06$ &  $1.35\pm0.07$ \\
S2-21   &  $11.94\pm0.03$ &  $2.11\pm0.05$ &  $1.38\pm0.06$ \\
S2-22   &   $11.9\pm0.04$ &  $2.04\pm0.05$ &  $1.07\pm0.06$ \\
S2-23   &  $12.63\pm0.05$ &  $2.64\pm0.06$ &  $1.83\pm0.07$ \\
S2-24   &  $12.05\pm0.03$ &   $2.4\pm0.05$ &  $1.64\pm0.06$ \\
S2-25   &  $12.37\pm0.04$ &  $2.37\pm0.05$ &  $1.41\pm0.07$ \\
S2-259  &  $14.71\pm0.15$ &  $2.21\pm0.07$ &  $1.44\pm0.17$ \\
S2-261  &  $14.16\pm0.05$ &  $2.42\pm0.08$ &  $1.59\pm0.07$ \\
S2-273  &  $14.38\pm0.05$ &  $2.41\pm0.06$ &  $1.71\pm0.07$ \\
S2-3    &  $12.72\pm0.05$ &  $2.41\pm0.05$ &  $1.52\pm0.07$ \\
S2-300  &  $14.34\pm0.07$ &  $2.52\pm0.06$ &   $1.7\pm0.09$ \\
S2-303  &  $13.98\pm0.08$ &  $2.46\pm0.05$ &  $1.97\pm0.09$ \\
S2-31   &  $11.49\pm0.03$ &  $2.45\pm0.05$ &   $1.5\pm0.05$ \\
S2-316  &  $13.57\pm0.04$ &  $2.62\pm0.06$ &  $1.63\pm0.08$ \\
S2-32   &  $10.82\pm0.05$ &  $2.16\pm0.05$ &  $1.45\pm0.07$ \\
S2-4    &  $10.52\pm0.05$ &  $2.35\pm0.05$ &  $1.46\pm0.07$ \\
S2-40   &  $13.92\pm0.04$ &  $2.16\pm0.08$ &  $1.56\pm0.07$ \\
S2-42   &  $13.94\pm0.05$ &  $2.42\pm0.06$ &  $1.74\pm0.07$ \\
S2-5    &  $12.07\pm0.06$ &  $2.12\pm0.05$ &  $1.18\pm0.08$ \\
S2-51   &  $14.01\pm0.04$ &  $2.38\pm0.07$ &  $1.69\pm0.07$ \\
S2-55   &  $13.64\pm0.06$ &  $2.25\pm0.06$ &   $1.54\pm0.1$ \\
S2-56   &  $14.21\pm0.06$ &  $2.34\pm0.06$ &  $1.66\pm0.08$ \\
S2-57   &  $12.66\pm0.06$ &  $2.28\pm0.05$ &  $1.56\pm0.09$ \\
S2-59   &  $14.13\pm0.08$ &  $2.41\pm0.06$ &   $1.26\pm0.1$ \\
S2-6    &  $10.55\pm0.03$ &  $2.24\pm0.05$ &  $1.34\pm0.06$ \\
S2-62   &  $13.96\pm0.05$ &  $2.34\pm0.06$ &   $1.1\pm0.08$ \\
S2-63   &  $13.99\pm0.05$ &  $2.25\pm0.06$ &  $1.49\pm0.07$ \\
S2-67   &  $11.84\pm0.04$ &  $2.42\pm0.05$ &  $1.59\pm0.06$ \\
S2-71   &  $13.94\pm0.03$ &  $2.14\pm0.05$ &  $1.26\pm0.06$ \\
S2-73   &  $13.34\pm0.05$ &  $2.22\pm0.06$ &  $1.55\pm0.07$ \\
S2-74   &  $11.67\pm0.04$ &  $2.29\pm0.05$ &  $1.62\pm0.07$ \\
S2-75   &  $12.97\pm0.06$ &  $2.29\pm0.05$ &  $1.48\pm0.08$ \\
S2-78   &   $12.0\pm0.04$ &  $2.45\pm0.05$ &  $1.59\pm0.07$ \\
S2-80   &  $14.17\pm0.05$ &  $2.18\pm0.06$ &  $1.42\pm0.07$ \\
S2-81   &  $13.86\pm0.04$ &  $2.36\pm0.07$ &  $1.61\pm0.06$ \\
S2-82   &  $13.43\pm0.11$ &  $2.28\pm0.08$ &  $1.86\pm0.12$ \\
S2-84   &  $14.19\pm0.09$ &  $2.29\pm0.06$ &  $1.07\pm0.11$ \\
S3-10   &  $10.86\pm0.04$ &  $1.95\pm0.06$ &  $1.27\pm0.07$ \\
S3-103  &  $14.43\pm0.05$ &  $2.43\pm0.06$ &  $1.76\pm0.07$ \\
S3-108  &  $14.12\pm0.06$ &   $2.2\pm0.06$ &  $1.51\pm0.07$ \\
S3-11   &  $13.45\pm0.04$ &  $2.46\pm0.05$ &  $1.55\pm0.07$ \\
S3-116  &  $14.58\pm0.06$ &   $2.2\pm0.06$ &  $1.33\pm0.09$ \\
S3-123  &  $14.86\pm0.11$ &  $2.18\pm0.06$ &  $0.92\pm0.13$ \\
S3-13   &   $12.2\pm0.05$ &  $2.28\pm0.05$ &  $1.36\pm0.07$ \\
S3-134  &  $12.02\pm0.04$ &  $2.32\pm0.05$ &  $1.77\pm0.07$ \\
S3-149  &  $11.53\pm0.04$ &  $2.53\pm0.07$ &  $1.79\pm0.08$ \\
S3-155  &  $14.28\pm0.08$ &  $1.86\pm0.05$ &  $0.89\pm0.09$ \\
S3-156  &  $11.46\pm0.04$ &   $2.2\pm0.05$ &   $1.6\pm0.06$ \\
S3-159  &  $13.77\pm0.07$ &  $2.45\pm0.06$ &  $1.65\pm0.09$ \\
S3-169  &  $13.88\pm0.05$ &  $2.31\pm0.06$ &  $1.68\pm0.07$ \\
S3-171  &  $14.29\pm0.05$ &  $2.19\pm0.05$ &  $1.66\pm0.07$ \\
S3-178  &  $11.57\pm0.04$ &  $2.27\pm0.05$ &  $1.48\pm0.06$ \\
S3-207  &  $12.41\pm0.05$ &  $2.26\pm0.06$ &  $1.64\pm0.08$ \\
S3-208  &  $13.42\pm0.06$ &   $1.9\pm0.05$ &  $1.05\pm0.08$ \\
S3-216  &  $13.93\pm0.05$ &  $2.07\pm0.06$ &  $1.15\pm0.07$ \\
S3-22   &   $9.57\pm0.03$ &  $2.34\pm0.05$ &  $1.57\pm0.07$ \\
S3-23   &  $14.04\pm0.04$ &   $2.3\pm0.08$ &  $1.35\pm0.06$ \\
S3-25   &  $12.57\pm0.04$ &  $2.09\pm0.06$ &  $1.46\pm0.06$ \\
S3-260  &  $14.98\pm0.05$ &  $2.45\pm0.06$ &  $1.34\pm0.07$ \\
S3-263  &  $14.23\pm0.07$ &  $2.19\pm0.07$ &  $1.39\pm0.09$ \\
S3-268  &   $14.16\pm0.1$ &   $2.3\pm0.05$ &  $1.13\pm0.11$ \\
S3-27   &  $12.18\pm0.05$ &  $2.43\pm0.06$ &  $1.75\pm0.08$ \\
S3-28   &  $12.12\pm0.05$ &  $2.49\pm0.06$ &  $1.81\pm0.08$ \\
S3-286  &  $13.96\pm0.05$ &  $2.17\pm0.08$ &  $1.34\pm0.07$ \\
S3-288  &  $12.35\pm0.06$ &  $3.11\pm0.06$ &  $1.93\pm0.09$ \\
S3-29   &   $12.4\pm0.04$ &  $2.08\pm0.06$ &  $1.23\pm0.07$ \\
S3-294  &  $14.11\pm0.05$ &  $2.36\pm0.09$ &  $1.68\pm0.07$ \\
S3-3    &  $13.31\pm0.06$ &  $2.08\pm0.05$ &  $1.81\pm0.08$ \\
S3-30   &  $10.96\pm0.04$ &  $2.29\pm0.06$ &  $1.54\pm0.06$ \\
S3-314  &  $13.74\pm0.05$ &  $2.06\pm0.07$ &   $1.8\pm0.12$ \\
S3-319  &   $13.8\pm0.04$ &  $2.12\pm0.06$ &  $1.54\pm0.06$ \\
S3-32   &   $13.9\pm0.06$ &  $2.24\pm0.06$ &  $1.31\pm0.13$ \\
S3-33   &  $13.94\pm0.09$ &  $2.12\pm0.07$ &    $1.4\pm0.1$ \\
S3-331  &  $12.27\pm0.08$ &  $2.15\pm0.07$ &   $1.52\pm0.1$ \\
S3-334  &  $14.18\pm0.05$ &  $2.62\pm0.15$ &  $1.88\pm0.08$ \\
S3-344  &  $13.94\pm0.07$ &  $2.16\pm0.05$ &  $1.35\pm0.08$ \\
S3-348  &  $14.15\pm0.05$ &  $2.13\pm0.06$ &  $1.17\pm0.07$ \\
S3-35   &  $11.94\pm0.05$ &  $2.73\pm0.06$ &  $1.99\pm0.08$ \\
S3-36   &  $13.36\pm0.05$ &  $2.07\pm0.05$ &  $1.26\pm0.07$ \\
S3-364  &  $13.12\pm0.06$ &   $2.15\pm0.1$ &  $1.44\pm0.09$ \\
S3-370  &   $12.2\pm0.03$ &   $2.2\pm0.05$ &  $1.41\pm0.08$ \\
S3-383  &  $14.01\pm0.04$ &  $2.26\pm0.05$ &  $1.48\pm0.06$ \\
S3-385  &   $13.7\pm0.05$ &  $2.12\pm0.06$ &  $1.54\pm0.08$ \\
S3-437  &  $14.38\pm0.13$ &  $2.36\pm0.07$ &  $1.06\pm0.14$ \\
S3-48   &  $14.45\pm0.09$ &  $2.33\pm0.08$ &  $1.41\pm0.11$ \\
S3-5    &  $10.02\pm0.04$ &  $2.34\pm0.06$ &  $1.92\pm0.07$ \\
S3-51   &  $13.88\pm0.05$ &   $2.1\pm0.06$ &  $1.37\pm0.07$ \\
S3-6    &  $11.22\pm0.04$ &   $2.4\pm0.05$ &  $1.62\pm0.07$ \\
S3-62   &  $13.92\pm0.07$ &   $2.2\pm0.06$ &  $1.41\pm0.08$ \\
S3-8    &  $12.41\pm0.04$ &  $2.25\pm0.05$ &   $1.5\pm0.06$ \\
S3-88   &  $12.78\pm0.03$ &  $2.16\pm0.05$ &  $1.47\pm0.07$ \\
S3-91   &  $14.23\pm0.05$ &  $2.26\pm0.05$ &  $1.38\pm0.06$ \\
S4-1    &  $11.77\pm0.05$ &   $2.3\pm0.05$ &  $1.51\pm0.07$ \\
S4-122  &   $14.5\pm0.08$ &  $2.67\pm0.07$ &  $1.89\pm0.09$ \\
S4-123  &  $14.13\pm0.06$ &  $2.35\pm0.07$ &   $1.6\pm0.08$ \\
S4-126  &  $15.07\pm0.07$ &  $2.34\pm0.06$ &  $1.16\pm0.08$ \\
S4-129  &  $10.54\pm0.04$ &  $2.43\pm0.05$ &  $1.61\pm0.07$ \\
S4-131  &  $14.42\pm0.04$ &   $2.5\pm0.06$ &  $1.41\pm0.07$ \\
S4-139  &  $13.08\pm0.04$ &  $2.29\pm0.05$ &  $1.32\pm0.07$ \\
S4-140  &  $13.24\pm0.07$ &  $2.31\pm0.05$ &   $2.4\pm0.09$ \\
S4-143  &  $11.89\pm0.04$ &  $2.56\pm0.06$ &  $1.59\pm0.06$ \\
S4-152  &  $14.36\pm0.06$ &   $2.3\pm0.06$ &  $1.61\pm0.08$ \\
S4-154  &  $14.13\pm0.06$ &  $2.41\pm0.07$ &  $1.77\pm0.08$ \\
S4-158  &  $12.76\pm0.05$ &  $2.12\pm0.07$ &  $1.32\pm0.07$ \\
S4-161  &  $12.06\pm0.05$ &  $2.43\pm0.07$ &  $1.67\pm0.07$ \\
S4-167  &  $14.09\pm0.06$ &  $2.25\pm0.06$ &  $1.55\pm0.08$ \\
S4-169  &  $12.33\pm0.05$ &  $1.95\pm0.05$ &  $1.31\pm0.07$ \\
S4-170  &  $12.74\pm0.05$ &  $2.15\pm0.05$ &  $1.45\pm0.08$ \\
S4-176  &  $14.14\pm0.07$ &  $2.21\pm0.06$ &  $1.22\pm0.09$ \\
S4-180  &  $11.42\pm0.05$ &   $2.7\pm0.06$ &  $1.75\pm0.07$ \\
S4-182  &  $14.46\pm0.05$ &  $2.33\pm0.06$ &   $1.7\pm0.07$ \\
S4-197  &  $12.38\pm0.05$ &  $2.54\pm0.05$ &  $1.74\pm0.08$ \\
S4-2    &  $10.96\pm0.04$ &  $2.36\pm0.05$ &   $1.7\pm0.06$ \\
S4-22   &  $13.93\pm0.05$ &  $2.34\pm0.05$ &   $1.4\pm0.07$ \\
S4-23   &  $14.33\pm0.05$ &  $2.39\pm0.06$ &  $1.47\pm0.07$ \\
S4-236  &  $12.97\pm0.05$ &  $2.24\pm0.05$ &  $1.62\pm0.07$ \\
S4-262  &  $14.42\pm0.05$ &  $2.12\pm0.05$ &  $1.27\pm0.06$ \\
S4-273  &   $14.18\pm0.1$ &  $2.25\pm0.06$ &  $1.45\pm0.12$ \\
S4-279  &  $13.73\pm0.06$ &  $2.31\pm0.06$ &  $1.62\pm0.07$ \\
S4-280  &  $14.35\pm0.04$ &  $2.39\pm0.05$ &  $1.25\pm0.07$ \\
S4-283  &  $14.59\pm0.07$ &  $2.31\pm0.06$ &   $1.3\pm0.09$ \\
S4-284  &  $13.94\pm0.07$ &   $2.2\pm0.06$ &  $1.84\pm0.08$ \\
S4-287  &  $12.29\pm0.06$ &  $2.14\pm0.05$ &  $1.42\pm0.08$ \\
S4-3    &  $11.55\pm0.03$ &  $2.17\pm0.05$ &  $1.42\pm0.07$ \\
S4-31   &  $14.53\pm0.06$ &  $2.31\pm0.05$ &  $1.09\pm0.08$ \\
S4-312  &  $13.94\pm0.05$ &  $2.23\pm0.06$ &  $1.38\pm0.07$ \\
S4-314  &  $14.17\pm0.05$ &  $2.22\pm0.06$ &  $1.24\pm0.07$ \\
S4-319  &  $12.73\pm0.05$ &  $2.23\pm0.06$ &  $1.42\pm0.07$ \\
S4-34   &  $14.27\pm0.05$ &  $2.39\pm0.06$ &  $1.65\pm0.15$ \\
S4-341  &  $14.42\pm0.04$ &  $2.19\pm0.06$ &  $1.55\pm0.05$ \\
S4-342  &  $12.08\pm0.06$ &   $2.6\pm0.08$ &  $1.77\pm0.08$ \\
S4-344  &  $11.01\pm0.05$ &  $2.47\pm0.06$ &  $1.64\pm0.08$ \\
S4-375  &  $13.91\pm0.06$ &  $2.31\pm0.07$ &  $1.36\pm0.09$ \\
S4-40   &   $14.0\pm0.04$ &  $2.32\pm0.05$ &   $1.6\pm0.06$ \\
S4-46   &  $13.72\pm0.05$ &  $1.84\pm0.05$ &  $1.04\pm0.07$ \\
S4-50   &  $14.28\pm0.08$ &  $2.47\pm0.06$ &   $1.61\pm0.1$ \\
S4-55   &  $14.45\pm0.05$ &  $2.29\pm0.06$ &   $1.5\pm0.07$ \\
S4-56   &  $14.15\pm0.06$ &  $2.31\pm0.19$ &  $1.41\pm0.09$ \\
S4-59   &  $12.48\pm0.05$ &  $2.05\pm0.07$ &  $1.37\pm0.09$ \\
S4-63   &  $14.32\pm0.05$ &   $2.0\pm0.06$ &   $1.6\pm0.06$ \\
S4-71   &  $10.93\pm0.04$ &   $2.2\pm0.05$ &   $1.4\pm0.07$ \\
S4-73   &  $14.73\pm0.06$ &  $2.59\pm0.08$ &  $1.57\pm0.08$ \\
S4-74   &  $14.29\pm0.05$ &  $2.31\pm0.05$ &  $1.59\pm0.07$ \\
S4-75   &  $14.79\pm0.05$ &   $3.2\pm0.14$ &  $2.37\pm0.09$ \\
S4-8    &  $14.05\pm0.05$ &  $2.17\pm0.05$ &  $1.29\pm0.07$ \\
S4-86   &  $13.94\pm0.04$ &  $2.07\pm0.05$ &  $1.32\pm0.06$ \\
S4-98   &  $13.12\pm0.04$ &  $2.41\pm0.07$ &  $1.62\pm0.07$ \\
S5-12   &  $12.61\pm0.05$ &  $2.64\pm0.06$ &  $2.29\pm0.09$ \\
S5-131  &  $13.95\pm0.04$ &  $2.34\pm0.06$ &  $1.44\pm0.06$ \\
S5-134  &   $11.9\pm0.04$ &   $2.4\pm0.06$ &  $1.53\pm0.06$ \\
S5-141  &  $13.96\pm0.06$ &  $2.08\pm0.06$ &   $1.4\pm0.08$ \\
S5-146  &  $14.54\pm0.06$ &   $2.5\pm0.07$ &  $1.68\pm0.09$ \\
S5-155  &  $13.41\pm0.08$ &  $2.45\pm0.07$ &    $2.5\pm0.1$ \\
S5-156  &  $13.59\pm0.05$ &  $2.33\pm0.06$ &  $1.64\pm0.07$ \\
S5-165  &  $13.98\pm0.05$ &   $2.2\pm0.06$ &  $1.47\pm0.06$ \\
S5-168  &  $12.84\pm0.06$ &  $2.41\pm0.08$ &  $1.71\pm0.09$ \\
S5-174  &  $12.76\pm0.07$ &  $2.49\pm0.06$ &   $1.6\pm0.09$ \\
S5-175  &  $14.95\pm0.09$ &  $2.44\pm0.06$ &   $1.49\pm0.1$ \\
S5-178  &  $13.85\pm0.05$ &   $2.2\pm0.07$ &  $1.49\pm0.07$ \\
S5-181  &  $14.01\pm0.11$ &  $2.54\pm0.07$ &  $2.02\pm0.13$ \\
S5-183  &  $10.25\pm0.04$ &  $1.93\pm0.06$ &  $1.31\pm0.07$ \\
S5-185  &  $12.55\pm0.04$ &  $2.29\pm0.06$ &  $1.26\pm0.07$ \\
S5-191  &  $11.54\pm0.03$ &  $2.02\pm0.06$ &  $1.33\pm0.05$ \\
S5-199  &   $12.5\pm0.04$ &  $2.52\pm0.06$ &  $1.68\pm0.07$ \\
S5-211  &  $12.25\pm0.05$ &  $1.71\pm0.06$ &  $1.05\pm0.07$ \\
S5-212  &  $14.08\pm0.04$ &  $2.12\pm0.07$ &  $1.29\pm0.07$ \\
S5-213  &  $11.35\pm0.04$ &  $2.23\pm0.07$ &  $1.51\pm0.07$ \\
S5-243  &  $13.29\pm0.04$ &    $2.3\pm0.1$ &  $1.52\pm0.08$ \\
S5-25   &  $13.43\pm0.09$ &  $2.32\pm0.05$ &   $2.23\pm0.1$ \\
S5-270  &   $14.8\pm0.07$ &  $2.57\pm0.05$ &   $1.5\pm0.09$ \\
S5-28   &  $14.36\pm0.07$ &  $2.53\pm0.09$ &   $1.87\pm0.1$ \\
S5-31   &  $13.78\pm0.06$ &   $1.85\pm0.1$ &  $1.22\pm0.08$ \\
S5-34   &   $12.2\pm0.06$ &  $2.45\pm0.06$ &  $1.53\pm0.08$ \\
S5-43   &  $12.43\pm0.08$ &  $2.59\pm0.05$ &    $1.8\pm0.1$ \\
S5-44   &  $14.22\pm0.05$ &  $2.23\pm0.05$ &  $1.41\pm0.07$ \\
S5-55   &  $14.03\pm0.03$ &  $2.35\pm0.05$ &  $1.51\pm0.06$ \\
S5-7    &  $14.66\pm0.04$ &  $2.18\pm0.05$ &  $1.32\pm0.06$ \\
S5-71   &  $12.19\pm0.06$ &  $2.24\pm0.05$ &  $1.69\pm0.08$ \\
S5-79   &  $14.33\pm0.04$ &  $2.21\pm0.06$ &  $1.48\pm0.07$ \\
S5-80   &  $14.08\pm0.06$ &  $2.17\pm0.06$ &  $1.49\pm0.07$ \\
S5-83   &  $13.25\pm0.06$ &  $2.18\pm0.07$ &  $1.42\pm0.08$ \\
S5-89   &   $9.16\pm0.04$ &   $2.4\pm0.06$ &  $1.72\pm0.07$ \\
S5-94   &  $13.82\pm0.07$ &  $2.45\pm0.06$ &   $1.71\pm0.1$ \\
S5-95   &  $11.77\pm0.04$ &  $2.33\pm0.06$ &  $1.57\pm0.07$ \\
S5-98   &  $13.86\pm0.05$ &  $2.11\pm0.05$ &  $1.36\pm0.06$ \\
S5-99   &  $13.49\pm0.05$ &  $2.24\pm0.06$ &  $1.41\pm0.07$ \\
S6-128  &  $13.46\pm0.08$ &  $2.53\pm0.07$ &   $1.65\pm0.1$ \\
S6-181  &  $14.27\pm0.07$ &  $2.28\pm0.07$ &   $1.4\pm0.08$ \\
S6-27   &   $10.8\pm0.05$ &  $2.28\pm0.06$ &  $1.53\pm0.07$ \\
S6-76   &  $12.15\pm0.05$ &  $2.33\pm0.07$ &  $1.52\pm0.07$ \\
S6-77   &  $12.91\pm0.04$ &   $2.2\pm0.06$ &  $1.48\pm0.07$ \\
S6-80   &  $13.94\pm0.05$ &  $2.35\pm0.06$ &   $1.4\pm0.07$ \\
IRS 16NW &    $8.6\pm0.04$ &  $2.05\pm0.06$ &  $1.68\pm0.07$ \\
IRS 33E  &    $8.6\pm0.04$ &   $2.2\pm0.06$ &  $1.66\pm0.07$ \\
\enddata
\end{deluxetable}